\documentclass[12pt]{article}
\usepackage{mheckII}

\usepackage{manfnt}
\usepackage{longtable}

\usepackage{blkarray}

\usepackage{tikz} 
\usetikzlibrary{matrix}

\theoremstyle{theorem}
\newtheorem*{thm}{Theorem}

\newtheorem{fact}{Fact}

\newtheorem*{corl}{Corollary}

\newtheorem*{lem}{Lemma}
\newtheorem*{cla}{Claim}

\theoremstyle{definition}
\newtheorem*{defn}{Definition}
\newtheorem*{rem}{Remark}

\newtheorem{exe}{Example}

\def\Dsl{\,\raise.15ex\hbox{/}\mkern-13.5mu D}
\def\dsl{\,\raise.25ex\hbox{/}\mkern-10.5mu \partial}

\date{October, 2019}

\title{FQHE and $tt^*$ geometry
}

\authors{Riccardo Bergamin\footnote{e-mail: {\tt rbergamin@sissa.it}} and Sergio Cecotti\footnote{e-mail: {\tt cecotti@sissa.it}}\vskip 9pt

\centerline{SISSA, via Bonomea 265, I-34100 Trieste, ITALY}}

\abstract{Cumrun Vafa \cite{cumrun} has proposed a microscopic description of the Fractional Quantum Hall Effect (FQHE) in terms of a many-body Hamiltonian $H$ invariant under four supersymmetries. The non-Abelian statistics of the defects (quasi-holes and quasi-particles) is then  determined by the monodromy representation of the associated $tt^*$ geometry. In this paper we study the monodromy representation of the Vafa 4-\textsc{susy} model. Modulo some plausible assumption, we find that the monodromy representation factors through
a Temperley-Lieb/Hecke algebra with $q=\pm\exp(\pi i/\nu)$ as predicted in \cite{cumrun}. The emerging picture agrees with the other predictions of \cite{cumrun} as well.

The bulk of the paper is dedicated to the development of new concepts, ideas, and techniques in $tt^*$ geometry which are of independent interest. We present several examples of these geometric structures in various contexts.}

\begin{document}

\maketitle
\newpage

\tableofcontents

\vskip40pt

\section{Introduction: the Vafa proposal for FQHE}

The Fractional Quantum Hall Effect (FQHE) describes some peculiar quantum phases of a system of a large number $N$ of electrons moving in a two-dimensional surface $S$ in presence of a strong normal magnetic field $B$ at very low temperature (for background see \cite{background}). These quantum phases are classified by a 
rational number $\nu\in\bQ_{>0}$, called the \emph{filling fraction,} which measures the fraction of states in the first Landau level which are actually occupied by the electrons
\be
\nu=\frac{2\pi\,N}{\Phi},\qquad\Phi=\int_S B\ \ \text{(magnetic flux).}
\ee
The quantum phase for a given $\nu$ is characterized by a specific topological order of the ground state(s). The topological order  is 
captured by the (possibly non-Abelian) generalized statistics of the topological defects
(quasi-holes and quasi-particles)  
which may be inserted at given points $\{w_k\}$ in the surface $S$ where the electrons move.
The generalized statistics of quasi-holes is  the main
object of interest in the theory of such phases.

In principle, the microscopic description of the system is provided by the Schroedinger equation governing the dynamics of the $N$ electrons: 
\be\label{schrcz}
H|\psi\rangle=E_0|\psi\rangle.
\ee  
The system is described in first quantization: the microscopic degrees of freedom entering in the Hamiltonian $H$ are the positions $z_i\in \R^2\cong \C$ of the $N$ electrons,
their conjugate momenta $p_i$, and their   discrete spin d.o.f.\! $s_i$.
The precise details of the Hamiltonian $H$ are unimportant: what matters is that the Hamiltonian under consideration belongs to the correct \textit{universality class}. We say that two gapped Hamiltonians, $H$ and $H^\prime$, belong to the same \emph{strict} universality class if their ground state(s) have the same topological order or, in more technical terms, iff we can find a continuous family of interpolating Hamiltonians $H(t)$ for $t\in[0,1]$ such that
\be 
H(0)=H,\qquad  H(1)=H^\prime,\qquad \text{and $H(t)$ is gapped for all $t\in[0,1]$.}
\ee
The dependence of $H$ on some continuous parameters is however interesting, even if their deformation does not close the gap and leaves the Hamiltonian in the same strict topological class. A basic example are the positions $w_k\in\R^2$ where one inserts the defects. If we keep track of the dependence on these parameters in solving the Schroedinger equation \eqref{schrcz}, we may follow how the ground state(s) change when we take one defect around another, thus determining their generalized statistics. Morally speaking, the solution to
\eqref{schrcz} defines a connection on the space of defect configurations, and parallel transport
along closed loops in this space defines the general statistics.\footnote{\ This is quite rough. In general, parallel transport defines a holonomy which depends on the actual loop, not just on its homotopy class. To get a cleaner definition of the generalized statistics one should be able to show that the relevant connection is flat, so that the generalized statistics coincides with its monodromy representation. 
The existence of a flat connection holds automatically in the Vafa context, and in facts it was one of the motivations for the proposal of \cite{cumrun}.}
Then, out of the infinite-dimensional space of possible deformations of the Hamiltonian $H$, all of which locally preserve the energy gap\footnote{\ Being gapped is an open condition in parameter space.}, there is a finite-dimensional sub-space of deformations which may be used to probe the quantum order; the corresponding couplings 
 are essential to understand the nature of the topological phase. All other couplings are pretty irrelevant, and we are free to deform them in any convenient way in order to make the analysis easier.

Thus, pragmatically, a microscopic description consists of a family of (gapped) Hamiltonians $H(w_k)$ for the $N$ electron system, where the $w_k$ are the essential parameters which take value in some essential coupling space $\mathscr{X}$.  $H(w_k)$ is unique up to an equivalence relation given by arbitrary deformations of all inessential parameters while preserving the gap.

In a given FQHE topological phase, from 
 the dynamics of the microscopic degrees of freedom there emerges at low-energy  an effective 2d QFT $\cq$ for the (non-local) quasi-hole ``field'' operators
$h(w)$; the topological phase is then captured by the braiding properties
of their multi-point correlators 
\be\label{opppq}
\big\langle h(w_1)\,h(w_2)\cdots h(w_n)\big\rangle_\cq,
\ee 
as we transport the $h(w)$'s around each other in closed loops. One of the goals of the theory is to understand
the effective QFT of quasi-holes for a given value of the filling fraction $\nu$.
\medskip

Starting from M-theory considerations, Vafa \cite{cumrun} puts forward the remarkable proposal  
that the relative universality class of Hamiltonian families which describes  FQHE with given filling fraction $\nu$ contains explicit  families $\{H(w)\}_{w\in\mathscr{X}}$ which are invariant under  extended supersymmetry with four-supercharges (4-\textsc{susy}).
As we review in \S.\,\ref{basictt*},
this means that the action of the braid group $\cb_n$ on the
topological defects $h(w_j)$ coincides with the monodromy representation of the flat connection 
of the 4-\textsc{susy} supersymmetric Quantum Mechanics (SQM), and then 
the topological order of the FQHE system may be
studied with the powerful tools of $tt^*$ geometry \cite{tt*,ising,CV92,twistor,tt*3}.
\medskip

The purpose of the present paper is to study the $tt^*$ monodromy representation of the 4-\textsc{susy} SQM Hamiltonians which represent the FQHE relative universality classes, and 
determine the properties of their quantum topological phase. The observables one computes this way may potentially be tested in actual experiments in the laboratory.
\medskip

Before going to that, in section 2 we argue from the first principles of Quantum Mechanics that the Vafa Hamiltonian is the physically correct one to describe many electrons, moving in a plane which interact with each other, in presence of a parametrically large magnetic field.
\medskip  

Ref.\!\cite{cumrun} discusses FQHE from several viewpoints besides the microscopic one based on the 4-\textsc{susy} SQM model, all of them inspired by M-/string theory consideration. The results of the effective approach in \S.\! 2 of \cite{cumrun} then constitute predictions of the results one is expected to obtain from the microscopic description (\S.\! 3 of 
\cite{cumrun}). In this paper we get full agreement with Vafa expectations:
the way they arise from the microscopic theory looks quite elegant and deep 
from the geometrical side. 
We find that the 
4-supercharge Hamiltonian proposed by Vafa
describes FQHE for the following series of filling fractions\footnote{\ Warning: the table may contain repetitions, i.e.\! one $\nu$ may appear more than once.} $\nu$:
\be
\begin{tabular}{c|c|c}\hline\hline
$\nu$ & & \\\hline
$\frac{m}{2m\pm 1}$ & $m\geq 1$ & principal series\\
$\frac{m}{m+2}$ & $m\geq 2$ & Moore-Read-Rezayi series\\
$\frac{m}{2(m\pm 1)}$ & $m\geq3$ odd\\
$\frac{m}{3m-2}$ & $m\geq2$\\\hline\hline
\end{tabular}
\ee   

\medskip

Our results are ``exact'' in the sense that no asymptotic limit is implied: we do not assume any particular regime of the discrete or continuous parameters of the quantum model besides the defining assumption of the FQHE that the magnetic field $B$ is parametrically large.
Our computations do not rest on some approximation scheme, but on subtle general properties of $tt^*$ geometry and some plausible assumption. While the geometric statements are beautiful, plausible, and supported by explicit examples, the arguments we present fall short of being proofs.

\medskip

The bulk of the paper is devoted to the study of advanced topics in $tt^*$ geometry required for the analysis of FQHE. Most of these developments have not appeared before in print, and some look rather surprising.
In this direction there is still work to do.

\medskip

The idea that the IR physics of some concrete physical system, actually realized in the laboratory - as the FQHE materials - does have a microscopic description in terms of a Lagrangian with extended supersymmetry may seems rather odd at first. In itself supersymmetry is not a problem since, for a gapped \textsc{susy} system, the supercharges just vanish in the IR sector. But \emph{extended} supersymmetry is a subtler story.
There are obvious obstructions to the uplift of
the IR sector of a gapped quantum system to a 4-\textsc{susy} Hamiltonian model.  
We conclude this introduction by showing that these obstructions are avoided in the FQHE case. This is quite remarkable in its own right. In \S.\,\ref{physics} we shall give detailed arguments to the effect that the real-world microscopic FQHE Hamiltonian $H_\text{FQHE}$
does have a \emph{canonical}  4-\textsc{susy} uplift of the form proposed by Vafa.

\subparagraph{Obstructions to {\rm 4-\textsc{susy}} uplift}\label{jjasqwi0}

The Lagrangian of a 4-\textsc{susy} SQM is the sum of two pieces, called the $D$-term and the $F$-term.
Couplings entering in the $D$-term are inessential for the IR sector, but there are finitely many $F$-term couplings which do are essential in the IR: they take value in some manifold\footnote{\ Basic $tt^*$ geometry is reviewed in \S.\,3 below. In that section $\mathscr{X}_{tt^*}$ is written simply $\mathscr{X}$.} $\mathscr{X}_{tt^*}$ (the $tt^*$ manifold). Therefore, in order to have a 4-\textsc{susy} uplift, our quantum system should satisfy
a necessary condition:  its essential coupling space $\mathscr{X}$
should match 
the $tt^*$ one $\mathscr{X}_{tt^*}$.
This is a formidable restriction since $\mathscr{X}_{tt^*}$ is a very special kind of manifold: \textit{a)} it is a complex analytic space, \textit{b)} it admits a complete K\"ahler metric with a global K\"ahler potential, and \textit{c)} if all infinitesimal deformations are unobstructed, $\mathscr{X}_{tt^*}$ has the structure of a Frobenius manifold \cite{frob}.
The fact that the essential parameters of FQHE satisfy all these peculiar conditions looks quite remarkable in itself, and gives confidence on the proposal put forward by Vafa. While this  correspondence may look quite unlikely at first, it is pretty  natural from the M-theory perspective \cite{cumrun}.
More direct physical arguments to believe Vafa's supersymmetric picture is correct will be discussed in section \S\,.\ref{physics}.

\subparagraph{Organization of the paper.} The paper is organized as follows: 
in section \ref{physics} we shall discuss the physics of the FQHE and present the reasons to believe in the 4-supercharges  description. Here we fill in the details of various deep arguments sketched in \S.\,3 of \cite{cumrun}. In section \ref{basictt*} we review the basics of $tt^*$ geometry mainly to fix the language and notation.
In section \ref{advanced1} we introduce a first block of new developments in $tt^*$ geometry: here the focus is on the natural and deep interconnection between $tt^*$ geometry and subjects like the Knizhnik-Zamolodchikov equation \cite{KZ}, (Iwanori-)Hecke algebras \cite{braid}, the Gaudin integrable model \cite{gaudin} and all that. Section \ref{advanced2} contains a second block of special $tt^*$ topics: here we consider the interplay between $tt^*$ geometry and statistics from the viewpoint of $tt^*$ functoriality, and connect these issues to the Heine-Stieltjes theory. In this section we also introduce the notion of \emph{$tt^*$ dualities,} i.e.\! correspondences between different looking quantum systems with four supercharges which have \emph{identical} $tt^*$ geometry (i.e.\! same brane amplitudes, metrics, new indices \emph{etc.}). In section \ref{vvvmmm} the ideas developed in \S\S.\,\ref{advanced1},\ref{advanced2} are applied to the Vafa model of FQHE to get the monodromy representation we look for.
We present our conclusions in section 7.

\section{The Vafa model vs.\! the microscopic physics of FQHE} \label{physics}

The fractional quantum Hall effect arises from the quantum dynamics of a large number $N$ of electrons moving in a two-dimension surface $S$ subject to a strong  external  magnetic field $B$. In principle the quantum physics may be determined by solving the Schroedinger equation \eqref{schrcz} for the many electron system.
The actual Hamiltonian $H$ contains a large number of degrees of freedom, it is involved and poorly known, so the direct approach from the microscopic side may seem totally hopeless. However, as far as the \textit{\underline{only}} observables we wish to compute from the Schroedinger  equation \eqref{schrcz} are the ones which control the topological order of the quantum phase, the problem becomes tractable under some mild assumptions.

\subsection{Generalities}

The basic assumption is that the strong magnetic field is really strong, so that there is a parametrically large energy-gap between the low-lying energy levels and the rest of the Hilbert space $\boldsymbol{H}$. More precisely, the Hamiltonian is
assumed to have the schematic form
\be\label{genH}
H=\sum_{i=1}^N\left(\frac{1}{2m} \big |\vec p_i-e\,\vec A(\vec y_i)\big|^2+g\, \vec B\cdot\vec \sigma_i+\text{const}\right)+H_\text{int}\equiv H_B+H_\text{int}
\ee
where $\vec y_i$ is the position of the $i$-th electron in the plane $\R^2\cong\C$
(we shall set $z_i\equiv y_{i,1}+iy_{i,2}$),
$\vec p_i$ its conjugate momentum, $\vec\sigma_i$ its spin d.o.f., and $\vec A$ the background gauge field $\nabla \times \vec A=\vec B$.
The interacting Hamiltonian $H_\text{int}$ describes all other interactions; its crucial property is that it is $O(1)$ as $B\to\infty$. The additive constant in the large parenthesis is chosen so that the ground state energy vanishes.

We assume $\nu\leq1$. Let $\boldsymbol{H}_\Phi\subset\boldsymbol{H}$ be the subspace of the Hilbert space consisting of states whose energy is bounded in the limit $B\to\infty$; the orthogonal complement $\boldsymbol{H}_\Phi^\perp$ is separated from $\boldsymbol{H}_\Phi$ by a large $O(B)$ energy-gap. One has
\be
\dim \boldsymbol{H}_\Phi={\Phi/2\pi\choose N}={N/\nu\choose N}.
\ee
Note that in 
$\boldsymbol{H}_\Phi$ the electrons are polarized and the spin d.o.f.\! get frozen in their Clifford vacua. Thus, if we are only interested in the physics at energies $\ll B/m$ we may forget these degrees of freedom. Acting on the vector space $\boldsymbol{H}_\Phi$, the operator $H_B$ is identically zero; we are reduced to a quantum system with
a finite-dimensional Hilbert space $\boldsymbol{H}_\Phi$ with Hamiltonian 
$H_\text{eff}=P_\Phi H_\text{int}P_\Phi$ where $P_\Phi$ is the projection on $\boldsymbol{H}_\Phi$. The fact that $\boldsymbol{H}_\Phi$ is finite-dimensional is not a significative simplification (unless $\nu=1$), since in realistic situations the dimension of the ``small'' space $\boldsymbol{H}_\Phi$  is something like $10^{10^{14}}$  and gets  strictly infinite in the thermodynamic limit.

To proceed forward one needs new physical insights. In ref.\!\!\cite{cumrun}
two novel ideas were proposed\footnote{\ Cfr.\! the discussion at the end of \S.\,3.3 in \cite{cumrun}.}:
\begin{itemize}
\item[1.] the low-lying Hilbert space $\boldsymbol{H}_\Phi$ is isomorphic to the space of supersymmetric vacua of a certain 4-\textsc{susy} SQM model;
\item[2.] the SQM system has a unique preferred vacuum $|\textsf{vac}\rangle$  which is identified with the vacuum of the physical FQHE under the isomorphism in 1.
\end{itemize}

Our first goal is to flesh out the above two ideas in some detail.

\subsection{Charged particles in magnetic fields
$\equiv$ 4-supercharge {\rm\textsc{susy}}}
\label{basisoxxx}

%

\subsubsection{Electrons in a finite box with magnetic flux} 
 
To get a clean problem, we work in a finite box, i.e.\! we replace the plane $\C$ in which the electrons move with a very large flat 2-torus $E$. The complex structure $\tau$ on the elliptic curve $E$ is immaterial in the infinite volume limit:  we fix it to any convenient value.
Also the spin structure is irrelevant; it is convenient to pick up an \emph{even} one,\footnote{\ \textbf{Notation and conventions:} \textbf{1)} $\Gamma(X,V)$ stands for the space of holomorphic sections of the coherent sheaf $V$ on the complex space $X$. $\co$ is the structure sheaf of $X$ and $\mathscr{M}$ the sheaf of germs of meromorphic functions. An asterisque denote the sub-sheaf of invertible elements of the given sheaf. \textbf{2)} If $D=\sum_i n_i p_i$ is a divisor on a \emph{smooth} curve $X$, we fix a Cartier representative of it, i.e.\! we take a sufficiently fine open cover $\{U_i\}$ of $X$
and fix $\psi_{0,i}\in\Gamma(U_i,\mathscr{M}^*)$ such that
$\psi_{0,i}/\psi_{0,j}\in \Gamma(U_i\cap U_j, \co^*)$.
We write $\co(D)$ for the associated line
 bundle ($\equiv$ invertible sheaf) with transition functions $\psi_{0,i}/\psi_{0,j}$.
The \textit{defining section} $\psi_0$ of   $\co(D)$ is the one given by $\psi_{0}|_{U_i}=\psi_{0,i}$. We write $\sim$ for linear equivalence of divisors.} 
$\co(S)$, associated to a divisor $S=p_0-q$
 where $p_0,q\in E$ are distinct points which satisfy $2p_0=2q$. We are free to translate $p_0\in E$ according convenience.

 In a holomorphic gauge, $A_{\bar z}=0$, an Abelian gauge field $A$ on $E$ is determined by two data: \textit{i)}
a holomorphic line bundle $\cl\to E$ with Chern class $c_1(\cl)=\Phi/2\pi$, where $\Phi>0$ is the magnetic flux through the surface $E$,
and \textit{ii)} a Hermitian metric $h$
on the fibers of $\cl$. Locally
\be
eA_z=h^{-1}\partial_z h,\qquad eA_{\bar z}=0.
\ee
In such a holomorphic gauge, the low-lying wave functions $\psi$ of the one-particle Hamiltonian\footnote{\ We do not write the spin d.o.f.\! since their are frozen in their vacua.} 
\be
H_\text{1-par}=\frac{1}{2m} \big |\vec p_i-e\,\vec A\big|^2+\text{const}
\ee
 are simply the  holomorphic sections of the line bundle $\cl$ twisted by $\co(S)$,
\be
\boldsymbol{H}_{\Phi,\text{1-par}}=\Gamma(E,\cl(S)),\qquad \dim \boldsymbol{H}_{\Phi,\text{1-par}}=
\deg\cl(S)=\frac{\Phi}{2\pi},
\ee
and for the $N$ electron system
\be
\boldsymbol{H}_{\Phi}=\bigwedge\nolimits^{\!N}\Gamma(E,\cl(S)),\qquad \dim \boldsymbol{H}_{\Phi}=
{\Phi/2\pi\choose N}.
\ee
In this gauge the low-lying wave-function $\Psi$ are independent of $h$; this does not mean that $h$ is irrelevant for the low-energy physics, because the inner product in the space $\boldsymbol{H}_\Phi$
\be
\langle\Psi_1|\Psi_2\rangle=\int_E h^N\,\Psi_1^*\,\Psi_2\;d\upsilon,
\ee
depends on $h$.
 \medskip

 To be very explicit, we choose an effective divisor $D=\sum_{i=1}^\ell n_i p_i$ such that $\cl=\co(D)$. Then $\cl(S)=\co(D+S)$.
 The divisor $D+S$, unique up to linear equivalence, has a defining {meromorphic} section $\psi_0$
with a zero of order $n_i\geq1$ at each point $p_i\in E$, a simple zero at $p_0$, and no other zeros. In addition, $\psi_0$ has a single pole at $q$ and no other poles. Because of the pole $\psi_0\not\in\Gamma(E,\co(D+S))$.

The map $\psi\mapsto \psi/\psi_0\equiv \phi$ sets an isomorphism
\be
\begin{split}
\boldsymbol{H}_{\Phi,\text{1-par}}&\equiv\Gamma\!\big(E,\co(D+S)\big)\xrightarrow{\!\sim\!}\\
&\xrightarrow{\!\sim\!}
\Big\{\text{$\phi\in\Gamma(E,\mathscr{M})$ with polar divisor $D_\infty\leq D+p_0$ vanishing at $q$}\Big\}.
\end{split}
\ee
Composing with the map\footnote{\ Here $z_i$ is a local parameter at $p_i\in E$, and $P\!P_p(\phi)$ stands for the \emph{principal part} of the meromorphic function $\phi$ at $p\in E$.}
\be
\phi\longmapsto \Big\{ z_1^{n_1}\, P\!P_{p_1}(\phi), z_2^{n_2}\, P\!P_{p_2}(\phi),\cdots, z_\ell^{n_\ell}\, P\!P_{p_\ell}(\phi)\Big\}, 
\ee
we get the linear isomorphism
\be\label{lllasqw}
\boldsymbol{H}_{\Phi,\text{1-par}}\xrightarrow{\!\sim\!} \prod_{i=1}^\ell  \C[z]\big/(z^{n_i}).
\ee

 \subsubsection{Magnetic system $\to$ 4-{\rm\textsc{susy}} SQM: the linear isomorphism}\label{isommmr}

On the other side of the correspondence,
we consider a 4-\textsc{susy} SQM with a single chiral field $z$ taking value in $\ck\equiv E\setminus \mathsf{supp}\, F$,
where $E$ is the elliptic curve on which the electrons move,
$dz$ is a holomorphic differential on $E$, and $F$ an effective divisor.\footnote{\  If $\mathsf{supp}\, F\neq\emptyset$, the target space $\ck$ is Stein \cite{curvestein}. This ensures that the elements of the chiral ring $\mathscr{R}$ may be represented (non-uniquely) by global holomorphic functions
\cite{iosqm}, see also \textbf{Hilfssatz C} in 
\cite{curvestein2}. The results of the latter paper imply that these nice properties hold  even when $\dim\mathscr{R}=\infty$ (i.e.\! for \emph{infinite} degree divisors) a fact we shall need in \S.\,\ref{covR} (for an exposition of these results, see \S.\,26 of \cite{Rbook}). } 
We choose the one-particle  superpotential\footnote{\ We stress that we require only the derivative $W^\prime$ to be univalued in $\ck$, \underline{not} the superpotential $W(z)$ itself which is typically multivalued.} $W(z)$ such that its derivative, $W^\prime(z)$, is a meromorphic function on $E$ whose zero-divisor $D\equiv\sum_{i=1}^\ell n_i p_i$ is the one describing the magnetic background in which the electrons move. The polar divisor of $W^\prime(z)$ is $F\sim D$. In making the dictionary between the two quantum  models, we  
use our freedom in the choice of $p_0$ to set $p_0\in \mathsf{Supp}\,F$, i.e.\! $p_0\not\in\ck$.

By the Chinese remainder theorem,\footnote{\ \label{f12}We stress that the ring of holomorphic functions on a one-dimensional Stein manifold is a Dedekind domain. Then (say) \textbf{Theorem 4} of \cite{ANT} applies.} the chiral ring $\mathscr{R}$ of this 4-\textsc{susy} model is 
\be
\mathscr{R}\cong \prod_{i=1}^\ell  \C[z]\big/(z^{n_i}).
\ee
Comparing with \eqref{lllasqw} we get
\be
\boldsymbol{H}_{\Phi,\text{1-par}}\cong \mathscr{R},\qquad \boldsymbol{H}_{\Phi}\cong\bigwedge\nolimits^{\!N}\mathscr{R}
\ee
as vector spaces.
On the other hand, in a 4-\textsc{susy} theory we have a linear isomorphism between the chiral ring $\mathscr{R}$ and the space $\mathscr{V}$ of \textsc{susy} vacua \cite{Lerche:1989uy}\!\!\cite{iosqm}. Composing the two we get a natural isomorphism between the low-lying states of the two quantum systems
\be
\boldsymbol{H}_{\Phi,\text{1-par}}\cong\mathscr{V},\qquad \boldsymbol{H}_{\Phi}\cong \mathscr{V}_N\equiv \bigwedge\nolimits^{\!N}\mathscr{V}.
\ee 
At the level of explicit Schroedinger wave-functions the isomorphism reads
(for the one-particle theory)
\be\label{whatsayyss}
\psi\mapsto \psi_\text{susy}\equiv\frac{\psi}{\psi_0}\,dW+\overline{Q}(\text{something}),
\ee
where in the \textsc{rhs} we wrote the supersymmetric wave-functions as differential forms on $\ck$, as it is customary
\cite{Witten:1982im}\!\!\cite{iosqm}. $\overline{Q}$ is a nilpotent supercharge, $\overline{Q}^2=0$, which acts in the Schroedinger representation as the differential operator
 \cite{iosqm}
 \be
 \overline{Q}=\overline{\partial}+dW\wedge.
 \ee 
 The space of \textsc{susy} vacua $\mathscr{V}$ (and $\mathscr{R}$) is isomorphic to the $\overline{Q}$-cohomology with $L^2$-coefficients.

Eqn.\eqref{whatsayyss} says that, up to a boring  factor, the low-lying wave-functions for the original magnetic system and the ones for the 4-\textsc{susy} SQM models are \underline{identical} in $\overline{Q}$-cohomology.  To see that \eqref{whatsayyss} is an isomorphism 
note that the elliptic function $\psi/\psi_0$ is holomorphic for $\psi\in\Gamma(E,\co(D+S))$ if and only of it is identically zero, that is, the \textsc{rhs} of \eqref{whatsayyss} is $\overline{Q}$-exact iff $\psi=0$.
The identification of the actual Schroedinger wave-functions on the two sides of the correspondence, if not fully canonical, is pretty  natural.

\subsubsection{Motion in the plane}\label{motplkane}

When the electron moves on $\C$ instead of a torus, the corresponding 4-\textsc{susy} SQM is defined by a one-form $dW$ which is a rational differential  on $\bP^1$ with a pole of order $\geq 2$ at $\infty$
\be
dW(z)=\frac{\prod_i(z-z_i)^{n_i}}{P(z)}\,dz,\qquad D=\sum_i n_i z_i,\quad \deg P(z)\leq \sum_i n_i=\frac{\Phi}{2\pi}.
\ee
With this prescription on the behaviour at $\infty$, the scalar potential
$|W^\prime|^2$ is bounded away from zero at infinity for all complete K\"ahler metrics on $\bP^1\setminus\{\infty\}$. This makes the quantum  problem well-defined in the following senses:
\begin{itemize}
\item[A.] if we consider the 2d (2,2) Landau-Ginzburg model with superpotential $W(z)$, this condition guarantees the absence of   
run-away vacua;
\item[B.] if we consider the 1d 4-\textsc{susy} SQM obtained by dimensional reduction from the above 2d model,
it guarantees the presence of a finite energy-gap, and also normalizability of the vacuum wave-functions.
\end{itemize}
We mentioned both 2d and 1d models since the $tt^*$ geometry is the same for the two theories \cite{tt*}, and it is convenient to pass from one language to the other, since some arguments are more transparent in 2d and some other  in 1d.
\medskip

 The minimal regular choice is $dW$ having a double pole at $\infty$; we shall mostly focus on this case\footnote{\ The double pole at $\infty$ just compensates the non-trivial canonical divisor $K_{\bP^1}$, so effectively cancels the curvature of $\bP^1$.}.
 The same argument as in the torus geometry  gives the linear isomorphism $\boldsymbol{H}_\Phi\cong\mathscr{V}$
 also in the plane. The magnetic flux is
 $2\pi \deg D$, $D$ being the zero divisor of $dW$. One writes the spin structure in the form $\co(-q)$ for some reference point $q\not\in\mathsf{Supp}\,D\cup\{\infty\}$. The low-lying magnetic wave-functions are $\psi\in\Gamma(\bP^1,\co(D-q))$, 
 $\dim\Gamma(\bP^1,\co(D-q))=\Phi/2\pi$. 
 \medskip
 
In conclusion: for $N$ non-interacting electrons in presence of a magnetic field, the low-lying Hilbert space is
\be\label{jjjjza12}
\boldsymbol{H}_\Phi\cong \wedge^N \mathscr{R}\cong \mathscr{V}_N.
\ee  
This is a mere linear isomorphism:  the Hermitian structures on the two sides of the correspondence depend on additional data:
in the original magnetic system on the fiber metric $h$, while in the 4-\textsc{susy} SQM on the detailed form of $dW(z)$ which determines the ground-state Hermitian metric through the $tt^*$ equations \cite{tt*}.
Our next task is to find the explicit form of $dW(z)$ which best mimics the Hilbert structure of $\boldsymbol{H}_\Phi$ for the magnetic system.

\subsubsection{Comparing Hermitian structures on $\boldsymbol{H}_\Phi$}\label{herhermm}

For simplicity, we consider a single electron moving in $\C\equiv\bP^1\setminus\{\infty\}$ in presence of a strong magnetic field $B$ macroscopically uniform along the surface. The extension to the case of $N$ electrons is straightforward.

\medskip

 In the magnetic side, the Hermitian structure is defined by the fiber metric  $h=e^{-B|z|^2}$, so that in a unitary gauge
the low-level wave functions read
\be\label{lauherm}
\psi(z)_\text{uni}=\psi(z)_\text{holo}\,e^{-B|z|^2/2}\qquad B>0.
\ee
In the 4-\textsc{susy} side we have the rational differential $dW(z)$ with $\Phi/2\pi$ zeros and a polar divisor of the form $F=F_f+2\infty$. 
Generically, such a differential has the form
\be\label{xxxkkksa}
dW(z)=\left(\mu+\sum_{i=1}^{\Phi/2\pi}\frac{a_i}{z-\zeta_i}\right)\!dz,\qquad F_f=\sum_i\zeta_i
\ee
with $a_i\in\C^\times$ and $\zeta_i\in\C$ all distinct. 

An exact identification of the microscopic Hilbert space structures is a requirement a bit too strong. 
We content ourselves with equality after averaging over small but macroscopic domains $U\Subset \C$. In the present context $U$ being \textit{macroscopic} means $\int_U B/2\pi\gg1$. This weaker condition is all we need if we are interested only in predicting long-wave observables of the kind which characterize the quantum topological order.

Let $U\Subset \C$ be such a domain. For $B$ and $\mu$ large,
\be\label{kz1o}
\frac{1}{2\pi}\,\big(\text{magnetic flux through $U$}\big)\approx
\#\big(\text{\textsc{susy} classical vacua in $U$}\big)
\approx
 \#\big\{\zeta_i\in U\big\},
\ee    
so a large macroscopically uniform $B$ corresponds (non surprising) to a roughly homogeneous distribution in $\C$ of the points $\zeta_i$; the domain $U$ is macroscopic iff it is much larger than the typical separation of the $\zeta_i$'s. 
After taking the $\zeta_i$ to be regularly distributed in the plane,
matching the Hermitian structures on the two sides of the correspondence  boils down to fixing the residues $a_i$ of $dW$ so that the probability of finding the electron in the macroscopic domain $U\Subset \C$ in the original magnetic system is the same as in the supersymmetric model.

It is clear that a homogenous field should correspond to the residues being all equal. By a rotation of the Grassman coordinates $\theta$ we may assume the $a_i$ to be all real.

In the magnetic system the probability of finding the electron at $z$ is 
\be\label{pro1}
\big|\psi(z)_\text{uni}\big|^2=e^{-B|z|^2+\text{subleading as $|z|\to\infty$}}.
\ee 
The \textsc{susy} wave-functions have the form \cite{iosqm} \be\label{mmmnbvz}
\psi_\text{susy}= \Phi(z)\,dz+\widetilde{\Phi}(z)\,d\bar z,
\ee
and the probability distribution is
\be\label{pro2}
\big|\Phi(z)\big|^2+\big|\widetilde{\Phi}(z)\big|^2.
\ee
The two probabilities \eqref{pro1} and \eqref{pro2} should agree when averaged over a macroscopic region $U$. Let us give a rough argument suggesting that this holds iff $a=\pm1$.
We can choose a ``real'' basis of vacua such that the two terms in \eqref{pro2} are equal. Then eqn.\eqref{pro1} yields
\be\label{aaad12}
\log\!\big|\Phi(z)\big|^2=-B\,|z|^2+\text{subleading as $|z|\to\infty$.}
\ee
From the Schroedinger equation of the supersymmetric system, one has \cite{iosqm,Cecotti:1989gv}
\be
\left(-\frac{\partial^2}{\partial \bar z\,\partial z}+ \left|\frac{dW}{dz}\right|^2\right)\!\frac{\Phi}{W^\prime}=0.
\ee
A possible large-field asymptotics consistent with this equation is 
\be\label{comp222}
\Phi(z)= \exp\!\Big(\pm 2\,\mathrm{Re}\,W(z)+\text{subleading}\Big), \quad\text{as $|z|\to\infty$}
\ee
provided the function in the \textsc{rhs}: \textit{a)} is univalued in the large $|z|$ region, and \textit{b)} it goes to zero rapidly at infinity, so that $\psi_\text{susy}$ has a chance to be normalizable. For a superpotential as in eqn.\eqref{xxxkkksa} with $a_i\in\R$ for all $i$ the first condition holds
\be
2\,\mathrm{Re}\,W(z)= \mu z+\bar \mu\bar z+\sum_i a_i \log|z-\zeta_i|^2.
\ee
This function is the electrostatic potential of a system of point charges of size $a_i$ at positions $\zeta_i$ superimposed to a constant background electric field $\bar\mu$.
When averaged over a macroscopic region $U$, it looks like the potential for a continuous charge distribution with density $\sigma(z)$ such that
\be
\int_U d^2z\;\sigma(z)= \sum_{\zeta_i\in U} a_i,\qquad \text{for all }U\subset\C.
\ee
Comparing eqns.\eqref{aaad12},\eqref{comp222} we conclude that,
for all macroscopic domain $U\Subset \C$,
we have\footnote{\ To get the factors 2 right, recall that $i\,dz\wedge d\bar z=2\,dx\wedge dy$ is \emph{twice} the volume form on $\R^2$.}
\be\label{mmmafgnet}
\text{(magnetic flux through $U$)}=
\frac{i}{2}\int_U\bar \partial \partial \log|\Phi|^2 \approx
\pm i \int_U \bar\partial\partial \big(2\,\mathrm{Re}\,\cw\big)=\mp 2\pi \sum_{\zeta_i\in U}a_i,
\ee
where in the last equality we used the
Poisson equation of electrostatics\footnote{ Or the Poincar\'e-Lelong formula \cite{GH}.}. Comparing eqn.\eqref{mmmafgnet} with eqn.\eqref{kz1o}, which also should be true for all macroscopic domain $U$, we get that either all $a_i=-1$ or all $a_i=+1$, the two possibilities being related by a change of orientation. We fix conventions so that the external magnetic field is modelled in the \textsc{susy} side by \eqref{xxxkkksa} with $a_i=-1$ for all $i$.

\subsubsection{Introducing defects}

From the \textsc{susy} side there is a natural way to introduce topological defects in the systems. One flips sign to a small number $h$ of the residues $a_i$. Now there is a small mismatch between the number of vacua and the effective magnetic field as measured by the fall-off of the wave-function at infinity: we have two extra vacua per defect. The extra vacua are localized near the position of the corresponding defect in the plane and may be interpreted as ``internal states'' of the defect.  
We identify these defects with the quasi-holes of FQHE.

\subsubsection{The Vafa superpotential  emerges}

We return to Schroedinger equation with 
Hamiltonian \eqref{genH}. In the large $B$ limit, the low-energy physics is described by a quantum system with 
Hilbert space $\boldsymbol{H}_\Phi$
and Hamiltonian $\widehat{H}\equiv P_\Phi H_\text{int}P_\Phi$. Under 
the isomorphism discussed above,
this system may be seen as a deformation of the 4-\textsc{susy} model
with superpotential $\cw$ the sum of $N$ copies of the above one-particle superpotential, i.e.\! $\cw=\sum_{i=1}^NW(z_i)$. The additional terms in the Hamiltonian describe the interactions between the electrons. We can split these interactions in two groups: the ones which preserve supersymmetry and the ones which do not. The first ones may be inserted in the superpotential $\cw$ (or in the $D$-term, these ones being IR irrelevant). One is led to a superpotential of the form 
\be
d\cw=\sum_{i=1}^N\left(dW(z_i)+\sum_{a=1}^h \frac{dz_i}{z_i-x_a}\right)+\sum_iU_i(z_1,\cdots, z_N)\,dz_i
\ee
where $dW(z_i)$ models the background magnetic field and $x_a$ are the positions of the topological defects.
As a function of the position $z_i$ of the $i$-electron at fixed $z_{j\neq i}$, the meromorphic one-form $U_i\,dz_i$ can have poles only when 
$z_i=z_j$ for some $j\neq i$. Generically $U_i dz_i$ has only simple poles (including at $\infty$): we assume this to be the case. The residues are entire functions bounded at $\infty$, hence constants. Since $\cw$ must be symmetric under permutations of the electrons, the most general superpotential differential is
\be\label{kkkasqweru}
d\cw(z_i;x_a) =\sum_{i=1}^N\left(dW(z_i)+\sum_{a=1}^h \frac{dz_i}{z_i-x_a}\right)+2\beta\!\!\!\sum_{1\leq i<j\leq N} \frac{d(z^i-z^j)}{z^i-z^j},
\ee 
for some complex constant $\beta$.   
The Vafa model \cite{cumrun} has a superpotential of this form.
Ref.\!\cite{cumrun} proposes to model the magnetic field  by
\be\label{kazqw56}
dW(z)_\text{Vafa}=-\sum_{k=1}^{\Phi/2\pi-h} \frac{dz}{z-\zeta_k},
\ee  
where $\zeta_k$ are points forming some regular ``lattice''.
Working on the plane $\C$ we prefer to add a constant contribution to $dW(z)_\text{Vafa}$
\be
dW(z)=dW(z)_\text{Vafa}+\mu\,dz,\qquad \mu\neq0,
\ee 
 in order to get the regularizing double pole at infinity (cfr.\! \S.\,\ref{motplkane}). Note that the added term has no effect on the computations in \S.\,\ref{herhermm}:
indeed, it may be seen as an integration constant for the Poisson equation satisfied by the ``electrostatic potential'' $2\,\mathrm{Re}\,\cw$.

\subsubsection{The coupling $\beta$}
\label{cobeta} 

In Vafa's proposal the last term in eqn.\eqref{kkkasqweru} models the most relevant part of the 2-electron interactions. In the electrostatic language, its real part is proportional to the electron-electron Coulomb potential. From the point of view of the 4-\textsc{susy} model with target space $\C$, the coupling $\beta$ may be any complex number. However, $\beta$ gets quantized to a rational number when we study the model more carefully in a finite box $E$ and we insist that the residues of $d\cw$ have the correct values in \eqref{kkkasqweru}. In this case 
$d\cw$ is a meromorphic one-form in
the K\"ahler space $E^N$; its restriction to the $i$-th factor space at fixed $z_j$ ($j\neq i$) is a meromorphic one-form on the elliptic curve $E$ with single poles of residue $-1$ at the $\zeta_k$, residue $+1$ at the 
$x_a$, and residue $2\beta$ at the $z_{j\neq i}$ and no extra pole at $p_0$.
Since the total residue of a meromorphic one-form vanishes
\be\label{jjjazq1}
0=-\left(\frac{\Phi}{2\pi}-h\right)+h+2\beta (N-1)\approx \big(2\beta\,\nu-1\big)\frac{\Phi}{2\pi},
\ee  
i.e., as $N, \Phi\to\infty$,
\be\label{vafaqua}
2\beta=1/\nu\in \bQ_{>0}
\ee 
which is the value given in ref.\!\cite{cumrun}. If \eqref{jjjazq1} is (exactly) satisfied, $d\cw$ on $E^N$ reads
\begin{gather}
d\cw= \sum_{i=1}^N\left(\sum_a U(z_i;x_a)-\sum_k U(z_i;\zeta_k) +\frac{1}{\nu}\sum_{j\neq i}U(z_i;z_j)\right)\\
\text{where}\quad U(z;w)\equiv \frac{\wp^\prime(w/2)\,dz}{\wp(z-w/2)-\wp(w/2)}.
\end{gather}
In ref.\!\cite{cumrun} the equality $2\beta=1/\nu$ was obtained by comparing the 4-\textsc{susy} brane amplitudes \cite{iqbal} in the (unphysical) asymmetric limit with the Laughlin phenomenological wave-functions \cite{Laughlin}.
However that argument does not fix $\beta$ unambigously\footnote{\ The brane amplitudes in the asymmetric limit have the general form $\int_\Gamma e^{\cw}\phi$ where $\phi$ is a holomorphic $N$-form which represents the (cohomology class of) a \textsc{susy} vacuum.
Clearly we are free to redefine $\phi\to h\phi$ and $\cw\to \cw-\log h$ for $h$ a holomorphic function, leading to an ambiguity in reading the superpotential out of the integral $\int_\Gamma e^{\cw}\phi$. \textbf{Remark:} knowing the set of allowed integration cycles $\Gamma$ reduces (or eliminates) the ambiguity.} since the superpotential is not univalued and one should go to a cover (see \S.\,\ref{nonunivaluedppp}); then  the effective coupling $\beta_\text{eff}$ appearing in the brane amplitudes is a ``renormalized'' version of the superpotential coupling $\beta$  \cite{tt*,twistor,tt*3}.
\vskip10pt

In the rest of this paper we shall work on the plane $\C$ 
and keep $\beta$ generic.
We shall identify the filling fraction $\nu$ with $(2\beta_\text{eff})^{-1}$. 

\subsection{Conclusion of the argument: emergence of a unique vacuum} \label{univacc} 

Under the isomorphism of \S.\ref{isommmr} the many-body Hamiltonian \eqref{genH} takes the form
\be
H=H_\cw+H_\text{su.br.}
\ee 
where $H_\cw$ is the 4-\textsc{susy} Hamiltonian corresponding to the Vafa superpotential $\cw$ in eqn.\eqref{kkkasqweru}
(supplemented by an appropriate $D$-term) while 
$H_\text{su.br.}$ contains the \textsc{susy} breaking interactions. For large magnetic fields the first term is $O(B)$ while the second one is $O(1)$, and hence a small perturbation. However this does not mean that we are allowed to neglect $H_\text{su.br.}$ when studying the quantum topological order of the FQHE system. $H_\text{su.br.}$ lifts the huge degeneration of the ground-states of $H_\cw$ producing 
a unique true vacuum $|\textsf{vac}\rangle$. The FQHE topological order is a property of this particular state.

We may formalize the situation as follows. The zero-energy eigenvectors of the supersymmetric Hamiltonian $H_\cw$ define a vacuum bundle
over the space $\mathscr{X}$ of couplings entering in the superpotential
\be
\mathscr{V}\to\mathscr{X},\qquad
\mathrm{rank}\,\mathscr{V}={\Phi/2\pi\choose N},
\ee
whose fiber $\mathscr{V}_x$ is the space of \textsc{susy} vacua for the model with couplings $x$. $\mathscr{V}$ is equipped with a flat connection $\nabla$ extending the holomorphic Berry connection $D$ (see next section). The quantum topological order of the supersymmetric model $H_\cw$ is captured by the monodromy representation of $\nabla$.

Switching on the interaction $H_\text{su.br.}$ selects one vacuum $|\textsf{vac}\rangle_x\in \mathscr{V}_x$.
The states  $|\textsf{vac}\rangle_x$ span the fibers of a \emph{smooth} line sub-bundle
\be
\mathscr{L}\subset \mathscr{V}.
\ee  
$\mathscr{L}$ is endowed with two
canonical sub-bundle connections, $\nabla^\textsf{vac}$ and $D^\textsf{vac}$,
inherited from $\nabla$ and $D$, respectively.
In general the sub-bundle curvature is quite different from the curvature of the original vector bundle; the discrepancy is measured by the torsion\footnote{\ \textbf{\underline{Notations}:} In this paper $\Lambda^k$ stands for the space of smooth $k$-forms, while $\Omega^k$ for the space of holomorphic ones. We use the same symbols for the corresponding sheaves.} \cite{GH,vhs}
\be
\ct\colon \mathscr{L}\to \Lambda^1\otimes \mathscr{V}/\mathscr{L},\qquad \ct\colon\eta\mapsto \nabla\eta\bmod \Lambda^1\otimes\mathscr{L}.
\ee
Correspondingly, \emph{a priori} the monodromy of
$\nabla^\textsf{vac}$ is neither well-defined nor simply related to the one of $\nabla$. \emph{A priori} there is no simple relation between the quantum order of the FQHE Hamiltonian $H$ and the quantum order of the \textsc{susy} model with Hamiltonian $H_\cw$.
In order to have a useful relation two ``miracles'' should occur: 
\begin{itemize}
\item[\bf M1]
the monodromy representation $\mathscr{V}$ of the flat connection $\nabla$
should be reducible with an invariant sub-bundle $\widetilde{\mathscr{L}}\subset\mathscr{V}$ of rank 1. Then the
4-\textsc{susy} SQM has a unique \textit{preferred} vacuum which spans the fiber $\widetilde{\mathscr{L}}_x$ of the line sub-bundle;  
\item[\bf M2] the physical FQHE vacuum
$|\textsf{vac}\rangle$ is mapped by the isomorphism in \S.\,\ref{isommmr} to the  preferred vacuum of ``miracle''  \textbf{M1} (up to corrections which vanish as $B\to\infty$).
In other words, $\mathscr{L}=\widetilde{\mathscr{L}}$.
\end{itemize}

Whether \textbf{M1} happens or not is purely a question about the supersymmetric model $H_\cw$.
The question may phrased as asking whether 
$H_\cw$ has an unique preferred vacuum. Ref.\!\cite{cumrun} suggests that such a preferred vacuum exists and is the spectral-flow\footnote{\ See \S.\,3 for a review of the spectral-flow isomorphism.} of the identity operator. While this sounds as a natural guess, it is certainly not true that in a general $tt^*$ geometry  the spectral-flow of the identity spans a monodromy invariant subspace of $\mathscr{V}$. That \textbf{M1} holds for the special class of 4-\textsc{susy} models
\eqref{kkkasqweru} appears to be a genuine miracle.

The validity of \textbf{M2} then rests on the fact that the preferred vacuum -- if it exists at all -- is bound to be the most symmetric one. Then one may argue as follows \cite{cumrun}: as long as the \textsc{susy} breaking interaction $H_\text{su.br.}$ is symmetric under permutations of electrons/quasi-holes,  translations and rotations, the true vacuum $|\textsf{vac}\rangle$ will also be the (unique) maximal symmetric one.
\medskip

The conclusion is that -- under our mild assumptions -- the quantum order of the FQHE is captured by the 4-\textsc{susy} SQM model proposed in \cite{cumrun}.

%
%
%
%
%
%
%

\section{Review of basic $tt^*$ geometry}\label{basictt*}

We review the basics of $tt^*$ geometry in a language convenient for our present purposes. Experts may skip to the next section.

\subsection{4-supercharge LG models: vacua and branes}
Even if $tt^*$ geometry is much more general, we describe it in a specific context, namely Landau-Ginzburg (LG) models with four supercharges (4-\textsc{susy}). By a (family of) LG models  we mean the following data:
a Stein manifold\footnote{\ For properties of the Stein spaces see \cite{stein1,stein2,g}\!\!\cite{GH}. We recall that: 1) a non-compact Riemann surface \cite{Rbook} is automatically Stein \cite{curvestein}; 2) all affine varieties are Stein.} $\ck$ and a  family of non-degenerate holomorphic functions
\be
\cw(z;x)\colon \ck\to\C,\qquad z\in\ck\quad x\in\mathscr{X},
\ee
 parametrized holomorphically by a connected complex manifold $\mathscr{X}$ of 
``coupling constants''\footnote{\ It is often convenient to see the $x$'s as a fixed background of additional chiral superfields.} $x$.  \emph{Non-degenerate} means that, for all $x\in\mathscr{X}$, the set of zeros of the differential\footnote{\ $d$ is the exterior derivative in $\ck$. It acts trivially on the constant couplings $x$.} $dW(z;x)$ is discrete in $\ck$; for technical reasons it is also convenient to assume that
the square-norm of the differential $\|dW(z;x)\|^2$ is bounded away from zero outside a (large) compact set (cfr.\! \S.\,\ref{motplkane}).
\medskip 

In the LG model the coordinates\footnote{\ In a Stein manifold, in the vicinity of each point there is a complex coordinate system made of \emph{global} holomorphic functions \cite{stein1,stein2,g}. I.e.\!\! we may choose the chiral fields $z$'s so that they are well-defined quantum operators.} $z$ of $\ck$ are promoted to chiral superfields, and we have a family of Lagrangians of the form
\be\label{lagran}
\mathscr{L}_x=\int\! d^4\theta\, K+\left(\int\! d^2\theta\, \cw(z;x)+\text{h.c.}\right),\quad x\in\mathscr{X}.
\ee  
The details of the $D$-terms are immaterial for us; we  only need that there \textit{exists} some K\"ahler potential $K$ yielding a complete K\"ahler metric: this is guaranteed since $\ck$ is Stein.\footnote{\ $K\colon \ck\to\R$ may be chosen to be a global exhaustion \cite{stein1,stein2,g}.} 

Out of the data $(\ck,\cw)$ we can construct two related 4-\textsc{susy} LG theories: 
a two-dimensional (2d) $(2,2)$ QFT
and a one-dimensional 4-supercharges supersymmetric Quantum Mechanical (SQM) system, the latter being the dimensional reduction of the first one  by compactification on a circle $S^1$. The physics of the two situations is quite different (e.g.\! mirror symmetry \cite{mirror} holds only in 2d), but the $tt^*$ geometries of the two theories are identical
 \cite{tt*,family}. 
 Thus, in studying the $tt^*$ geometry we may use the quantum-mechanical and the field-theoretical language interchangeably. Some aspects of the geometry may be physically obvious in one language but not in the other. Hence,
while most of the literature uses the 2d perspective, in this paper we feel free to change viewpoint according convenience.
Of course, the universality class of FQHE is described by the SQM LG model.
\medskip

The Hilbert space $\mathbf{H}$ of the SQM model is the space of differential forms on $\ck$ with $L^2$-coefficients \cite{Witten:1982im}\!\!\cite{iosqm}.
The Lagrangian $\mathscr{L}_x$ is invariant under a supercharge $\overline{Q}_x$ which acts
on forms as 
\be\label{Qbar}
\overline{Q}_x\psi=\overline{\partial}\psi+dW(z;x)\wedge\psi.\ee
$\overline{Q}_x$ is obviously nilpotent, $\overline{Q}_x^2=0$, and it commutes with multiplication by holomorphic functions. The vacuum vector space
\be\label{fiber}
\mathscr{V}_x:=\Big\{\psi\in \mathbf{H}\; :\; \overline{Q}_x\psi=\overline{Q}^\dagger_x\psi=0\Big\}\subset\mathbf{H}
\ee is isomorphic to the cohomology of $\overline{Q}_x$ in $\mathbf{H}$. Under the present assumptions, the vacuum space $\mathscr{V}_x$ consists of primitive forms of degree $N\equiv\dim_\C\ck$ \cite{iosqm}. In particular, the vacua are invariant under the Lefshetz R-symmetry $SU(2)_R$ \cite{GH}. $d\equiv\dim \mathscr{V}_x$ is the Witten index \cite{Witten:1982df}, invariant under continuous deformations of $x\in\mathscr{X}$ such that $\|d\cw\|^2$ remains bounded away from zero outside a large compact set $C\Subset \ck$. 
The cohomology of $\overline{Q}_x$ in the space of operators acting on $\mathbf{H}$ is called the \emph{chiral ring} $\mathscr{R}_x$. A simple computation \cite{iosqm} yields\footnote{\ Here and below $\co_\ck$ denotes the structure sheaf of the complex manifold $\ck$.}
\be\label{xring}
\mathscr{R}_x=\Gamma(\ck, \co_\ck/\cj_x)\overset{\text{Stein}}{=}\Gamma(\ck,\co_\ck)/\Gamma(\ck,\cj_x)
\ee  
where $\cj_x\subset\co_\ck$ is the sheaf of ideals 
whose stalks are generated by the germs of the partials $\partial_{z^i}\cw(z;x)$. 
In the present framework, $\mathscr{R}_x$ is a finite-dimensional, commutative, associative, unital $\C$-algebra which in addition is Frobenius, i.e.\! endowed with a \textit{trace map} $\langle-\rangle_x\colon \mathscr{R}_x\to\C$ such that $\langle \phi_1\phi_2\rangle_x$ is a non-degenerate bilinear form on $\mathscr{R}_x$.   From the definitions we have an obvious linear isomorphism
(the ``spectral flow'') \cite{Lerche:1989uy}
\be\label{ddqw}
\varkappa\colon\mathscr{R}_x\cong\mathscr{V}_x,\qquad
\varkappa\colon\phi\mapsto \phi\,dz^1\wedge\cdots\wedge dz^n+\overline{Q}_x \eta_x
\ee
which in the 2d context can be understood as the state-operator correspondence for the Topological Field Theory (TFT) obtained by twisting the physical model
\cite{tt*,mirror}.
Eqn.\eqref{ddqw} extends to an isomorphism of $\mathscr{R}_x$-modules.
The Frobenius bilinear form (the topological two point function in the 2d language) is \cite{iosqm}\footnote{\ Note that, as $\varkappa(\phi_a)$ are primitive $N$-forms, the \textsc{rhs} does not depend on the chosen K\"ahler metric in virtue of the Riemann bilinear relations \cite{GH,vhs}.}
\be
\langle \phi_1\phi_2\rangle_x=\int_\ck \varkappa(\phi_1)\wedge\ast\varkappa(\phi_2).
\ee 
A direct computation of the \textsc{rhs} \cite{iosqm} shows that the trace form is the Grothendieck residue \cite{GH} of $\phi_1\phi_2$ with respect to the regular sequence of the partials $\{\partial_{z_1}\cw,\cdots, \partial_{z_N}\cw\}$.
\medskip

As a matter of notation, we shall write $\langle\phi|$ for the vacuum state whose wave-function is $\varkappa(\phi)$ which we write as a bra. We stress that in our conventions $\langle \phi|$ is $\C$-\emph{linear} in $\phi$, not anti-linear.
\medskip
  
Eqn.\eqref{ddqw} implies that $tt^*$ geometry is functorial with respect to (possibly branched) holomorphic covers\footnote{\ If $\ck$ is Stein, $\ck^\prime$ is automatically also Stein \cite{stein1}.} $f\colon \ck^\prime\to \ck$
\cite{tt*} a property that will be crucial in section 5 below. 
\medskip

Let $\zeta\in\bP^1$ be a twistor parameter, and consider the smooth function
\be
F(z,\bar z;\zeta)=\mathrm{Re}\big(\cw(z;x)/\zeta+\overline{\cw}(\overline{z};\overline{x})\zeta\big)
\ee
Morse cobordism\footnote{\ See e.g.\! \textbf{Theorem 3.9} in \cite{morsecob}.}  implies the isomorphism \cite{iqbal}
\be\label{lllasqwp}
\mathscr{V}_x\cong H^*(\ck, \ck_{x;\zeta};\C)\qquad \zeta\in\bP^1,
\ee
where $H^*(\ck, \ck_{x;\zeta};\C)$ denotes the relative cohomology\footnote{\ The space $H^*(\ck, \ck_{x;\zeta};\C)$ is non-zero only in degree $N$.} with complex coefficients, and 
\be
\ck_{x;\zeta}:=\Big\{z\in\ck\;:\; F(z,\bar z;\zeta)> \Lambda\Big\}\subset \ck
\ee
for some sufficiently large\footnote{\ $\Lambda$ should be larger than the image  of all critical values of $\cw$.} constant $\Lambda$. The dual relative homology
$H_*(\ck, \ck_{x;\zeta};\C)$ is called the space of \emph{branes}, because in 2d the corresponding objects have the physical interpretation of half-BPS branes \cite{iqbal}; the twistor parameter $\zeta$ specifies which linear combinations of the original 4 supercharges leave the brane invariant. The space of branes has an obvious integral structure given by homology with integral coefficients
\be\label{nnnncx}
\mathscr{V}_x^\vee\cong H_\ast(\ck,\ck_{x;\zeta};\Z)\otimes_\Z \C.
\ee
An integral basis of $H_\ast(\ck,\ck_{x;\zeta};\Z)$ may be explicitly realized by special Lagrangian submanifolds of $\ck$ and, more specifically, by Lefshetz timbles describing the gradient flow of $F(z,\bar z;\zeta)$ for generic $\zeta$ \cite{iqbal}. By abuse of notation, we write $|\alpha;\zeta\rangle_x$, ($\alpha=1,\dots,d$) for such an integral basis. Let $\{\phi_i\}$
($i=1,\dots, d$) be a basis of $\mathscr{R}_x$;
we write $\{\langle \phi_i|\}$ for the corresponding basis $\{\varkappa(\phi_i)\}$ of $\mathscr{V}_x$.
We may form the \emph{non-degenerate} 
$d\times d$ matrix
\be\label{jasqw0}
\Psi(x;\zeta)_{i\alpha}=\langle\phi_i|\alpha;\zeta\rangle_x
\ee
called the \emph{brane amplitudes}.
$\Psi(x;\zeta)_{i\alpha}$ is not uni-valued as a function of $\zeta$ due to the Stokes phenomenon \cite{CV92} (and, in the 2d language, the related issue of BPS wall-crossing). 

\subsection{$tt^*$ geometry}
On the coupling space $\mathscr{X}$ we have the vacuum vector bundle $\mathscr{V}$
\be
\begin{gathered}
\xymatrix{0\ar[r] & \mathscr{V} \ar[r]\ar[dr] & \mathbf{H}\times \mathscr{X} \ar[d] & \text{exact row,}
\\
&&\mathscr{X}}
\end{gathered}
\ee
namely the sub-bundle of the trivial Hilbert  bundle
 $\mathbf{H}\times \mathscr{X}$ whose fiber $\mathscr{V}_x$ is the vacuum space \eqref{fiber} for the model with couplings $x\in \mathscr{X}$. The differential operator $\overline{Q}_x$ depends holomorphically on $x$ (cfr.\! eqn.\eqref{Qbar}); then the
  isomorphism 
 \be
 \mathscr{V}_x\cong \mathrm{ker}\,\overline{Q}_x\big/\mathrm{im}\,\overline{Q}_x,
 \ee  
implies that the bundle $\mathscr{V}\to\mathscr{X}$ is holomorphic. The vacuum Berry connection, i.e.\! the sub-bundle connection on $\mathscr{V}$ induced by the trivial connection on $\mathbf{H}\times \mathscr{X}\!$, is then both metric and holomorphic. There is a unique such connection, the Chern one \cite{GH}, whose (1,0) and (0,1) parts are respectively  
\be
D=\partial+g\partial g^{-1}\quad\text{and}\quad \overline{D}=\overline{\partial}, 
\ee
where $g$ is the $tt^*$ (Hermitian) metric matrix \cite{tt*}
\be\label{mmmetr*}
g_{i\bar j}=\int_\ck \varkappa(\phi_i)\wedge \ast \overline{\varkappa(\phi_j)}.
\ee
Clearly, the (2,0) and (0,2) parts of the vacuum Berry curvature vanish
\be\label{tt1}
D^2=\overline{D}^2=0.
\ee
We have a canonical (holomorphic) sub-bundle $\mathscr{R}$ of $\mathrm{End}(\mathscr{V})$: 
\be
\begin{gathered}
\xymatrix{0\ar[r] & \mathscr{R} \ar[r]\ar[dr] & \mathrm{End}(\mathscr{V}) \ar[d] & \text{exact row,}\\
&&\mathscr{X}}
\end{gathered}
\ee
whose fiber $\mathscr{R}_x$ is the chiral ring of the theory with coupling $x$. Spectral-flow (or 2d topological twist) then yields the bundle isomorphism $\mathscr{R}\cong\mathscr{V}$. Note that $tt^*$ geometry defines \emph{two} (distinct) natural Hermitian metrics on $\mathscr{V}$:
the one induced by the monomorphism
$\mathscr{V}\hookrightarrow \boldsymbol{H}\times \mathscr{X}$ and the one induced by $\mathscr{V}\cong\mathscr{R}\hookrightarrow \mathrm{End}(\mathscr{V})\hookrightarrow (\boldsymbol{H}\otimes
\boldsymbol{H})\times\mathscr{X}$.
\medskip 

A superpotential $\cw$ produces a $(1,0)$-form $C$ on $\mathscr{X}$
with coefficients in $\mathscr{R}$, i.e.\! 
\be
C:=\big[\partial_{x^i}\cw\big]_x dx^i \in\Gamma(\mathscr{X}, \mathscr{R}\otimes \Omega^1)\subset \Gamma(\mathscr{X}, \mathrm{End}(\mathscr{V})\otimes \Omega^1),
\ee 
where $[\phi]_x$ stands for the class of the holomorphic function $\phi$ in $\mathscr{R}_x$ (cfr.\! \eqref{xring}). 
Since we are free to add to $\cw(z;x)$ a $x$-dependent constant, we may assume without loss that the coefficients of $C$ belong to the trace-less part of
$\mathrm{End}(\mathscr{V})$.  
$C$ is manifestly nilpotent, and both holomorphic and covariantly-closed \cite{tt*}
\be\label{rel}
C\wedge C=\overline{D}C=DC=0.
\ee
We write $\overline{C}$ for the  (0,1)-form which is the Hermitian conjugate of $C$ with respect to the metric \eqref{mmmetr*}. $\overline{C}$ satisfies the conjugate of relations
\eqref{rel}. It remains to specify the (1,1) part of the curvature of the Berry connection; one gets \cite{tt*}
\be\label{tt2}
D\overline{D}+\overline{D} D+C\wedge\overline{C}+\overline{C}\wedge C=0.
\ee
Eqn.\eqref{tt1},\eqref{rel} and \eqref{tt2} are the $tt^*$ equations \cite{tt*}. They are integrable \cite{Dubrovin:1992yd}\!\!\cite{CV92} and, in fact most (possibly all)
integrable systems reduce to special instances of $tt^*$ geometry. For $\zeta\in\bP^1$ one considers the
(non-metric!) connection on the vacuum bundle $\mathscr{V}\to\mathscr{X}$
\be\label{uuure}
\nabla^{(\zeta)}=D+\frac{1}{\zeta} C,\qquad 
\overline{\nabla}^{(\zeta)}=\overline{D}+\zeta \overline{C}
\ee
The $tt^*$ equations can be neatly  summarized in the statement that this connection is flat identically in the twistor parameter $\zeta$
\be
(\nabla^{(\zeta)})^2=(\overline{\nabla}^{(\zeta)})^2=
\nabla^{(\zeta)}\overline{\nabla}^{(\zeta)}+
\overline{\nabla}^{(\zeta)}\nabla^{(\zeta)}=0\qquad\text{for all }\zeta\in\bP^1.
\ee
Hence the linear system (called the \emph{$tt^*$ Lax equations}) 
\be\label{linearsyst}
\nabla^{(\zeta)}\Psi(\zeta)=\overline{\nabla}^{(\zeta)}\Psi(\zeta)=0
\ee
is integrable for all $\zeta$. A fundamental solution to \eqref{linearsyst} is a 
$d\times d$ matrix $\Psi(\zeta)$ whose columns are linearly independent solutions, i.e.\!  a basis of linear independent flat sections of $\mathscr{V}$.
A deeper interpretation of the $tt^*$ Lax equations as describing $\zeta$-holomorphic sections of hyperholomorphic bundles in hyperK\"ahler geometry may be found in ref.\!\cite{tt*3}.
\medskip   

Given a fundamental solution $\Psi(\zeta)$ we may recover the $tt^*$ metric $g$.
This is best understood by introducing the \emph{real structure} (compatible with the rational structure induced by the branes) \cite{tt*,CV92,Dubrovin:1992yd}
\be
\overline{\Psi(1/\overline{\zeta})}=g^{-1}\,\Psi(\zeta).
\ee

\begin{rem} Jumping slightly ahead, we observe that when the chiral ring is semi-simple (\S.\,\ref{seeemis}) we may choose as an integral basis of branes the Lefschetz thimbles which originate from the (non-degenerate) critical points of $\cw$ \cite{iqbal}. In this case (for a certain canonical basis of $\mathscr{R}$ defined in \S.\,\ref{seeemis}) one has \cite{tt*,CV92,Dubrovin:1992yd}
\be\label{bin1}
\Psi(\zeta)\Psi(-\zeta)^t=1
\ee
where one should think of the \textsc{rhs} as the topological metric $\eta$ in the canonical basis. Hence
\be\label{bin2}
g=\Psi(\zeta)\,\Psi(-1/\overline{\zeta})^\dagger.
\ee
The interpretation of eqns.\eqref{bin1}\eqref{bin2} is that the brane spaces $H_*(\ck,\ck_{x,\zeta})$ and
$H_*(\ck,\ck_{x,-\zeta})$ are each other dual (with respect to the natural intersection pairing\footnote{\ This is Lefschetz duality.}) and both the topological and $tt^*$ metrics can be written in terms of the Lefschetz intersection pairing. This observation will be useful to clarify the $tt^*$-theoretical origin of most constructions in the theory of braid group representations \cite{braid}.
\end{rem}

\subsection{The $tt^*$ monodromy representation} Let $\Psi(\zeta)_{(\gamma)}$ be the analytic continuation of the fundamental solution $\Psi(\zeta)$ along a closed curve $\gamma\in\mathscr{X}$ in coupling space.
Both $\Psi(\zeta)$ and $\Psi(\zeta)_{(\gamma)}$ solve the  $tt^*$ Lax equations at $x\in\mathscr{X}$, hence there must be an invertible matrix $\varrho(\gamma)_\zeta$ such that
\be\label{sssxq}
\Psi(\zeta)_{(\gamma)}=\Psi(\zeta)\,\varrho_\zeta(\gamma).
\ee
This produces a representation
\be
\varrho_\zeta\colon \pi_1(\mathscr{X})\to GL(d,\C),
\ee
which is independent of the particular choice of the
fundamental solution $\Psi(\zeta)$ modulo conjugacy in $GL(d,\C)$.
 
\begin{cla} We may conjugate the representation
$\varrho_\zeta$ in $GL(d,\C)$ so that it lays in the \emph{arithmetic} subgroup $SL(d,\Z)$.
\end{cla}

To show the claim, we have to exhibit a preferred
fundamental solution which has a canonical  $\Z$-structure. This is provided by the branes. 
It is easy to check that the brane amplitudes \eqref{jasqw0} are a particular fundamental solution to the $tt^*$ Lax equation \cite{iqbal}.
This may be understood on general grounds: since the branes with given $\zeta\in\bP^1$ have well-defined \emph{integral} homology classes,  for each $\zeta\in\bP^1$ they  define a local system on $\mathscr{X}$ canonically equipped with a flat connection, the Gauss-Manin one. Dually, the branes define a $\bP^1$-family of flat connections on $\mathscr{V}$ which is naturally identified with the $\bP^1$-family of $tt^*$ Lax connections $\nabla^{(\zeta)}$, $\overline{\nabla}^{(\zeta)}$.

The parallel transport along the closed loop $\gamma$ should map a brane into a linear combination of branes with integral coefficients \cite{iqbal}.
Then the matrix $\varrho_\zeta(\gamma)$
in \eqref{sssxq}  and its inverse
$\varrho_\zeta(\gamma)^{-1}\equiv \varrho_\zeta(\gamma^{-1})$  should have integral entries, which entails $\det_\zeta\varrho(\gamma)=\pm1$.
The negative sign is not allowed.\footnote{\ \label{ffoot33}If we normalize $C$ to be traceless (as we are free to do), it follows from eqn.\eqref{uuure}  that the function $\det \Psi(\zeta)/\det g\equiv \det \Psi(\zeta)/|\det \eta|$ is constant in $\mathscr{X}$. In special coordinates (which always exist \cite{frob}) $\eta$ is a constant, so $\det\Psi(\zeta)$ is also a constant with these canonical choices.}

Since the entries of $\varrho_\zeta(\gamma)$ are integers, they are locally independent of $\zeta$.
The brane amplitudes are multivalued on the twistor sphere; going from one determination to another
the representation $\varrho_\zeta(-)$ gets conjugated in $SL(d,\Z)$. Then, modulo conjugation, the \textit{$tt^*$ monodromy representation}
\be\label{monrep}
\varrho\colon \pi_1(\mathscr{X})\to SL(d,\Z)
\ee 
is independent of $\zeta$. By the same token, the conjugacy class of $\varrho$ is also invariant under continuous deformation of the parameters $x$, i.e.\! changing the base point $\ast\in\mathscr{X}$ we use to define $\pi_1(\mathscr{X})$ will not change the conjugacy class of $\varrho$ (of course, this already follows from the properties of the fundamental group).
\medskip

The $tt^*$ equations \eqref{tt1},\eqref{rel},\eqref{tt2} 
then describe the possible deformations of the coefficients of the flat connection 
$\nabla^{(\zeta)}, 
\overline{\nabla}^{(\zeta)}$ which leave the monodromy representation $\varrho$ invariant, i.e.\! they are the equations of an \textit{isomonodromic problem.} 
In the special case that the chiral rings $\mathscr{R}_x$
($x\in\mathscr{X}$) are \emph{semi-simple} ($\equiv$ the 2d (2,2) model is gapped) the $tt^*$ isomonodromic problem is equivalent to the Miwa-Jimbo-Sato one \cite{MJS1,MJS2,MJS3,MJS4}, see ref.\!\cite{ising} for the detailed dictionary between the two subjects. 
\medskip

At the opposite extremum we have the situation in which $\mathscr{R}_x$ is a \emph{local ring} for all $x\in\mathscr{X}$.
In this case the 2d (2,2) model is superconformal, and $\mathscr{X}$ is its conformal manifold; the $tt^*$ geometry is equivalent to the Variations of Hodge Structure (VHS) in the sense of Griffiths \cite{vhs} and Deligne \cite{deligne}, see ref.\!\cite{family} for a precise dictionary between the two geometric theories. In the particular case of Calabi-Yau 3-folds the VHS is called ``special geometry'' \cite{mybook} in the  string literature. 

For a generic superpotential $\mathscr{R}_x$ is automatically semi-simple.\footnote{\ $\mathscr{R}_x$ is semi-simple iff,  for all $z\in\ck$, the stalks $(\cj_x)_z$ of the sheaf $\cj_x\subset\co_\ck$ are either the trivial ideal, i.e.\! $(\co_\ck)_z$, or a maximal ideal $\mathfrak{m}_z\subset (\co_\ck)_z$. Then the coherent sheaf $\co_\ck/\cj_x$ is a skyscraper with support on the (isolated) zeros of $d\cw$, the stalk at a zero being $\C$. Therefore $\mathscr{R}_x\equiv\Gamma(\ck,\co_\ck/\cj_x)\cong \prod_{\upsilon\in\mathsf{sup}\,\cj_x} (\C)_\upsilon$.} The locus in $\mathscr{X}$
where $\mathscr{R}_x$ is not semi-simple is an analytic subspace, hence it has\footnote{\ \label{foot}Assuming that $\mathscr{X}$ is not contained in that locus, as it is the case for the models of interest in this paper.}
 real codimension at least 2; therefore, for all element
 $[\gamma]\in\pi_1(\mathscr{X})$, we may find a representative closed path $\gamma$ which avoids the 
non-semi-simple locus, that is, we may effectively replace $\mathscr{X}$ with the open dense subspace where
$\mathscr{R}_x$ is semi-simple. 

\subsection{$\mathscr{R}_x$ local}

For completeness, we briefly mention the situation for $\mathscr{R}_x$ local, even if the main focus of this paper is the semi-simple case.
Historically, $tt^*$ geometry was created \cite{tt*,family} on the model of VHS,
thinking of it as a ``mass deformation'' of VHS which holds even off-criticality.  Hodge theory provides a good intuition about the properties of $tt^*$ geometry, and many Hodge-theoretical arguments may be extended to the wider $tt^*$ context.  Typical massive (2,2) systems have a UV fixed point which is a regular SCFT, whose $tt^*$ geometry is described by VHS. In this case VHS geometry supplies the boundary condition needed to specify the particular solution of the massive $tt^*$ PDEs which corresponds
to the given physical system: the correct solution is the one which asymptotes to the VHS one as the radius $R$ of the circle on which the 2d theory is quantized is sent to zero \cite{tt*,CV92,Cecotti:1992qh}.

\subsection{$\mathscr{R}_x$ semi-simple}\label{seeemis}
We recall some useful facts about semi-simple 
chiral rings. A commutative semi-simple $\C$-algebra of dimension $d$ is the product of $d$ copies of $\C$. Hence there is a complete system of orthogonal idempotents $e_i$ ($i=1,\dots,d$) which span the algebra $\mathscr{R}_x$ and have a very simple multiplication table\footnote{\ In refs.\!\cite{tt*,CV92} the basis $\{e_i\}$ of $\mathscr{R}_x$ was called the ``point basis''.}
\be
e_i e_j=\delta_{ij}\,e_i,\qquad 1=e_1+e_2+\cdots+e_d.
\ee
Explicitly, $e_i$ represents the class of holomorphic functions on $\ck$ with value 1 at the $i$-th zero of $d\cw$ and 0 at the other critical points (such functions exist since $\ck$ is Stein).  The Frobenius bilinear pairing has the form
\be
\eta_{ij}:= \langle e_i e_j\rangle =\delta_{ij}\,\langle e_i\rangle\qquad \langle e_i\rangle\neq0.
\ee
We write
 \be\label{jjasz}
 E_i= \langle e_i\rangle^{-1/2}\,e_i\qquad i=1,\dots,d.
 \ee
The basis $\{\langle E_i|\}$ of $\mathscr{V}_x$ yields the \emph{canonical} (holomorphic) trivialization of
$\mathscr{V}$; the \emph{natural} trivialization is the one associated to the non-normalized basis $\{\langle e_i|\}$.  
The canonical trivialization
is convenient since it makes the $tt^*$ equations model-independent and the connection with the isomonodromy PDEs transparent. But it has a drawback: the sign of the square-root in \eqref{jjasz} has no canonical determination. Going along a non-trivial loop in coupling space we may come back with the opposite sign. 
The unification comes at the price of a sign conundrum: getting the signs right in the present matter is a well-known headache. To simplify our live, we often study sign-insensitive quantities, such as squares, and be content if they have the correct properties, without bothering to fix the troublesome signs.
\medskip

Since $\langle E_i E_j\rangle =\delta_{ij}$, in the canonical trivialization we do not need to distinguish upper and lower indices. The reality constraint \cite{tt*}
implies that the canonical $tt^*$ metric
\be\label{cantriv}
G_{i\bar j}:=\langle E_j| E_i\rangle\in SO(d,\C)\cap \mathsf{Her}(d)_{+},
\ee
$\mathsf{Her}(d)_+$ being the set of positive-definite $d\times d$ Hermitian matrices. In the canonical trivialization the Berry connection
\be
A:=g\partial g^{-1}\in \mathfrak{so}(d)\otimes \Omega^1(\mathscr{X})
\ee
is anti-symmetric $A^t=-A$.
\medskip

Since $\{e_i\}$ form a basis of $\mathscr{R}$, we have
\be
\big[\cw(z;x)\big]_x= \sum_{i=1}^d w_i\, e_i,
\ee
for certain functions $w_i:\mathscr{X}\to\C$. The $\{w_i\}$'s are the critical values of $\cw(z;x)$. 
The map $w\colon \mathscr{X}\to \C^d$
given by $x\mapsto (w_1(x),\dots,w_d(x))$
is a local immersion. In facts, the
$w_i$ form a local coordinate system on the Frobenius manifold of all couplings of the TFT \cite{frob,Dubrovin:1992yd}
which contains the physical coupling space $\mathscr{X}$ as a submanifold.\footnote{\ In general it is a submanifold of positive codimension. Consider e.g.\! the 2d $\sigma$-model with target $\bP^n$ with $n>1$. Higher powers of the K\"ahler form are elements of the chiral ring and their 2-form descendents can be added to the TFT action. Adding them to the physical action would spoil UV completeness. The corresponding phenomenon in the $tt^*$ geometry is that the solutions to the PDEs become singular for $R$ small enough, i.e.\! at some large (but finite) energy scale. } We write $\mathring{\mathscr{X}}\subset \mathscr{X}$ for the dense open domain\footnote{\ The qualification in footnote \ref{foot} applies here too.} in which $\mathscr{R}_x$ is semi-simple and
\be
i\neq j\quad\Rightarrow\quad w_i\neq w_j.
\ee
Equivalently, $\mathring{\mathscr{X}}$ is the domain in which the function $\cw$ is strictly Morse.

 Let $\ce:= w^i\partial_{w^i}$ be the Euler vector in $\mathscr{X}$; the anti-symmetric matrix
\be
Q:=i_\ce A,
\ee
is called the \textit{new index} \cite{Cecotti:1992qh}. In 2d (2,2) models $Q_{ij}$ plays two roles. First \cite{Cecotti:1992qh}\!\!\cite{CV92} it is the index capturing the
half-BPS solitons in $\R$ which asymptote the $i$-th (resp.\! $j$-th) classical vacuum as $x\to-\infty$ (resp.\! $+\infty$)
\be\label{xxxq}
Q_{ij}=\lim_{L\to\infty} \frac{i\beta}{2L}\,\mathrm{Tr}_{(i,j)} \big[(-1)^F F e^{-\beta H}\big],
\ee 
where the theory is quantized in a strip of width $L$
with boundary conditions the classical vacua $i$, $j$ on the two boundary components. From \eqref{xxxq} one learns\footnote{\ The statement is less elementary that it sounds.} that the matrix $Q$ is Hermitian.
Second \cite{tt*} it is a generalization of Zamolodchikov $c$-function \cite{zamo} since
$Q_{ij}$ is stationary only at fixed points of the RG flow, where the eigenvalues of $Q_{ij}$ become the 
$U(1)_R$ charges of the Ramond vacua of the fixed point (2,2) SCFT\footnote{\ If the 2d (2,2) model is asymptotically free the statement requires some specification, see \cite{CV92}.} which determine the conformal dimension of the chiral primaries and, in particular the Virasoro central charge $c$. 

The new index is a central object in $tt^*$ geometry also in 1d,
where the above physical interpretations do not hold.
Indeed, the full $tt^*$ geometry may be described in terms of the matrix $Q$ only as we now review. 
\medskip

\noindent We warn the reader that in the rest of this subsection the convention on the sum over repeated index does \underline{not} apply.

\begin{lem}[see \cite{ising,Dubrovin:1992yd}] Let $\mathscr{R}_x$ be semi-simple. The Berry connection in the canonical holomorphic gauge is antisymmetric with off-diagonal components
\be\label{basfor}
A_{kl}:= Q_{kl}\,\frac{d(w_k-w_l)}{
w_k-w_l}.
\ee
\end{lem}
The $tt^*$ equations may be written as a pair of differential equations for $Q_{kl}$ \cite{Dubrovin:1992yd}. The first one expresses the fact that the (2,0) part of the curvature vanishes
\be\label{kasqw}
\partial Q_{kl}\wedge u_{kl}+\sum_h Q_{k h}Q_{h l}\,u_{kh}\wedge u_{hl}=0,
\ee
where the symbol $u_{kl}$ stands for the Arnold form \cite{arnold} in configuration space
\be\label{arnold}
u_{kl}:=\frac{d(w_k-w_l)}{
w_k-w_l}.
\ee
The other equation for $Q$ is obtained by contracting \eqref{tt2} with $\ce$
\be\label{kkkkkxzm}
-\overline{\partial}Q+[\cw]_x\, \overline C-\overline C\, [\cw]_x=0.
\ee

\subsection{Computing the monodromy representation}

We study the $tt^*$ monodromy representation $\varrho$, eqn.\eqref{monrep}, for $\mathscr{R}_x$ generically semi-simple.

Since $tt^*$ is an isomonodromic problem we are free to continuously deform the model in coupling constant space $\mathscr{X}$; the only effect is to get matrices $\varrho(\gamma)$ which possibly differ by an irrelevant overall conjugation. In particular, the eigenvalues of the
monodromy matrices 
\be\label{kkasqzll4}
\Big\{\varrho(\gamma), \quad\gamma\in\pi_1(\mathscr{X})\Big\}
\ee 
(which are algebraic numbers of degree at most $d$) and the dimension of their Jordan blocks are invariant under any \emph{finite} continuous deformation.

However, typically, to really simplify the computation
we need to take the limit to a point at infinite distance in parameter space, i.e.\! a point in the closure $\overline{\mathring{\mathscr{X}}}$ of the ``good''
space. In this case the limiting monodromy may be related by a singular conjugation to the original one.
The eigenvalues of the monodromy matrices $\varrho(\gamma)$ are continuous in the limit but the Jordan blocks may decompose into smaller ones.\footnote{\ Having a determinate spectrum is a closed condition in the matrix space, while the having a Jordan block of size $>1$ is open.} 
This happens, for instance, when we take the UV limit of an asymptotically free model (see \cite{CV92}).

Therefore, monodromy eigenvalues are typically easy to compute, while the Jordan structure is subtler. However in many situations we know \emph{a priori} that the monodromy matrix is semi-simple and so we do not loose any information. In the case relevant for the FQHE, when $\pi_1(\mathscr{X})$ is a complicate
non-Abelian group, the Jordan blocks are severely restricted by the group relations, so it is plausible that they can be recovered from the knowledge of the eigenvalues.
\medskip

There are three obvious limits in which the computation is expected to simplify; in the $tt^*$ literature they are called: \textit{i)} the IR limit, \textit{ii)} the asymmetric limit \cite{iqbal,twistor}, and \textit{iii)} the UV limit.
In a related math context \textit{ii)} is called the
homological approach and \textit{iii)} the CFT approach \cite{lawrence}.

The IR and asymmetric approaches are widely known and used \cite{CV92,twistor,iqbal,gaiottoknot}.
They essentially reduce to the combinatorics of 2d wall-crossing \cite{CV92,iqbal}
(equivalently, of 1d BPS instantons \cite{rtwist}).
The UV approach seems less known, and we are not aware of a good reference for it, so we shall develop it in some detail in \S.\,\ref{uvappr} below. Of course, the three approaches
yield equivalent monodromy representations 
(at least when we have a good UV point as in the CFT context) and this statement summarizes many results in the math literature. From this point of view, the wall-crossing formulae are consistency conditions required for the monodromy representation, as computed in the IR/homological approach, to be a well-defined invariant of the UV fixed-point theory.
\medskip  

We briefly review the asymmetric approach for the sake of comparison.   

\subsubsection{Asymmetric approach (homological)}
One starts by rescaling the critical values
\be\label{redfff} 
\w^i\to R\, w^i
\ee
where $R$ is some positive real number\footnote{\ In the context of the 2d (2,2) LG model quantized in a cylinder, $R$ is identified with the radius of the cylinder \cite{tt*}. Alternatively, $R$ is the 2d inverse temperature if we look to the path integral on the cylinder as the theory quantized on the line at finite temperature $R^{-1}$.}.
The $tt^*$ flat connection becomes
\be\label{uuure2}
\nabla^{(\zeta)}=D+\frac{R}{\zeta} C,\qquad 
\overline{\nabla}^{(\zeta)}=\overline{D}+\zeta\,R\, \overline{C}
\ee
and the Berry curvature
\be\label{brryecru}
F= -R^2[C,\overline{C}].
\ee
Then one takes the
unphysical
 limit $R\to0$ with $R/\zeta$ fixed and large.
The Berry curvature vanishes in the limit, so the metric connection $A$ is pure gauge.  The $tt^*$ linear problem \eqref{linearsyst} then (formally\,!) reduces to
\be\label{linasy}
\big(\partial_{w^i}+A_i+\frac{R}{\zeta}\,e_i\big)\Psi= \overline{\partial}\Psi=0.
\ee
A solution to this equation, asymptotically for $R/\zeta$ large, is
\be
\Psi(w)_{i\alpha}=\int_{\Gamma_\alpha} E_i(z)\,e^{- R\,\cw(z;w)/\zeta}\; dz^1\wedge\cdots\wedge dz^n,
\ee
where the cycles $\Gamma_\alpha$ are the supports of 
an integral basis of branes, say Lefshetz thimbles,
and $E_i(z)$ holomorphic functions representing the rescaled unipotents $E_i$ in the chiral ring. Computing the \textsc{rhs} by the saddle point method, one checks that it is indeed a fundamental solution to \eqref{linasy}.

The homology classes of the branes (with given $\zeta$) are locally constant in coupling constant space $\mathscr{X}$, but jump
at loci where (in the 2d language) there are BPS solitons which preserve the same two supercharges as the branes. The jump  in homology at such a locus is given  the Picard-Lefshetz (PL) transformation \cite{CV92,iqbal,witten1,witten2}. Taking into account all the jumps in homology  one encounters along the path (controlled by the 2d BPS spectrum), one gets the monodromy matrix which is automatically integral of determinant 1. The full
monodromy representation is given by the combinatorics of the PL transformations.

Dually, instead of the action of the monodromy group on the homology of branes we may consider its action in the cohomology of the (possibly multivalued) holomorphic $n$-forms
\be
E_i(z)\,e^{- R\,\cw(z;w)/\zeta}\; dz^1\wedge\cdots\wedge dz^n.
\ee 

The method is conceptually clear and often convenient. On the other side,
the fact that we consider a limit which do not correspond to any unitary quantum system tends to make the physics somewhat obscure. 
For our present purposes the UV approach seems more natural.

\subsubsection{The UV approach (``CFT'')}\label{uvappr}

This is the physical UV limit of the 2d model.
Again one makes the redefinition \eqref{redfff} and
sends the length scale $R\to0$. But now $\zeta$ is kept fixed at its original value, which may be a unitary one $|\zeta|=1$. From eqn.\eqref{uuure2} we see that in this limit the flat connection reduces to the Berry one
\be
\nabla^{(\zeta)}+\overline{\nabla}^{(\zeta)}\xrightarrow{\ R\to0\ } D+\overline{D}. 
\ee
The Berry connection then becomes flat in the limit, as eqn.\eqref{brryecru} indeed shows. Since the monodromy of the flat connection $\nabla^{(\zeta)}+\overline{\nabla}^{(\zeta)}$ is  independent of $R$, the flat 
UV Berry connection should have the same monodromy
modulo the subtlety with the size of the Jordan blocks mentioned after eqn.\eqref{kkasqzll4}.

While the monodromy matrices $\varrho(\gamma)$ as computed in the asymmetric (or IR) approach are manifestly \textit{integral,} the monodromy matrices $\varrho^\flat(\gamma)$ computed in the UV approach are manifestly \textit{unitary}
(since the Berry connection is metric).

This observation is a far-reaching generalization to the full non-Abelian $tt^*$ monodromy representation $\varrho\colon \pi_1(\mathscr{X})\to SL(d,\C)$
of the formula for the relation between the 2d quantum monodromy as computed in the UV and in the IR, i.e.\! for the monodromy representation of the Abelian subgroup $\Z\subset\pi_1(\mathscr{X})$ associated to the overall phase of the superpotential $\cw$ \cite{CV92}
\be\label{classmon}
\overbrace{\:e^{2\pi i\, \boldsymbol{Q}}\:}^{\text{UV monodromy}}\xymatrix{
\ar@{<->}[rrr]^{\text{same spectrum}} &&&
}\overbrace{(S^{-1})^tS}^{\text{IR monodromy}}
\ee
where $\boldsymbol{Q}$ is the $U(1)_R$ charge acting on the Ramond vacua of the UV fixed point SCFT 
and $S$ is the integral Stokes matrix of the $tt^*$  Riemann-Hilbert problem \cite{CV92}
\be
|S_{ij}+S_{ji}|=2\,\delta_{ij}+\#\Big\{\text{2d BPS solitons connecting vacua $i$ and $j$}\Big\}. 
\ee
Instead the Jordan block structure is, in general, different between the two sides of the correspondence \eqref{classmon} as the examples in ref.\!\cite{CV92} illustrate.

In particular,

\begin{corl} The integral monodromy matrices $\varrho(\gamma)$ realizing the Picard-Lefshetz homological monodromy on thimbles are quasi-unipotent.
\end{corl}
This statements implies but it is much stronger than the  strong monodromy \cite{CV92}.\footnote{\ The strong monodromy theorem is the same statement but restricted to a special element of the monodromy group, i.e.\! the quantum monodromy. The \textbf{Corollary} claims that the property extends to the full group.} Indeed arithmetic subgroups of $SL(d,\Z)$ such that the spectrum of all elements consists of roots of unity have a very restricted structure.
\medskip

To give an explicit description of the UV Berry connection $\varrho^\flat$ we need additional details on $tt^*$ geometry
which we are going to discuss.

\section{Advanced $tt^*$ geometry I}
\label{advanced1}

To compute the monodromy representation of the Vafa model in the UV approach we need a more in-depth understanding of $tt^*$ geometry. A first block of advanced $tt^*$ topics is discussed 
in this section. Most material is either new or presented in a novel perspective.
The crucial issue is the notion of a \emph{very complete $tt^*$ geometry}.
\medskip

 What makes the UV approach so nice is its relation to the Kohno connections \cite{kohno0,kohno,kohnoB,kohnoN} in the theory of the braid group representation \cite{braid,braidintro}. Flatness with respect to a Kohno connection may be seen as a generalization of the Knizhnik-Zamolodchikov equations \cite{KZ}. 

In this section we go through the details of this beautiful relation.
Since some statement may sound a bit unexpected to the reader, we present several explicit examples.

\subsection{$tt^*$ monodromy vs.\! the universal pure  braid representation}

Following the strategy outlined in \S.\,\ref{uvappr}, we rescale
$w^i\to R w^i$ and send $R\to0$
(note that if $w^i\in\mathring{\mathscr{X}}$
also $R w^i\in\mathring{\mathscr{X}}$ for all $R>0$, so the limiting point indeed lays in the closure $\overline{\mathring{\mathscr{X}}}$ of the semi-simple domain).  
As we approach a fixed point of the RG flow
 the element   $[\cw]_x\in\mathscr{R}_x$ becomes a multiple of the identity operator
 \cite{ising,CV92} and eqn.\eqref{kkkkkxzm} implies $\overline{\partial}Q\to 0$. 
Since $Q$ is Hermitian, $\partial Q\to0$ as well, so that $\lim_{R\to0} Q$ is a constant matrix. 
Naively,  to get the UV Berry connection
 we just replace this constant matrix in the
the basic $tt^*$ formula
\eqref{basfor}. 
However, this is not the correct way to define the $R\to0$ limit. The point is that the canonical trivialization becomes too singular in the UV limit: the chiral ring $\mathscr{R}$ is believed to be regular (even as a Frobenius algebra) in the UV limit but, since the limit ring is no longer semi-simple, its generators are related to the canonical ones by a singular change of basis. A trivialization  which is better behaved as $R\to0$ is the natural one. We write $\boldsymbol{A}$ for the natural gauge connection. Starting from eqn.\eqref{basfor}, and performing the diagonal gauge transformation, we get
\be\label{hhhhhzq}
\boldsymbol{A}_{kl}= h_k Q_{kl}h_l^{-1}\,\frac{d(w_k-w_l)}{w_k-w_l}-\delta_{kl}\, d\log h_l
\ee
where $h_l= \langle e_l\rangle^{1/2}$.
Taking the limit $R\to0$, the second term in the \textsc{rhs} of
\eqref{hhhhhzq} becomes (locally) a meromorphic one-form  $f_{l,i}(w_j)\,dw_i$ invariant under $w_j\to w_j+c$ and $w_j\to \lambda w_j$ with at most single poles when $w_l=w_j$ for some $l\neq j$.
In addition, its contraction with the Euler vector $\ce$ has no poles. Thus as $R\to0$ the Berry connection
$\cd\equiv D+\overline{\partial}$  
should locally take the form
\be\label{uniffr}
\cd=d+\sum_{1\leq i<j\leq n} B_{ij}\,\frac{d(w_i-w_j)}{w_i-w_j}
\ee
where the entries of the matrix $B_{ij}$ are holomorphic functions of the $w_k-w_l$ homogeneous of degree zero. They should produce the  correct quantum monodromy  \eqref{classmon}, so that
\be\label{kkasqw}
\exp\!\left[2\pi i\sum_{i<j} B_{ij}\right]=\exp\!\Big(2\pi i\,\boldsymbol{Q}\Big)\quad \text{up to conjugacy} 
\ee
where $\boldsymbol{Q}$ is the same matrix as in eqn.\eqref{classmon} but now written in a different basis which makes it symmetric trace-less.

The simplest solution to these conditions is given by constant $B_{ij}$ matrices.
However, the $B_{ij}$ cannot be just constant in general; 
indeed
the matrices $B_{ij}$
are restricted by a more fundamental condition i.e.\! that 
the connection \eqref{uniffr} is flat
\be
\cd^2=0,
\ee
how predicted by the $tt^*$ equations.
Note that this constraint on $B_{ij}$ arises from setting to zero the (2,0) part of the curvature, which vanishes for all $R$.

\subsection{Complete and very complete $tt^*$ geometries}

Let $\mathscr{X}$ be the essential coupling space of a 4-\textsc{susy} LG model with Witten index $d$. We write $\mathring{\mathscr{X}}\subset\mathscr{X}$ for the open domain
(assumed to be non-empty and connected) in which the superpotential is a Morse function\footnote{\ In this sub-section we use the term \textit{``Morse function''} in the strong sense i.e.\! all critical points are non-degenerate and the critical values $w_k$ are all distinct.}. 

\subsubsection{Configuration spaces}
The configuration space $\cc_d$ of $d$  ordered distinct points in the plane is
\be
\cc_d:=\Big\{ (w_1,w_2,\cdots, w_d)\in\C^d\;\Big|\; w_i\neq w_j\ \text{for }i\neq j\Big\}.
\ee
The cohomology ring $H^*(\cc_d,\Z)$ is the ring 
in the ${d\choose 2}$ generators
\be
\omega_{ij}=\omega_{ji}=\frac{1}{2\pi i}\,\frac{d(w_i-w_j)}{w_i-w_j}
\ee  
subjected to the ${d\choose 3}$ relations
\cite{arnold}
\be
\omega_{ij}\wedge \omega_{jk}+\omega_{jk}\wedge \omega_{ki}+\omega_{ki}\wedge \omega_{ij}=0.
\ee
The fundamental group $\mathscr{P}_d=\pi_1(\cc_d)$ is called the \textit{pure braid group on $d$ strings}.

The configuration space of $d$ \emph{unordered} distinct points is the quotient space 
\be\label{qqquaspa}
\cy_d=\cc_d/\mathfrak{S}_d.
\ee
Its fundamental group $\cb_d=\pi_1(\cy_d)$ is the \textit{(Artin) braid group in $d$ strings} \cite{braid}.
It is an extension of the symmetric group
$\mathfrak{S}_d$ by the pure braid group
\be\label{extennnsion}
1\to\mathscr{P}_d\xrightarrow{\;\,\iota\,\;} \cb_d\xrightarrow{\;\beta\;}\mathfrak{S}_d\to 1.
\ee
$\cb_d$ has a presentation with $d-1$ generators $g_i$ ($i=1,\dots, d-1$) and relations
\be\label{bdrel}
g_ig_{i+1}g_i=g_{i+1}g_i g_{i+1},\qquad g_ig_j=g_jg_i\ \text{for }\ |i-j|\geq 2.
\ee

\subsubsection{(Very) complete $tt^*$ geometries}

The critical value map
\be\label{wmap}
w\colon\mathring{\mathscr{X}}\to \cy_d,\qquad x\mapsto \{w_1,\cdots, w_d\}
\ee
is a holomorphic immersion by definition of ``essential'' couplings. We say that the $tt^*$ geometry is \emph{complete} if, in addition,
$w$ is a submersion, hence a covering map. 
The notion of completeness is akin to the one for 4d $\cn=2$ QFTs \cite{Cecotti:2011rv,Alim:2011ae}; in particular, the 2d correspondent \cite{Cecotti:2010fi}
 of a 4d complete theory has a complete $tt^*$ geometry in the present sense. Equivalently, we may say that a $tt^*$ geometry is complete iff it is defined over the full Frobenius manifold $\mathscr{X}_\text{Frob}$ of the associated TFT \cite{frob}.\footnote{\ In general the perturbations of the model by elements of the chiral ring, $L\to L+(\epsilon\int\! d^2\theta\,\phi+\text{h.c.})$, $\phi\in\mathscr{R}$ are ``obstructed'' in the sense that the coupling is UV relevant and the perturbed theory develops Landau poles. In this case the TFT theory is still  well-defined, but the $tt^*$ metric gets singular for $R$ less than a certain critical values $R_c$ (from the formulation of $tt^*$ in terms of integral equations, it is clear that a smooth solutions always exists for large enough \cite{Dubrovin:1992yd}, but nothing prevents a singularity to appear at finite $R$). In practice,
 the $tt^*$ being complete means that all chiral ``primary'' operators are IR relevant or marginal non-dangerous.} Completeness is a strong requirement.
 \vskip6pt

The category of coverings of $\cy_d$ is equivalent\footnote{\ See e.g. \S.II.2.9 of \cite{gelfand}.} to the category $\cb_d\textsf{-sets}$. 
We say that a complete $tt^*$ geometry is \textit{very complete} iff the action of $\cb_d$ on the  $\cb_d$-set $S$ which corresponds to the cover  $\mathring{\mathscr{X}}\to\cy_d$ factors through
the map $\beta$ in \eqref{extennnsion} or, equivalently, if the canonic projection $p\colon \cc_d\to \cy_d$ factors through the critical-value map $w$
\be\label{kkxz00irr}
\xymatrix{\cc_d\ar[rr]\ar@/^1.5pc/[rrrr]^p &&\mathring{\mathscr{X}}\ar[rr]_w&&\cy_d}
\ee  
In this case we may view $\cc_d$ as the coupling space on which it acts a group of ``$S$-dualities'' given by the deck group of the cover
$\cc_d\to\mathring{\mathscr{X}}$.
We may pull-back the vacuum bundle $\mathscr{V}\to\mathring{\mathscr{X}}$ to a bundle over $\cc_d$, which we denote by the same symbol $\mathscr{V}$, and consider the $tt^*$ geometry on the configuration space $\cc_d$.
\medskip

In a very complete $tt^*$ geometry, pulled back to $\cc_d$, the local expression \eqref{uniffr} becomes global, since in this case the $w_i$ are global coordinates and the partials $\partial_{w_i}\cw$ define a global trivialization of the bundle $\mathscr{R}\to\cc_d$.  

In the very complete case the entries of the matrices $B_{ij}$ are holomorphic functions on $\cc_d$, homogenous of degree zero and invariant under overall translation, which satisfy \eqref{kkasqw}. Assuming the \textsc{rhs} of that equation to be well-defined (we take this as part of the definition of \textit{very complete),} the matrices $B_{ij}$ should be constant. They are further constrained by the flatness condition $\cd^2=0$.
This leads to the theory of Kohno  connections \cite{kohno0,kohno} that  we briefly review.

\begin{rem} An especially important class of very special $tt^*$ geometries are the \emph{symmetric} ones; in this case  
the $d\times d$ matrices $B_{ij}$ satisfy
\be\label{jjjxxjxjx}
(B_{\pi(i)\pi(j)})_{\pi(k)\pi(l)}= (B_{ij})_{kl}\quad \text{for all }\pi\in\mathfrak{S}_d.
\ee 
This holds automatically when the map $w$
in eqn.\eqref{kkxz00irr} is an isomorphism.
\end{rem}

\subsubsection{Khono connections}

In the very complete case the flat UV Berry connection $\cd$ has the form \eqref{uniffr}
with $B_{ij}$ constant $d\times d$ matrices.
We recall the 

\begin{lem}[see e.g.\! \cite{kohno0,kohno,kohnoB}] 
Let $\cd=d+\ca$ be a connection of the general form \eqref{uniffr} where $B_{ij}$ are constant $d\times d$ matrices. Then $\cd$ is flat if and only if the following relations hold
\begin{align}\label{first}
&[B_{ij},B_{ik}+B_{jk}]=[B_{ij}+B_{ik}, B_{jk}]=0 && \text{for }i<j<k,\\
&[B_{ij},B_{kl}]=0 &&\text{for distinct }i,j,k,l.
\label{second}
\end{align}
Eqns.\eqref{first},\eqref{second} are called
the \emph{infinitesimal pure braid relations.} A connection of the form \eqref{uniffr} where the constant matrices $B_{ij}$ satisfy the relations \eqref{first},\eqref{second} is called a rank-$d$ \emph{Kohno connection.} 
\end{lem}

\noindent A rank-$d$ Kohno connection defines  
a representation of  the pure braid group $\mathscr{P}_d$ in $d$ strings
\be\label{reddde}
\sigma\colon \mathscr{P}_d\to GL(d,\C)
\ee 
via parallel transport with the connection $\cd$ on the configuration space $\cc_d$
\be
\sigma\colon\mathscr{P}_d\ni [\gamma]\longmapsto P\exp\!\left(-\int_\gamma\ca\right)\in GL(d,\C).
\ee
The family of representations $\sigma$ parametrized by the matrices $B_{ij}$ satisfying the infinitesimal pure braid relations is called the \emph{universal monodromy} \cite{kohno}.
If, in addition, eqn.\eqref{jjjxxjxjx} holds the connection $\cd$ descends to a flat connection on a suitable bundle $\mathscr{V}\to \cy_d$
and yields a universal monodromy representation of the full braid group $\cb_d$ \cite{kohno0,kohno,kohnoB}. 
\medskip

We conclude:
\begin{fact} For a \emph{very complete $tt^*$ geometry}
the UV Berry monodromy representation $\varrho^\flat$ is the universal monodromy representation of the pure braid group $\mathscr{P}_d$ in $d$ string specialized to the Khono matrices
$B_{ij}$ computed from the grading of the UV chiral ring $\mathscr{R}_\text{UV}$
in superconformal $U(1)_R$ charges. If, in addition, the $tt^*$ geometry is \emph{symmetric} the UV Berry monodromy representation extends to a representation of the full braid group $\cb_d$. 
\end{fact}

The condition of being very complete is really very restrictive for a $tt^*$ geometry. So, to convince the reader that we are not concerned with properties of the empty set, we present a few examples. They will be used later to illustrate various aspects of the theory.

\subsection{First examples of very complete $tt^*$ geometries}

We omit \textbf{Example 0}, the free (Gaussian) theory.
Note that its superpotential $W(z)$ satisfies the ODE $(d_zW(z))^2=a\, W(z)+b$.

\begin{exe} All massive models with Witten index 2 are trivially very complete. The simplest instance is the cubic model $W(z)=x_1(z^3-3z)/2+x_2$
whose critical values are $w_{1,2}=x_2\mp x_1$ and 
$h_{i}=(-3/2)^{-1/2}(w_i-w_j)^{-1/2}$ ($j\neq i$).
\end{exe}

\begin{exe}[Mirror of 2d $\bP^1$ $\sigma$-model]
This LG model has a superpotential $W(z)$ which is a solution to the ODE
\be\label{p2}
\left(\frac{dW}{dz}\right)^{\!2}=P_2(W),\qquad P_2(z)\ \text{a monic quadratic polynomial.}
\ee
The general form is
\be
W(z)= x_1 \sinh z+x_2,\quad \text{where }z\sim z+2\pi i,
\ee
with couplings $(x_1,x_2)\in \C^\times\times\C$.
The map  $w$ in eqn.\eqref{wmap} becomes
\be
\begin{gathered}
w\colon (x_1,x_2)\mapsto (x_1+x_2, -x_1+x_2)\equiv (w_1,w_2),
\\
w\colon\C^\times\times\C\xrightarrow{\sim} \cc_2\equiv\Big\{(w_1,w_2)\in\C^2\;\Big|\; w_1\neq w_2\Big\},
\end{gathered} 
\ee
which shows that the model is very complete.
\end{exe}

\begin{exe}[The Weierstrass LG model] Eqn.\eqref{p2} is replaced by 
\be\label{p3}
\left(\frac{dW}{dz}\right)^{\!2}=P_3(W),\qquad P_3(z)\ \text{cubic polynomial.}
\ee
and the superpotential becomes
\be
W(z)=x_1\,\wp(z\,|x_2)+x_3, \quad \text{where } z\sim z+m+ x_2 n,\ \ m,n\in\Z.
\ee
The coupling constant space is
$\mathscr{X}=\tilde{\mathscr{X}}$
where\footnote{\ \textbf{Notation}: $\ch$ is the upper half-plane, $\Gamma(2)$ the principal congruence subgroup of the modular $SL(2,\Z)$ of level 2. We recall that $SL(2,\Z)/\Gamma(2)\cong\mathfrak{S}_3$, the symmetric group in three letters.}
\be
(x_1,x_2,x_3)\in \C^\times\times \ch/\Gamma(2)\times \C\equiv \tilde{\mathscr{X}}.
\ee
The critical values are 
\be
w_a=x_1\,e_a(x_2)+x_3,
\ee
where $e_a(x_2)$ $(a=1,2,3$) are the three roots of the Weierstrass cubic polynomial as a function of the period $x_2$ which are globally defined for  $x_2\in\ch/\Gamma(2)$. Note that for $x_2\in\ch/\Gamma(2)$ the $e_a(x_2)$ are all distinct, so $w_a\neq w_b$ for $a\neq b$. Hence the critical-value
yields the isomorphism $\tilde{\mathscr{X}}\xrightarrow{\!\sim\!}\cc_3$. The group $\mathfrak{S}_3\cong SL(2,\Z)/\Gamma(2)$ permutes the three roots $e_a(x_2)$,
so the map $w$ is an isomorphism
\be
w\colon \mathscr{X}\xrightarrow{\!\sim\!} \cy_3\equiv \cc_3/\mathfrak{S}_3.
\ee
The model is very complete and symmetric. 
\end{exe}

\begin{exe}[A $d=4$ model]
We consider the elliptic superpotential
\begin{gather}
W(z)= \frac{x_1}{\wp(z\,|x_2)-x_3}+x_4,\\ 
\intertext{with parameter space covered by}
\tilde{\mathscr{X}}=\Big\{(x_1,x_2,x_3,x_4)\in\C^\times\times \ch/\Gamma(2)\times \C\times \C\;\Big|\; x_3\not\in
\big\{e_1(x_2),e_2(x_2),e_3(x_2)\big\}\Big\}.
\end{gather} 
The four critical points correspond to the 2-torsion subgroup $(\Z_2)^2\cong E[2]\subset E$ of the elliptic curve $E$ of periods
$(1,x_2)$. The critical values are
\be\label{wqer}
(w_1,w_2,w_3,w_4)=\left(\frac{x_1}{e_1(x_2)-x_3}+x_4,
\frac{x_1}{e_2(x_2)-x_3}+x_4,\frac{x_1}{e_3(x_2)-x_3}+x_4,x_4\right)
\ee 
same notation as in the previous example. Since the $e_a(x_2)$ are distinct, we have the isomorphism 
\be
w\colon\tilde{\mathscr{X}}\xrightarrow{\sim}\cc_4\equiv\Big\{(w_1,w_2,w_3,w_4)\in\C^4\;\Big|\: w_i\neq w_j\ \text{for }i\neq j\Big\}.
\ee
One checks that $\mathfrak{S}_4$ is a ``duality'', so the actual coupling space is
$\tilde{\mathscr{X}}/\mathfrak{S}_4\cong\cy_4$. The superpotential satisfies an ODE of the form
\be
\left(\frac{dW}{dz}\right)^{\!2}= P_4(W),\ \text{a polynomial of order 4.}
\ee
\end{exe}

\begin{exe}[Hyperelliptic models]\label{hhhyperr} The above three examples may be easily generalized to the case of
LG models \cite{CV92} whose superpotential $W(z)$ satisfies the ODE
\be\label{pn}
\left(\frac{dW}{dz}\right)^{\!2}=P_d(W),\qquad P_d(z)=\prod_{i=1}^d(z-w_i),\quad d\geq 1.
\ee
Note that
\be
h_i^{-2}=\langle e_i\rangle^{-1}\equiv W^{\prime\prime}|_\text{$i$-th vacuum}=\frac{1}{2} P_d^\prime(w_i)=\frac{1}{2}\,\prod_{j\neq i}(w_i-w_j).
\ee
Comparing with eqn.\eqref{hhhhhzq}, we see
that the diagonal components of the UV Berry connection are
\be
\boldsymbol{A}_{kk}=-d\log h_k= \frac{1}{2} \sum_{j\neq k}\frac{dw^k-dw^j}{w^k-w^j}.
\ee
while the off-diagonal components are given by the entries of the UV matrix $\boldsymbol{Q}$.
Since this matrix is constant, we may compute $\boldsymbol{Q}_{kl}$ in the limit $w_k-w_l\to0$
whose effective theory is \eqref{p2}. The $2\times 2$ matrix
$(\boldsymbol{Q}_{ab})_{a,b=k,l}$ 
($k\neq l$) is symmetric with zeros on the diagonal and eigenvalues $\pm \hat c/2$ where
$\hat c$ is the Virasoro central charge of the effective theory, in this case $\hat c=1$; hence  
\be
\boldsymbol{A}_{kl}=-\frac{1}{2}\sum_{i<j}\big(\delta_{ki}\delta_{l j}+\delta_{kj}\delta_{li}\big)\frac{dw_i-dw_j}{w_i-w_j}\quad\text{for } k\neq l.
\ee
The matrix $B_{ij}$ is
\be
B_{ij}=J_{ij}(\tfrac{1}{2})\qquad\text{where}\quad J_{ij}(\lambda)\equiv\text{\begin{footnotesize}$\bordermatrix{ &&& k && l &&&\cr
\cr
\cr
k&&& \lambda&& -\lambda\cr
\cr
\cr
l &&&-\lambda&&\lambda\cr
\cr
\cr
}$\end{footnotesize}},\quad\lambda\in\C.
\ee
$J_{ij}(\lambda)$ is called the \textit{Jordan-Pochhammer} $d\times d$ matrix \cite{kohno}. It is well-known that the Jordan-Pochhammer matrix satisfies the infinitesimal braid relations for all $\lambda$. Since it also satisfies the symmetry conditions, $B_{ij}=J_{ij}(\lambda)$ defines a (reducible) representation of the braid group $\cb_d$. This representation is conjugate to the usual \textit{Burau} representation \cite{burau,braid} over the 
ring $\Z[t,t^{-1}]$ where $t=-e^{2\pi i\lambda}$.
For $t=1$, i.e.\! $\lambda=\tfrac{1}{2}\bmod1$, 
the Burau representation factors through the symmetric group $\mathfrak{S}_d$, so the braid representations from the UV Berry connection
of the hyperelliptic models is somehow ``trivial'':
\be
g_i \mapsto \begin{pmatrix}
1_{i-1} &&\\
& -\sigma_1 &\\
&& 1_{n-i-1}\end{pmatrix},
\ee
where $1_k$ is the $k\times k$ unit matrix and $\sigma_1$ the usual Pauli matrix. 
\end{exe}


\begin{rem} Notice that very complete $tt^*$ geometries corresponds to 2d (2,2) models
with no non-trivial wall-crossing phenomena. 
\end{rem}

\subsection{A fancier viewpoint: $\cq$-reconstruction}

\subsubsection{Generalities}

$tt^*$ geometry (in the domain where $\mathscr{R}_x$ is semi-simple) may be  stated in a fancier language \cite{ising,Cecotti:1992xg}.
 Let $F(x,\bar x)$ ($x\in\mathscr{X}$) be a physical quantity which is \textsc{susy} protected, i.e.\! invariant under continuous deformations of the $D$-terms: all $tt^*$ quantities have this property. Since the $w_i$ are local coordinates in $\mathscr{X}$, we may rewrite\footnote{\ We omit writing the dependence on the barred parameters $\bar x$, with the understanding that the functions are not necessarily holomorphic.} $F(x)$ in the form
$F(w_1,\cdots,w_d)$
where $d$ is the Witten index which we assume to be finite.
The functions $
F(w_1,\cdots,w_d)$ enjoy intriguing properties. First of all
\be\label{jjaqw}
F(e^{i\varphi} w_1+a,e^{i\varphi} w_2+a,\cdots,e^{i\varphi} w_d+a)=F(w_1,w_2\cdots,w_d)\quad\forall\;a\in\C,\ \varphi\in\R/2\pi i\Z,
\ee  
since $w_i\to e^{i\varphi}w_i+a$ corresponds to the trivial  deformation of the $F$-terms 
\be
\int d^2\theta\,\cw\to \int d^2\theta\big(e^{i\varphi}\,\cw+a\big)=\int d^2\theta^\prime\, \cw\qquad\text{(where $\theta^\prime=e^{i\varphi/2}\theta$),}
\ee 
which leaves invariant all physical quantities. The group $\C\rtimes\R/2\pi i\Z$ in \eqref{jjaqw} is the 2d Euclidean Poincar\'e symmetry: in this regard the protected functions $F(w_1,\cdots,w_d)$ behave as Euclidean $d$-point functions
\be\label{iasqw1}
F(w_1,\cdots,w_d)\leadsto \big\langle \co_1(w_1)\, \co_2(w_2)\cdots \co_n(w_d)\big\rangle_F.
\ee
This idea may be made more sound if we 
choose the function $F(w_i)$ in a clever way;
we may find $tt^*$ quantities $F$ which obey all the Osterwalder-Scharader axioms for the correlators of an Euclidean QFT except \emph{locality} and \emph{statistics,} i.e.\! univaluedness of the $d$-point functions.

In other words, for an appropriate choice of the $tt^*$ quantity $F$, the only unusual feature of the would-be ``operators'' $\co_k(w)$ in \eqref{iasqw1}
 is that, in general, they are not mutually local but rather have non-trivial braiding properties. The origin of this peculiar fact 
  is easy to understand: $tt^*$ geometry states that suitable combinations of the \textsc{susy} protected quantities satisfy exactly the \emph{same} PDEs as the correlation functions for (the scaling limit of) the off-critical  Ising model (the Sato-Miwa-Jimbo isomonodromic PDEs \cite{MJS1,MJS2,MJS3,MJS4}). The $tt^*$ functions differ from the actual Ising correlators only because the solutions of the PDEs relevant for a given \textsc{susy} model are specified by a different set of  boundary conditions \cite{ising,CV92}. For an isomonodromic system of PDEs, the boundary conditions are encoded in the monodromy representation, i.e.\! in the braiding properties of the $\co_k(w)$. Given the braid group action on the $\co_k(w)$, the $tt^*$ geometry is fully determined.

There is an obvious necessary condition for the existence of a fancy  correspondence like \eqref{iasqw1}.
The $tt^*$ quantities $F_A(w^i)$ must have the following property: they should be regular when the $w_i$ are all distinct, but get singularities of the form 
\be\label{kkasqwyy}
(w_i-w_j)^{-\Delta_{ij}}(\bar w_i-\bar w_j)^{-\bar\Delta_{ij}}
\ee
as two critical values coalesce together, $w_i-w_j\to 0$. 
More precisely, when two critical values collide in 
the coupling constant space $\mathscr{X}$ we should see an emergent OPE algebra 
\be\label{emope}
\co_i(w_i)\,\co_j(w_j)\sim \sum_\ell C(w_i-w_j)_{ij,\ell}\,\co_\ell(w_i)
\ee
for the ``operators'' $\co_j(w_j)$.
To check that the condition holds, it is convenient to adopt the 2d perspective and work in the set-up where the (2,2) LG is quantized on the line $\R$ at a finite temperature $T=R^{-1}$. The infinite volume Hilbert space decomposes into subspaces $\ch_{i,j}$ of definite \textsc{susy} central charge $Z$: 
the sector $\ch_{i,j}$ is defined by imposing the boundary condition that the field configuration approaches the $i$-th classical vacuum (resp.\! the $j$-th one) as $x\to-\infty$ (resp.\! $x\to+\infty$).  In $\ch_{i,j}$ the \textsc{susy} central charge is $Z_{ij}=2(w_i-w_j)$ \cite{tt*,CV92,Cecotti:1992qh}, and 
the BPS states in $\ch_{i,j}$ ($i\neq j$), if any, have masses $2|w_i-w_j|$. 
A typical protected quantity $F_A(w_i)$ may be computed by a periodic Euclidean path integral over the cylinder,
and hence has the schematic form
\be
F_A(w_i)=\mathrm{Tr}\Big[(-1)^F \cf_A\,e^{-RH}\Big]
\ee 
for some operator $\cf_A$.
Only BPS configurations contribute to \textsc{susy} protected quantities, so that 
\be\label{pppsum}
F_A(w_i)=\sum_{|m\rangle\in\text{M-BPS}}\langle m|(-1)^F\cf_A|m\rangle\langle m|e^{-RH}|m\rangle
\ee
where M-BPS stands for the set of BPS multi-particle $H$-eigenstates. The matrix element 
$\langle m|e^{-RH}|m\rangle$ is suppressed by
a factor $\prod e^{-2R|w_i-w_j|}$ where the product is over the BPS particles in the state $|m\rangle$. The sum \eqref{pppsum} is absolutely convergent if all the masses $|Z_{ij}|$ are non-zero, but
it may get singular as $w_i-w_j\to0$, producing a power-law IR divergence in $F_A(w_i)$ of the general form \eqref{kkasqwyy}. This is the only mechanism which may spoil regularity of $F_A(w_i)$
as a (multivalued) function of the $w_i$.
Since the coupling space OPEs \eqref{emope} encode the monodromy representation,
they fully determine the $tt^*$ geometry. Understanding the leading singularity as $w_i-w_j\to0$ amounts to know how many BPS species become massless in the limit $w_i-w_j\to0$ together with some tricky signs\footnote{\ These signs are akin of the tricky sign in the OPE of $\beta$, $\gamma$ commuting ghosts with respect to the conventional fermion $b$, $c$.} (or, more generally, phases) in the comparison between $\co_i(w_i)\,\co_j(w_j)$ and
$\co_j(w_j)\,\co_i(w_i)$. In particular,  
the limit $\lim_{w_j\to w_i}\co_j(w_j)\co_i(w_i)$ is regular  if and only if the net number of BPS solitons connecting the $i$-th and $j$-th vacua is zero.
\medskip

The idea of the reconstruction approach to $tt^*$ geometry is that {in principle} we can reconstruct a non-local QFT $\cq$ on
the $w$-plane from the $tt^*$ quantities interpreted as certain combinations of (multi-valued) correlation functions. Conversely, if we have a putative non-local QFT $\cq$ we may compute the $tt^*$ quantities by standard field-theoretical technics.
The $\cq$-reconstruction strategy is potentially effective since we know that $\cq$ is a ``free'' theory in the sense that its amplitudes are computed by Gaussian path integrals \cite{ising}.

\subsubsection{$w$-plane OPEs} The $w$-plane theory $\cq$ is modelled on the QFT describing the Ising model off-criticality. The basic degree of freedom is an Euclidean 2d Majorana\footnote{\ Imposing the Majorana condition is equivalent to imposing that the vacuum wave-functions are real. While we may chose a real basis for the wave-function, this is different from the holomorphic basis one uses in $tt^*$ geometry. This change of basis makes the comparison of formulae a little indirect. The relation is
$$
|\psi^{(j)}\rangle_\text{real}=\frac{1}{2}\Big( |E_j\rangle+g_{\bar jk}|E_k\rangle\Big).
$$
From the viewpoint of $tt^*$ taking the fermion to be Dirac rather than Majorana may be more natural.}  \emph{free} spinor 
\be
\Psi(u)=\begin{pmatrix}\Psi_+(u)\\ \Psi_-(u)
\end{pmatrix},\qquad\ u\in\C
\ee 
 of mass $R$ \cite{ising}. 
Locally on the $w$-plane the Lagrangian of $\cq$ may be written simply 
\be\label{lagggr}
\overline{\Psi}(\dsl-R)\Psi.
\ee
What makes $\cq$ non-trivial is the fact that $\Psi$ is not univalued, but rather has complicated branching properties due to the insertion of topological defect operators $\co_k(w_k)$ at the points corresponding to critical values of the superpotential $\cw$ of the original LG model. 

Let us study the singularity in the OPE
\be
\Psi_\pm(u)\,\co_k(w_k)
\ee
when $u\to w^k$. Let $z_\ast\in\ck$ be a critical point of $\cw$
which is mapped to the $k$-critical value $w_k$ by
the map $\cw\colon \ck\to\C$. Since $\mathscr{R}_w$ is semi-simple,
the superpotential is weakly\footnote{\ I.e.\! its critical points are non-degenerate but the critical values are not necessarily all distinct.} Morse, so that in a neighborhood $U\ni z_\ast$ we may find local holomorphic coordinates $z^a$ such that
\be
\cw(z^a;w)=w_k+\frac{1}{2}\sum_a (z^a-z^a_\ast)^2\qquad \text{in }U.
\ee
Working in perturbation theory around the $k$-th classical vacuum $z_\ast$, the situation is indistinguishable from free field theory to \emph{all} orders. Hence locally we see the same singularities as in free field theory. The free-field behavior defines two possible defect insertions at $w_k$: $\mu_k$ and $\sigma_k$. Their OPEs are \cite{ising} 
\be\label{kkazaswe}
\begin{aligned}
\Psi_+(u)\sigma_k(w)&\sim \frac{i}{2}(u-w)^{-1/2}\mu_k(w) &&& \Psi_-(u)\sigma_k(w)&\sim -\frac{i}{2}(\bar u-\bar w)^{-1/2}\mu_k(w)\\
\Psi_+(u)\mu_k(w)&\sim \frac{1}{2}(u-w)^{-1/2}\sigma_k(w) &&& \Psi_-(u)\mu_k(w)&\sim \frac{1}{2}(\bar u-\bar w)^{-1/2}\sigma_k(w),
\end{aligned}
\ee 
up to $O(|u-w|^{1/2})$ contributions.

\subsubsection{Hurwitz data and defect operators}

Although the $tt^*$ defect operators $\mu_k(w)$, $\sigma_k(w)$ have the same OPE singularities with the fermion field $\Psi(u)$  as the Ising order/disorder operators, they are \emph{not} in general mere Ising order/disorder operators since globally they have different topological order/disorder properties. In other words, their insertion makes the multi-valued Fermi field $\Psi(u)$ of the $\cq$ theory to have different monodromy properties.
Let us see how this arises.
\medskip

The fermion $\Psi(u)$ is univalued on a suitable connected cover 
$\mathring{\Sigma}$
of the $w$-plane punctured at the positions $\{w_k\}$ of the defects
\be
\mathring{f}\colon \mathring{\Sigma}\to \C\setminus\{w_k\}. 
\ee 
By the Riemann existence theorem \cite{graph},
we may extend $\mathring{f}$ over the punctures to a \emph{branched} cover of Riemann surfaces,  $f\colon\Sigma\to\bP^1$, branched at $\{w_1,w_2,\cdots, w_d,\infty\}$. 
In ``good'' models the order of the monodromy at $\infty$ is finite.
Let us consider first the special case
that the cover
has a finite degree $m$. Then $f$ is specified by its Hurwitz data  at the
$(d+1)$ branching points \cite{graph}.  The Hurwitz data consist of an
element $\pi_k\in \mathfrak{S}_m$ for each finite
branch point $w_k$, while $\pi_\infty=(\pi_1\pi_2\cdots\pi_d)^{-1}$. 
The monodromy group\footnote{\ Also called the \emph{cartographic} group \cite{graph}.} of the cover, $\mathfrak{Mon}$,
(not to be confused with the $tt^*$ monodromy group $\mathsf{Mon}$\,!) is the subgroup of $\mathfrak{S}_m$ generated by the $\pi_k$'s
\be
\mathfrak{Mon}=\big\langle\pi_1,\pi_2,\cdots,\pi_d\big\rangle\subset\mathfrak{S}_m.
\ee 
Since $\Sigma$ is connected, $\mathfrak{Mon}\subset\mathfrak{S}_m$ acts transitively on $\{1,2,\dots,m\}$. In other words,  
$\{\pi_1,\dots,\pi_d,\pi_\infty\}$ is a \textit{constellation} in $\mathfrak{S}_m$ \cite{graph}. We recall that the \emph{passport} of a constellation \cite{graph} is the list of the conjugacy classes of its permutations $\pi_k$. From the OPEs
\eqref{kkazaswe} we see that, for $k\neq\infty$
the conjugacy class corresponds to the partition
\be
\text{conjugacy class of $\pi_k$}\ \leadsto\  \overbrace{2+2+\cdots +2+2}^{\#\text{critical points over $w_k$}}+1+1+\cdots+1=m.
\ee  
In particular, when $\mathscr{R}$ is semi-simple  $\pi_k$ is an involution for $k\neq\infty$. In a complete\footnote{\ For a non-complete $tt^*$ geometry the following assertion is false.} $tt^*$ geometry \underline{generically} we have just one critical point over each critical value, and $\pi_k$ acts as a reflection in the standard representation of $\mathfrak{S}_m$; then for a semi-simple, complete $tt^*$ geometry \textit{with $m$ finite,}    
$\mathfrak{Mon}$ is a finite rational reflection group, hence the Weyl group of a Lie algebra. 
In this case the order $h$ of $\pi_\infty$ is equal to the order of the adjoint action of the quantum monodromy, that is, to the smallest positive integer $h$ such that $h q_s\in\bN$
for all $s$, where $\{q_s\}$ is the set of $U(1)_R$ charges of the chiral primaries at the UV fixed point of the 2d (2,2) Landau-Ginzburg QFT. Note that the cover $f$ is Galois only for $m=2$. 
\medskip

When $m$ is infinite the geometry is a bit more involved. One still has
$\mathfrak{Mon}=\big\langle\pi_1,\pi_2,\cdots,\pi_d\big\rangle$ where (in the semi-simple case) the $\pi_k$ are involutions. But now 
$\mathfrak{Mon}$ is an infinite group.
In general,

\begin{fact} For a semi-simple $tt^*$ geometry, the topological defect operator $\co_k(w_k)$ inserted at the $k$-th critical value (cfr.\! \eqref{iasqw1})
is specified by the choice between $\sigma$-type and $\mu$-type and the involution $\pi_k\in\mathfrak{Mon}$. 
\end{fact}

\subsubsection{Complete $tt^*$ geometries and Coxeter groups} 
For a complete $tt^*$ geometry, generically\footnote{\ An instance of the non-generic situation is described in \textbf{Example \ref{exnongeneric}}.} $\pi_k$
($k\neq\infty$) consists of just one 2-cycle
$(i_k,j_k)$ interchanging the $i_k$-th sheet of $\Sigma$ with the $j_k$-th one. The $k$-th classical vacuum is to be identified geometrically with the  intersection of these two sheets.\footnote{\ The precise sense of the identification will be clarified momentarily in \S\S.\,\ref{onefield}, \ref{Nfields}.} The absolute number of 2d BPS solitons connecting the $k$-th and $h$-vacua is given
by the number of sheets they share
\be
|\mu_{kh}|=\#\Big(\{i_k,j_k\}\cap \{i_h,j_h\}\Big).
\ee
Indeed, the map $\cw\colon \ck\to\C$ factors through $\Sigma$, and the BPS solitons are just the lifts of the straight segment in $\bP^1$ with end points $w_k$ and $w_h$ to sheets of the cover $\Sigma$ which contain both classical vacua.
In particular, for a \emph{complete} theory the number of 2d BPS solitons between any two vacua is at most 2. 
This can be easily seen directly.  
One has
\be
\begin{array}{c}
|\mu_{kh}|=0 \Leftrightarrow \text{order }\pi_k\pi_h\ \text{is }2\\
|\mu_{kh}|=1 \Leftrightarrow \text{order }\pi_k\pi_h\ \text{is }3\\
|\mu_{kh}|=2 \Leftrightarrow \text{order }\pi_k\pi_h\ \text{is }1
\end{array}
\ee
The fermionic part of the wave-function introduce some extra tricky minus signs.
\medskip

\begin{fact} In a complete $tt^*$ geometry with $d<\infty$ vacua we have a group epimorphism
\be
\mathsf{Cox}(\Gamma)\to \mathfrak{Mon} 
\ee
where $\mathsf{Cox}(\Gamma)$ is the Coxeter group {\rm\cite{coxgr}} with graph $\Gamma$
whose $d-\kappa$ nodes are the equivalence classes $\{1,2,\cdots,d\}/\!\!\sim$ ($k \sim h$ iff $|\mu_{kh}|=2$) two nodes $k\not\sim h$ being connected by $|\mu_{kh}|$ edges.
\end{fact}

Let us illustrate our claims in some  examples.

\begin{exe}\label{isiexe} The simplest instance are the Ising $n$-point functions themselves. In this case $\sigma$, $\mu$ are $\Z_2$ order/disorder operators and we do not need to distinguish them with the subfix $k$; in facts, in the Ising case  $m=2$ and all $\pi_k$'s are the permutation $(12)$. $\Sigma$ is the hyperelliptic curve
\be
y^2=\prod_{i=1}^n(z-w^i)
\ee
of \textbf{Example \ref{hhhyperr}}.
Between any two classical vacua there are precisely two BPS solitons.
No wall-crossing phenomena. $\Gamma$ is the Dynking graph of $A_1$ (one node, no edge)
and $\mathfrak{Mon}=\mathrm{Weyl}(A_1)\cong\Z/2\Z$. 
\end{exe}

\begin{rem} Ising $n$-point functions are not just complete $tt^*$ geometries, they are \emph{very} complete. Indeed, \textbf{Examples 2}, \textbf{3} and \textbf{4} correspond, respectively, to Ising $n=2,3,4$ points.
\end{rem}

\begin{exe}\label{jjasqwe} A (2,2) minimal model of type $\mathfrak{g}\in ADE$ is complete but not very complete for $d\equiv\mathrm{rank}\,\mathfrak{g}\geq3$. The monodromy group $\mathfrak{Mon}\cong \mathrm{Weyl}(\mathfrak{g})$, while $\mathsf{Mon}$ differs by a $\Z_2$ flat bundle due to the aforementioned signs.
For $k\neq\infty$, the involution $\pi_k$ is a reflection
with respect to some root of $\mathfrak{g}$. The monodromy at infinity $\pi_\infty$ belongs to the (unique) conjugacy class of the Coxeter element; its order $h$ is the Coxeter number.\footnote{\ For comparison, the conjugacy class of $\varrho_\infty\in\mathsf{Mon}$, as acting on the root lattice of $\mathfrak{g}$, is $-\pi_\infty$.} The relation $\pi_1\cdots\pi_r\pi_\infty=1$
is the usual expression of the Coxeter element in terms of simple reflections. Vacua $k$ and $j$ are connected
by $s_{kj}-2$ BPS solitons, $s_{kj}$ being the order of
$\pi_k\pi_j$. When $d\geq 3$  we have non-trivial wall-crossing: the several inequivalent BPS chambers are in one-to-one correspondence with the
integral quadratic forms  \cite{ringel}
\be
q(x_i)=\sum_{i=1}^d x_i^2+\sum_{i<j}\mu_{ij}\,x_ix_j,\qquad \mu_{ij}\in\Z
\ee 
$\Z$-equivalent to the Tits form of $\mathfrak{g}$, $\tfrac{1}{2}C_{ij}x_ix_j$, where $C_{ij}$ is the Cartan matrix of $\mathfrak{g}$. In particular, there is a special BPS chamber with soliton multiplicities $|\mu_{ij}|=|C_{ij}|$ for $i\neq j$. In this special chamber $\Gamma$ is just the Dynkin graph of $\mathfrak{g}$.
\end{exe}

\begin{exe} In the previous example we may replace
the finite-dimensional Lie algebra $\mathfrak{g}$ by an affine (simply-laced) Lie algebra $\widehat{\mathfrak{g}}\in\widehat{A}\widehat{D}\widehat{E}$. The $tt^*$ monodromy group $\mathsf{Mon}$ is again a $\Z_2$ twist of $\mathrm{Weyl}(\widehat{\mathfrak{g}})$, and
the cover monodromy group $\mathfrak{Mon}$ is a quotient of $\mathrm{Weyl}(\widehat{\mathfrak{g}})$. 
For $\widehat{g}=\widehat{A}_1$ we get back the case $n=2$ of \textbf{Example \ref{isiexe}}
and $\mathfrak{Mon}=\mathrm{Weyl}(A_1)\cong\mathrm{Weyl}(\widehat{A}_1)/\Z$.
The discrepancy (modulo signs) between $\mathsf{Mon}$ and $\mathfrak{Mon}$ expresses the fact that the affine LG models are asymptotically-free instead of having a regular UV fixed point \cite{CV92}.
The conjugacy class of $\pi_\infty$ is the image of a Coxeter element $c\in\mathrm{Weyl}(\widehat{\mathfrak{g}})$; in facts $(\pi_\infty)_\text{s.s.}\cong e^{2\pi i\boldsymbol{Q}}$ is the element of $\mathrm{Weyl}(\mathfrak{g})$
obtained by reducing the action of $c$ on the root lattice modulo the imaginary root.\footnote{\ See \textbf{Lemma 3.2} in \cite{keller}.}
In the $\widehat{A}_r$ case there are $r$ inequivalent conjugacy classes of Coxeter elements \cite{coxeterB}, hence $r$ inequivalent LG models
whose $\pi_\infty\in GL(r+1,\Z)$ satisfies
\be
(\pi_\infty^p-1)(\pi_\infty^{r+1-p}-1)=0,\qquad p=1,2,\dots,r.
\ee 
Their superpotential reads \cite{Cecotti:2011rv}
\be
\cw(z)=y^p+y^{p-r-1},\qquad p=1,\dots,r.
\ee
The case $\widehat{A}_1$ coincides with \textbf{Example 1}. Again the BPS chambers are related to quadratics forms $\Z$-equivalent to the Tits form of the affine Kac-Moody algebra,
but the relation is no longer one-to-one.
\end{exe}

\begin{exe}\label{kasqwmmmc} We may generalize the construction even further by considering an \textit{extended affine Lie algebra} (EALA)  \cite{eala1,eala2}
\be
\mathfrak{g}_r^{\tiny (\overbrace{1,\dots,1}^{\kappa\ 1's})},\qquad d=r+\kappa,
\ee 
of nullity $\kappa$ and type $\mathfrak{g}_r\in ADE$. These Lie algebras are central extensions of the Lie algebra of maps $(S^1)^\kappa\to\mathfrak{g}_r$; for $\kappa=1$ we get back the affine Kac-Moody algebra $\widehat{\mathfrak{g}}$ and for $\kappa=2$ the toroidal Lie algebras.
The EALA 
$A_1^{(1,1,\dots,1)}$ corresponds to the Ising $(\kappa+1)$-point functions: in this special case the $tt^*$ geometry is very complete, not just complete, and $\Gamma$ reduces to the $A_1$ Dynkin graph.
 The role of the EALA's in the classification of complete 4d $\cn=2$ QFTs is outlined in ref.\!\cite{Cecotti:2015qha}. E.g.\! $D_4^{(1,1)}$ corresponds to $\cn=2$ $SU(2)$ SYM coupled to $N_f=4$ fundamentals; the corresponding 2d (2,2) complete model has superpotential
 \be
 \cw(z)=\frac{P_4(z)}{z^2(z-1)^2(z-\lambda(\tau))^2},
 \ee
 where $P_4(z)$ is a polynomial of degree 4 coprime with the denominator.
\end{exe}

\begin{exe}\label{exnongeneric} Let us consider the minimal $A_{2m-1}$ models at a maximally \emph{non-}generic point in coupling space. We take $\cw(z)$ to be proportional to the square of the $m$-th Chebyshev polynomial\footnote{\ Up to fields redefinitions, it is the same as the model with the Chebyshev superpotential $T_{2m}(z)$.}, $\cw(z)=\Delta w\; T_m(z)^2$.
The superpotential is a Belyj function with Grothendieck \emph{dessin d'enfants} \cite{des1,des2,des3}
\be\label{kkasweii}
\xymatrix{\bullet\ar@{-}[r]&\circ\ar@{-}[r]&\bullet\ar@{-}[r]&\circ\ar@{-}[r]&\bullet\ar@{-}[r]&\cdots\cdots\ar@{-}[r] & \circ\ar@{-}[r]&\bullet\ar@{-}[r]&\circ\ar@{-}[r]&\bullet}
\ee
Erasing the two black nodes at the ends -- which do not correspond to \textsc{susy} vacua -- we get back the Dynkin graph of $A_{2m-1}$.
The SQM monodromy group $\mathfrak{Mon}$ coincides with the cartographic group of the \emph{dessin} \eqref{kkasweii}: it is generated by two involutions $\pi_\bullet,\pi_\circ\in\mathfrak{S}_{2m}\equiv\mathrm{Weyl}(A_{2m-1})$ associated to the
black/white nodes
\be
\pi_\circ=\prod_{i=1}^m s_{2i-1},\qquad\pi_\bullet=\prod_{i=1}^{m-1}s_{2i}
\ee 
where $s_j\in \mathrm{Weyl}(A_{2m-1})$ is the $j$-th simple reflection. It is well-known that
$\pi_\infty=\pi_\circ\pi_\bullet\in \mathrm{Weyl}(A_{2m-1})$ is a Coxeter element.  
\end{exe}

 \subsubsection{The case of one chiral field} \label{onefield}

The relation between the non-local QFT $\cq$ on the $w$-plane and the 4-supercharge SQM is especially simple when the superpotential depends on a single chiral field $z$, $W(z; w_1,\cdots, w_d)$. The actual Schroedinger wave-function of the $i$-th vacuum
(the one which corresponds to the idempotent $e_i\in\mathscr{R}_w$ under the isomorphism $\mathscr{V}_w\cong\mathscr{R}$), 
written as a one-form through the $\zeta$-dependent identification
\be
\chi^\dagger|0\rangle\leadsto \frac{dz}{\zeta},\qquad \bar\chi^\dagger|0\rangle \leadsto
\zeta\,d\bar z,
\ee
is (for $|\zeta|=1$)
\be\label{schr}
\begin{split}
\psi_i(z;\zeta)&=\big\langle \Psi_+(W(z))\, \mu_1(w_1)\cdots \mu_{i-1}(w_{i-1})\sigma_i(w_i)\mu_{i+1}(w_{i+1})\cdots\mu_d(w_d)\big\rangle\, \frac{W^\prime(z)\,dz}{\zeta\,\tau(w_j)}+\\
&+\big\langle \Psi_-(W(z))\, \mu_1(w_1)\cdots \mu_{i-1}(w_{i-1})\sigma_i(w_i)\mu_{i+1}(w_{i+1})\cdots\mu_{d}(w_d)\big\rangle\, \frac{\overline{W}^\prime(\bar z)\,\zeta\,d\bar z}{\tau(w_j)}
\end{split}
\ee
where the normalization constant $\tau(w^j)$ is the Sato-Miwa-Jimbo $\tau$-function \cite{MJS1,MJS2,MJS3,MJS4,ising}
\be
\tau(w_k)=\big\langle \mu(w_1)\cdots \mu(w_d)\big\rangle.
\ee
It is easy to check that the free massive Dirac equation satisfied by $\Psi_\pm(z)$ is equivalent to the zero-energy Schroedinger equation for $\psi_i(z;\zeta)$. The Hurwitz data should be chosen so that $\psi_i(z;\zeta)$ is univalued in $z$ for the given $W(z)$.
The exact brane amplitudes then have the form
\be
\langle i | \Gamma,\zeta\rangle=\int_\Gamma e^{-(RW/\zeta+R\zeta \overline{W})}\,\psi_i(z;\zeta),
\ee
where the relative one-cycle $\Gamma$ is the support of the brane.
Notice that $\langle i | \Gamma,\zeta\rangle$ depends only on the image of 
$\Gamma$ in the curve $\Sigma$, i.e.\! in 
the smallest branched cover of the $W$-plane on which the wave-functions are well defined.
\medskip

If the Stein manifold $\ck$ is one-dimensional, we may lift the condition that the chiral ring $\mathscr{R}_x$ is semi-simple. In facts, by the Chinese remainder theorem,\footnote{\ See footnote \ref{f12}.} we  have
\be
\mathscr{R}\cong \Gamma(\ck,\co_\ck)/(\partial_zW)\cong\prod_k \C[z]/(z^{\upsilon_k-1}),
\ee
where the product is over the distinct zeros of $\partial_zW$ and $\upsilon_k$ are their orders.
In correspondence to the critical value $w_k$ we have to insert in the $w$-plane a topological operator which
introduces a $\upsilon_k$-th root cut instead of a square-root cut as in the semi-simple case ($\upsilon_k=2$). The two-fold choice of spin operators $\mu$, $\sigma$ gets replaced by
a $(\upsilon_k+1)$-fold choice of 
topological insertions $\tau_{s_k}$ ($s_k=1,\dots, \upsilon_k+1$),  to be supplemented by an element $\pi_k\in\mathfrak{S}_m$ of order $(\upsilon_k+1)$.
One easily checks that, with these prescriptions, eqn.\eqref{schr} reproduces the correct vacuum wave-functions whenever they are known from other arguments \cite{iosqm}.

\subsubsection{$N$-fields: a formula for the brane amplitudes}\label{Nfields}

Suppose now that we have a complete LG model with $N$ chiral fields, i.e.\! $\dim \ck=N$.
The inverse image of a point $w$ in the $W$-plane has the homotopy type of a bouquet of
$(N-1)$-spheres \cite{milnor}\!\!\cite{CV92,iqbal}; we fix a set of $(N-1)$-cycles $S_\alpha(w)$ ($\alpha=1,\dots, d\equiv$ the Witten index) which form a basis of the homology of the fiber. The SQM wave-function of a \textsc{susy} vacuum $\Psi$ is a $N$-form on $\ck$ so
\be\label{kasqwzxa}
\psi_\alpha(w)=\int\limits_{S_\alpha(w)}\Psi
\ee
is a $d$-tuple of one-forms on the $W$-plane.
If we transport the homology cycles $S_\alpha(w)$ along a closed loop in the $W$-plane (punctured at the critical values) we come back with a different (integral) basis of $(N-1)$-cycles $S^\prime_{\alpha}(w)= N_{\alpha\beta}\,S_\beta(w)$. The integral matrix $N_{\alpha\beta}$ is described by the Picard-Lefshetz theory \cite{CV92,witten1,witten2}. Thus \eqref{kasqwzxa} is best interpreted as a single but multi-valued wave-function $\psi(w)$ on the $W$-plane branched at the critical points whose monodromy representation is determined by the Picard-Lefshetz formula in terms of the intersection matrix $S_\alpha\cdot S_\beta$, i.e., in physical terms \cite{CV92,iqbal}, by the BPS spectrum of the corresponding 2d model.
Let $\Sigma$ be the minimal branched cover of
the $W$-plane such that $\psi(w)$ is uni-valued ($\Sigma$ is then automatically Stein \cite{stein1}). Clearly the map $W\colon\ck\to \C$ factorizes through $\Sigma$. Let $\Gamma$ be the image in $\Sigma$ of the support of the brane $\mathscr{B}$. Then
\be
\langle \Psi |\, \Gamma,\zeta\rangle=\int_\Gamma e^{-(R w/\zeta+R\zeta \overline{w})}\,\psi(w;\zeta)
\ee
where, in terms of $\cq$-amplitudes ($|\zeta|=1$),
\be\label{schr2}
\begin{split}
\psi(w;\zeta)&=\big\langle \Psi_+(w)\, \mu_1(w_1)\cdots \mu_{i-1}(w_{i-1})\sigma_i(w_i)\mu_{i+1}(w_{i+1})\cdots\mu_d(w_d)\big\rangle\, \frac{dw}{\zeta\,\tau(w_j)}+\\
&+\big\langle \Psi_-(w)\, \mu_1(w_1)\cdots \mu_{i-1}(w_{i-1})\sigma_i(w_i)\mu_{i+1}(w_{i+1})\cdots\mu_{d}(w_d)\big\rangle\, \frac{\zeta\,d\bar w}{\tau(w_j)}
\end{split}
\ee
In other words, an $N$-field LG model with a Morse superpotential may be replaced by a one-field LG model with $\ck=\Sigma$ and superpotential $w\colon \Sigma\to\C$ given by the factorization of
$W$ through $\Sigma$.  

\subsubsection{$\tau$-functions vs.\! brane amplitudes}

From the isomonodromic viewpoint, the most important \textsc{susy} protected function is the $\tau$-function \cite{MJS1,MJS2,MJS3,MJS4}\!\!\cite{ising} i.e.\! the $d$-point function of the would-be order operators 
\be
\tau(w_j)= \big\langle \mu(w_1)\mu(w_2)\cdots \mu(w_d)\big\rangle.
\ee
$\tau(w_j)$ is just the partition function of a free  fermion with the non-trivial monodromy properties implied by the insertions of the $\mu$'s at the points $w^j$. Stated in a different language, it is the partition function of a free massive fermion on $\Sigma$ with suitable boundary conditions at $w_k$ and infinity.
The $\tau$ function may be recover from the vacuum wave $\psi(u;w_j)$ by quadratures \cite{ising}; geometrically $\tau$ is given by
the formula
\be
\log\tau=-K+\log\det(1+g)+\sum_iw_i\bar w_i,
\ee 
where $g$ is the $tt^*$ metric in the canonical bais and $K$ is the K\"ahler potential for the  metric on the 
coupling space $\mathscr{X}$ \cite{ising}
\be
K_{i\bar j}=\partial_i\partial_{\bar j} K\equiv\mathrm{tr}\big(C_i\overline{C}_j\big).
\ee

\begin{exe}\label{lllasq04} Consider the case of just two vacua, and let 
\be
G_\pm=\exp\!\big(\pm u(L)\big),\qquad\text{where } L=2R|w_1-w_2|,
\ee 
be the $tt^*$ metric of the symmetric/antisymmetric linear combinations of the two vacua. By definition
\be
\text{eigenvalues of }\boldsymbol{Q}=\pm \frac{\hat c}{2}=\pm \lim_{L\to 0}\frac{L}{2} \frac{d}{dL} u(L)
\ee
In this case the $tt^*$ equations reduce to a special form of Painlev\'e III \cite{McTW,McTW2,its1,its2}
\be
\frac{d^2 u}{dL^2}+\frac{1}{L}\,\frac{d u}{dL}=2 \sinh(2 u),\qquad L>0
\ee
the solutions which are regular for $0<L<\infty$  satisfy 
\be
u(L)\sim -r\log L+\text{subleading}\quad\text{as }L\to0,
\ee
with $-1\leq r\leq 1$. $r$ characterizes the regular solution completely \cite{McTW,McTW2,its1,its2}.  Changing the sign of $r$ just interchange order and disorder; we fix our conventions so that $r$ is non-negative. For a regular \emph{arithmetic} solution $r$ is the rational number such that \cite{CV92}
\be\label{jjjjjaqb}
\Big|2\,\sin\frac{\pi r}{2}\Big|=\#\big(\text{BPS solitons in 2d}\big)\quad\Rightarrow\quad 0\leq r\leq 1.
\ee
Then $\hat c=r$. In particular, $r=1$ for the Ising two-point functions, that is, for the 2d $\bP^1$ $\sigma$-model.
For a 2-vacua model one has \cite{McTW,tracey}
\begin{align}\label{ise1}
\tau=\big\langle\mu(w_1)\,\mu(w_2)\big\rangle= R^{(2r-r^2)/4}\,e^{u(L)/2} &\exp\!\left[\frac{1}{4}\int_L^\infty\left(s \sinh^2 u(s)- s\!\left(\frac{d u(s)}{ds}\right)^{\!2}\right)\!ds \right]\\
\intertext{while}
\big\langle\sigma(w_1)\,\mu(w_2)\big\rangle=R^{(2r-r^2)/4} e^{-u(L)/2} &\exp\!\left[\frac{1}{4}\int_L^\infty\left(s \sinh^2 u(s)- s\!\left(\frac{d u(s)}{ds}\right)^{\!2}\right)\!ds \right]\!.\label{ise2}
\end{align} 
\end{exe}

\subsubsection{UV limit: the $\cq$ conformal blocks}

In the physical 2d (2,2) LG model, the UV limit consists in sending to zero the radius $R$ of the circle $S^1$ on which we quantize the theory. But $R$ is also the mass of the Majorana fermion in the $\cq$ theory, see eqn.\eqref{lagggr}. Hence the physical UV limit of the 2d LG model coincides with the UV limit of the $\cq$ theory on the $w$-plane.
\medskip

As $R\to 0$ the $\cq$ theory gets critical, the left and right modes of the fermion $\Psi(u)$ decouple,
and the multi-valued would-be correlation functions
\eqref{iasqw1} become sums of products of \emph{bona fide} left/right 
conformal blocks. 

The statement  holds (roughly) for all $tt^*$ quantities: in the UV they become some
complicate combination of conformal blocks. Then the differential equations they satisfy -- the $tt^*$ equations --
should be related in a simple way to the PDEs for the conformal blocks: the analogue of the BPZ equations for the conformal blocks of the $(p,q)$ minimal models \cite{BPZ} and Knizhnik-Zamolodchikov equations for the 2d current algebra \cite{KZ}.
Both sets of equations define flat connections and
monodromy representations\footnote{\ See e.g.\! chap.\!\! XIX of \cite{QG}; for the minimal model case, see \cite{fro}.}. As already mentioned, they are  specializations of the universal Kohno monodromy \cite{kohno}. From this viewpoint the fact that 
 in the UV limit the Berry connection
($\equiv$ the $tt^*$ Lax one in the limit) has the
Kohno form -- typical of the monodromy action on 
conformal blocks -- comes as no surprise.
That things should work this way is somehow obvious in the case of \textbf{Example \ref{isiexe}} where 
the $tt^*$ geometry describes correlations of the Ising model off-criticality: sending the mass to zero $R\to0$, we just get the critical Ising model ($\equiv$ the minimal (4,3) CFT), and
the PDEs of the massive case should reduce to the 
conformal ones. 

In connecting the $tt^*$ monodromy with the braid representation of $\cq$ blocks, we need to use the precise disctionary between the two. 
From eqn.\eqref{schr2} we see that brane amplitudes, being normalized, are to be seen as \emph{ratios} of $n$-point functions in $\cq$ theory
\be\label{rrratio}
\frac{\langle\Psi_\pm(z) \mu(w_1)\cdots\sigma (w_j)\cdots\mu(w_n)\rangle}{\langle \mu(w_1)\cdots \mu(w_n)\rangle}
\ee
rather than correlators. Hence the actual braid representation on the $\cq$ theory operators is the $tt^*$ one twisted by the one defined by the $\tau$-function. In this way one solves an apparent problem with \textbf{Example \ref{hhhyperr}}: there the $tt^*$ UV Berry monodromy factorizes through $\mathfrak{S}_n$, whereas the braiding action of Ising blocks do not. Taking into account the twist by $\tau$ solves the problem. Moreover, the ratio \eqref{rrratio} does not correspond to the amplitude written in a holomorphic trivialization of the vacuum bundle $\mathscr{V}$.
\medskip

We continue  \textbf{Example \ref{lllasq04}}.

\begin{exe}
The asymptotics of the amplitudes \eqref{ise1},\eqref{ise2} as $L\sim0$ is
\cite{tracey}
\be
\langle \mu(w_1)\mu(w_2)\rangle\sim \mathrm{const.}\left(\frac{R}{L}\right)^{\!(2r-r^2)/4} \qquad \langle \sigma(w_1)\mu(w_2)\rangle\sim  \mathrm{const.}\, R^r\!\left(\frac{R}{L}\right)^{\!-(2r+r^2)/4}.\label{ooo098p}
\ee
So in the limit $R\to0$ the second correlation vanishes, whereas the first one becomes
the CFT 2-point function
 \be
\langle \mu(w_1)\,\mu(w_2)\rangle= \frac{\mathrm{const.}}{|w_1-w_2|^{\!(2r-r^2)/4}}
\ee
which says that the conformal 2d fields $\mu$ and $\sigma$ have dimensions\footnote{\ From the OPEs \eqref{kkazaswe} we see that the two fields have the same dimension.}
\be
\big(h,\tilde h\big)=\left(\frac{2r-r^2}{16},\frac{2r-r^2}{16}\right)
\ee
whereas the second equation \eqref{ooo098p} says that two $\mu$'s have a second fusion channel, besides the identity,
\be
\mu\cdot\mu=[1]+[\co],
\ee 
the primary field $\co$ having dimension $h_\co$
\be
2 h-h_\co= -\frac{2r+r^2}{8}\quad\Rightarrow\quad h_\co=\frac{r}{2}.
\ee
Setting $r=1$ we recover well known properties of the Ising model. 

For the left-movers of the critical Ising model we have
\be
\frac{\langle\psi(z) \sigma(w_1)\mu(w_2)\rangle}{\langle \mu(w_1)\,\mu(w_2)\rangle}=\mathrm{const.}\frac{(z-w_1)^{-1/2}(z-w_2)^{-1/2} (w_1-w_2)^{1/2-1/8}}{(w_1-w_2)^{-1/8}}
\ee
and
\be
\int_\Gamma\!\! dz\;\frac{\langle\psi(z) \sigma(w_1)\mu(w_2)\rangle}{\langle \mu(w_1)\,\mu(w_2)\rangle}=\mathrm{const.}\;(w_1-w_2)^{1/2}\!\int_\Gamma\frac{dz}{\sqrt{(z-w_1)(z-w_2)}}
\ee
The integral is regular as $w_1\to w_2$ and is a constant. Hence
\be
\left(d-\frac{1}{2}\frac{d(w_1-w_2)}{w_1-w_2}\right)\int_\Gamma\!\! dz\;\frac{\langle\psi(z) \sigma(w_1)\mu(w_2)\rangle}{\langle \mu(w_1)\,\mu(w_2)\rangle}=0.
\ee
\end{exe}

The $w$-plane CFT method works better if the underlying $tt^*$ geometry is very complete (as in the Ising cases). This leads to the idea of
computing the $tt^*$ monodromy representation by guessing the ``CFT $\cq$'' on the $w$-plane. 
However to do so one has to establish a precise dictionary between correlators in the $\cq$ theory and $tt^*$ quantities.

\subsection{Relation with $\mathfrak{sl}(2)$ Hecke algebra representations}\label{lllz10n}

We need to look more in detail to the matrices $B_{ij}$ in the UV Berry connection \eqref{uniffr} for a very complete $tt^*$ geometry. We already computed them
for \textbf{Example \ref{hhhyperr}}.

We consider the $B_{ij}$'s from the point of view of the $\cq$ theory on the $w$-plane. We put ourselves in the generic case, where the critical values $w_k$ are all distinct, although the argument goes through even without this assumption.\footnote{\ Cfr.\! \textbf{Example \ref{exnongeneric}}.} Since $B_{ij}$ is the residue of the pole of $\ca$ as $w_i-w_j\to 0$, we  focus on this limit from the viewpoint of the 2d (2,2) LG model. Without loss, we may deforme the $D$-terms so that the only light degrees of freedom are the BPS solitons interpolating between vacua $i$ and $j$ of mass $2|w_i-w_j|$.
We may integrate out all other degrees of freedom, and we end up with an effective IR description with just these two \textsc{susy} vacua.\footnote{\ A theory with just 2 vacua is not UV complete if the number of BPS species connecting them is more than 2 \cite{Cecotti:2011rv}, but here UV completeness is not an issue since we use the two-vacua theory just as an effective low-energy description valid up to some non-zero energy scale.} From the viewpoint of
SQM, the 2d BPS solitons look BPS \textit{instantons.}
The effect of these BPS instantons is to split the two vacua not in energy as it happens in non-\textsc{susy} QM -- vacuum energy is \textsc{susy} protected\,! --  but in the  charge $q$ of the $U(1)_R$ symmetry which emerges in the  $w_i-w_j\to 0$ limit. In this limit there is also an emergent $\Z_2$ symmetry interchanging the two ``classical''\footnote{\ By ``classical'' vacua we mean the \emph{quantum} vacua which under the isomorphism $\mathscr{R}\cong\mathscr{V}$ correspond to the idempotents of the chiral ring.} vacua, so the $U(1)_R$ eigenstates should be the symmetric and anti-symmetric linear combinations of the two ``classical'' vacua.
Their charges $q$ should be opposite by PCT, and may be computed from eqn.\eqref{jjjjjaqb} 
\be
2\,\sin(\pi q)=\pm\, \#\big(\text{BPS instantons}\big).
\ee 
Two simplify the notation, we renumber the $w_k$ so that $w_i$, $w_j$ become $w_1$ and $w_2$.
Then, with a convenient choice of the relative phases of the two states, the upper-left $2\times 2$ block of the matrix $\boldsymbol{Q}$ takes the form
\be\label{kaqw}
\boldsymbol{Q}\Big|_{\text{upper-left}\atop\text{block}}=-\lambda\begin{pmatrix} 0 & 1\\ 1 & 0\end{pmatrix}\equiv -\lambda\,\sigma_1\!.
\ee
This formula holds for the canonical trivialization; in a ``natural'' trivialization we have a shift by a constant multiple of 1. To be fully general, we allow a shift of the \textsc{rhs} of \eqref{kaqw} by $\mu$
\be\label{kaqw2}
-\lambda\begin{pmatrix} 0 & 1\\ 1 & 0\end{pmatrix}\to
-\lambda\begin{pmatrix} 0 & 1\\ 1 & 0\end{pmatrix}+\mu\begin{pmatrix}1 & 0\\ 0 & 1\end{pmatrix}.
\ee

\begin{exe} For the mirror of the $\bP^1$ $\sigma$-model (\textbf{Example 1}) one has
\be
\lambda=\frac{\hat c_\text{uv}}{2}=\frac{1}{2},\qquad
\mu=\frac{\#\text{(chiral fields)}}{2}=\frac{1}{2}.
\ee
\end{exe}
\medskip

At each of the two critical points $w_{1,2}$  we have a two-fold choice: 
 we may insert either a $\sigma$-like defect or
 a $\mu$-like one.\footnote{\ It is convenient to make complex the Ising fermion $\Psi$ \cite{ising}; then the two-fold choices corresponds to the two components of the spin operator for the free fermion system.   } From eqn.\eqref{schr} we see different choices correspond to different vacua of the original LG model.
 The matrix $\sigma_1$ in \eqref{kaqw} has the effect of flipping $\sigma\leftrightarrow\mu$ in the two-vacua system.
It is therefore convenient to introduce a two-component notation
\be
\Sigma_{k,\alpha}(w)=\begin{pmatrix}\sigma_k(w)\\
\mu_k(w)\end{pmatrix},
\ee 
and write the UV $\cq$-amplitudes (conformal blocks) for the effective two vacua theory in the form
\be
\Big\langle \cdots\, \Sigma_{1,\alpha}(w_1)\, \Sigma_{2,\beta}(w_2)\Big\rangle \in \boldsymbol{V}_{\!\!(1)}\otimes \boldsymbol{V}_{\!\!(2)}
\ee
where $\boldsymbol{V}_{\!\!(a)}\cong \C^2$, $a=1,2$,
are two copies
of the representation space of $\mathfrak{sl}(2,\C)$.
Notice that the amplitudes span only a subspace of $\boldsymbol{V}_{\!\!(1)}\otimes \boldsymbol{V}_{\!\!(2)}$
of dimension 2, two linear combinations vanishing since they are bounded holomorphic functions on the cover $\Sigma$ which vanish at infinity.

Acting on these blocks, the matrix \eqref{kaqw} reads
\be
B_{12}=-\lambda\,\Big(\sigma_+^{(1)}\otimes \sigma_-^{(2)}+
\sigma_-^{(1)}\otimes \sigma_+^{(2)}\Big)-\mu\, \sigma^{(1)}_3\otimes \sigma^{(2)}_3
\ee 
where $\sigma^{(a)}_\ell$ is the Pauli matrix acting on the $(a)$-copy $\boldsymbol{V}_{\!\!(a)}$ of $\C^2$ and $\lambda$ is a constant. 
\medskip

We are led to conclude that the UV Berry  connection $\cd$ of a very complete $tt^*$ geometry with $d$ \textsc{susy} vacua   must have the general form 
\be\label{rrrrtc}
\cd=d-2\sum_{i<j} \Big(\lambda_{i,j}\, s_\ell^{(i)}\,s_\ell^{(j)}+\mu_{i,j}\,s_3^{(i)}\,s_3^{(j)}\Big) \frac{d(w^i-w^j)}{w^i-w^j}
\ee
acting on sections of a bundle $\boldsymbol{\cv}\to \mathscr{X}$
whose fibers are modelled on the vector space
\be
\boldsymbol{V}^{\otimes d}\equiv\boldsymbol{V}_{\!\!(1)}\otimes\boldsymbol{V}_{\!\!(2)}\otimes\cdots\otimes\boldsymbol{V}_{\!\!(d)},
\ee
$s_\ell^{(a)}$ being the $\mathfrak{su}(2)$ generators which act on
the
$\boldsymbol{V}_{\!\!(a)}\cong \C^2$ factor space, i.e.
\be
s_\ell^{(a)}=1\otimes\cdots\otimes 1\otimes \overbrace{\frac{1}{2}\sigma_\ell}^{\text{$a$-th}}\otimes 1\otimes\cdots\otimes 1,\qquad \ell=1,2,3.
\ee
The natural connection on the ``$\cq$ conformal blocks'' may differ from $\cd$ by a line bundle twisting; for instance, in the Ising case we have the normalization factor $\langle\mu(w_1)\cdots\mu(w_g)\rangle^{-1}$. This corresponds to replacing $\cd\to \cd+ \boldsymbol{1}\cdot d\log f$ for some multivalued holomorphic function $f$, the ``normalization factor''. We shall omit this term which may be easily recovered using the reality constraint.

The actual brane amplitudes live in a rank $d$ sub-bundle 
$\mathscr{V}$ of the rank $2^d$ bundle $ \boldsymbol{\cv}$.
The $tt^*$ Lax equations requires this sub-bundle to be preserved by parallel transport with the connection $\cd$. To see this, consider the total angular momentum
\be
L_\ell=\sum_a s_\ell^{(a)},\qquad \ell=1,2,3.
\ee
$L_3$ commutes with $\cd$, so the eigen-bundles 
$\mathscr{V}_{m}\subset\boldsymbol{\cv}$ of given
$L_3$ are preserved by parallel transport. The vacuum bundle corresponds to the eigen-bundle of $L_3$ with eigenvalue $m=1-d/2$
\begin{gather}\label{kafswe}
\mathscr{V}=\mathscr{V}_{1-d/2}.
\end{gather}
The constants $\lambda_{i,j}$, $\mu_{i,j}$ in eqn.\eqref{rrrrtc} are restricted by two conditions:
\begin{itemize} 
\item[1)]$\cd$ is flat acting on the sub-bundle $\mathscr{V}$;
\item[2)]  the monodromy representation of $\cd$ is ``arithmetic''.
\end{itemize}
If, in addition, the very complete $tt^*$ geometry is symmetric:
\begin{itemize}
\item[3)] the constants
$\lambda_{i,j}$, $\mu_{i,j}$ should be  independent of $i,j$.
\end{itemize}

\subsubsection{The Knizhnik-Zamolodchikov equation}
A well-known solution to condition 1) is \cite{kohno0,kohno}:
\be\label{kzsol}
\lambda_{i,j}=\frac{\lambda}{2}\quad \mu_{i,j}=0\quad\text{for all }i,j,
\ee
that is,
\be\label{jasqw2}
\cd=d-\lambda\sum_{i<j} \frac{s_\ell^{(i)}\, s_\ell^{(j)}}{w_i-w_j}\, d(w_i-w_j).
\ee
This connection
is automatically flat for all $\lambda$ when acting on sections of the big bundle $\boldsymbol{\cv}$
since its coefficients are given by the universal $\mathfrak{sl}(2)$ $R$-matrix \cite{Rmat}; 3) is also satisfied. We shall see momentarily that condition 2) reduces to $\lambda\in\bQ$.

On the other hand, it is easy to check that the only symmetric (i.e.\! independent of $i,j$) solution to the flatness condition for a connection of the form \eqref{rrrrtc} is given by
eqn.\eqref{kzsol}. We see this observation
 as the basic evidence than a symmetric 
 very complete $tt^*$ geometry has a UV Berry connection of the form \eqref{jasqw2}. 
 \medskip

Since $\cd$ is flat on the larger bundle $\boldsymbol{\cv}$,
 the (physical) UV limit of the $tt^*$ linear problem\footnote{\ The equations do not contain the twistor parameter $\zeta$ any longer. Indeed, $\zeta$ is essentially the phase of the \textsc{susy} central charge, but in the superconformal algebra which emerge in the  UV the central charge should be the zero operator by the Haag--Lopusza\'nski--Sohnius theorem \cite{HLS}.},
$\cd\Psi=0$ with $\Psi\in\Gamma(\mathscr{X},\mathscr{V})$,
may extended to the big bundle
\be\label{kzbig}
\cd\boldsymbol{\Psi}=0\qquad \boldsymbol{\Psi}\in\Gamma(\mathscr{X},\boldsymbol{\cv}).
\ee
This 
equation is the celebrated $\mathfrak{sl}(2)$ Knizhnik-Zamolodchikov for the $d$-point functions 
in the 2d WZW model with group $SU(2)$
\cite{KZ}.
In that context $\lambda$ is quantized in discrete values
\be
\lambda=\frac{2}{\kappa+2},\qquad \kappa\in\Z,\
\ee 
for the 2d $SU(2)$ current algebra at level $\kappa$.

Since the connection \eqref{jasqw2} is invariant under
the symmetric group $\mathfrak{S}_d$,  the representation $\sigma$ of $\mathscr{P}_d$ given by its monodromy extends to a representation of the full braid group $\cb_d$ in $d$ strings.
A representation of this form is called a \textit{Hecke algebra representation} of $\cb_d$ \cite{kohno,jones} since it factorizes through a 
(Iwahori--)Hecke algebra \cite{braid}. 

By the argument leading to eqn.\eqref{kafswe}, parallel transport by the
Knizhnik-Zamolodchikov connection $\cd$
preserves the $tt^*$ vacuum sub-bundle $\mathscr{V}\subset\boldsymbol{\cv}$.
In facts more is true: Indeed, $\cd$  preserves all eigen-bundles $\mathscr{V}_{l,m}\subset
\mathscr{V}$ of given total angular momentum
\begin{gather}
\psi\in \mathscr{V}_{l,m}\quad\Leftrightarrow\quad \big(L^2-l(l+1)\big)\psi=\big(L_3-m\big)\psi=0,\\
m=l,l-1,l-2,\cdots, -l,\qquad l\in\frac{1}{2}\,\mathbb{N}.
\end{gather} 
Comparing with \eqref{kafswe}
\begin{gather}
\mathscr{V}=\mathscr{V}_{d/2,1-d/2}\oplus \mathscr{V}_{d/2-1,1-d/2}\\
\mathrm{rank}\,\mathscr{V}_{d/2,1-d/2}=1,\quad \mathrm{rank}\,\mathscr{V}_{d/2-1,1-d/2}=d-1.
\end{gather}
The fiber of $\mathscr{V}_{d/2,1-d/2}$
is spanned by a \emph{unique} vacuum preserved by the monodromy representation. It has the properties expected for the preferred vacuum $|\textsf{vac}\rangle$ of \S.\,\ref{univacc}.
\medskip

As we shall see momentarily, the monodromy representation of the braid group $\cb_d$ defined by restricting the
Knizhnik-Zamolodchikov connection to  the $tt^*$ sub-bundle $\mathscr{V}$
is isomorphic to the Burau one \cite{braid,birm}.

\begin{rem} The identification of $\cd|_{\mathscr{V}}$
with the UV Berry connection of a very complete $tt^*$ geometry entails that its monodromy representation is unitary.
It is known that the Burau representation is unitary \cite{squ,braid}.
\end{rem}

\subsubsection{Hecke algebra representations}

The presentation of the Artin braid group $\cb_d$ is given in eqn.\eqref{bdrel}. Let $q\in \C^\times$.
The Hecke algebra of the symmetric group $\mathfrak{S}_d$, $H_d(q)$,
is the $\C$-algebra \cite{braid} with generators 
\be
1,\ g_1,\ g_2,\ \cdots,\ g_{d-1}
\ee
and relations
\begin{align}\label{hacck}
&g_ig_{i+1}g_i=g_{i+1}g_ig_{i+1}, && g_ig_j=g_jg_i\ \text{for }|i-j|\geq 2, &&
(g_i+1)(g_i-q)=0.
\end{align}
$H_d(1)$ is simply the group algebra $\C[\mathfrak{S}_d]$ of the symmetric group $\mathfrak{S}_d$. If $q$ is not a root of unity of order $2\leq n\leq d$, 
$H_d(q)$ is semisimple \cite{braid} and its simple modules are $q$-deformations of the irreducible representations of
$\mathfrak{S}_d$. If $q$ is a non-trivial root of unity new interesting phenomena appear \cite{atunit}.

Comparing eqns.\eqref{bdrel},\eqref{hacck} we see that the correspondence
$\sigma_i\mapsto g_i$ yields an algebra homomorphism 
\be
\varpi\colon\C[\cb_d]\to H_d(q).
\ee
 A linear representation $\varrho$ of the braid group $\cb_d$ is called a
\emph{Hecke algebra representation} if it factorizes through $\varpi$. In such a representation the generators $\varrho(g_i)$ have at most two distinct eigenvalues: $-1$ and $q$.

The Hecke algebra may be rewritten in terms of the generators $e_i=(q-g_i)/(1+q)$ subjected to the relations
\begin{align}\label{rrrrelll23}
&e_i^2=e_i && e_ie_j=e_je_i\quad \text{for }|j-i|\geq 2\\
& e_ie_{i+1}e_i-\beta^{-1} e_i=e_{i+1}e_ie_{i+1}-\beta^{-1}e_{i+1}, && \text{with }\beta=2+q+q^{-1}.
\label{wwwera}
\end{align}
The Temperley-Lieb algebra $A_d(q)$ \cite{tp1,tp2} is
the $\C$-algebra over the generators $1, e_1,\cdots, e_{d-1}$ satisfying the two relations \eqref{rrrrelll23} while relation   
\eqref{wwwera} is replaced by the stronger condition 
\be
e_ie_{i+1}e_i-\beta^{-1} e_i=0.
\ee 
A special class of Hecke algebra representations are the ones which factorize through the 
Temperley-Lieb algebra $A_d(q)$. 
One has 
\begin{thm}[see \cite{kohno0}] The monodromy representation of the flat connection \eqref{jasqw2} is a Hecke algebra representation of the braid group $\cb_d$ which factorizes through the Temperley-Lieb algebra $A_d(q)$ with 
\be
q=\exp(\pi i \lambda)
\ee 
given by the correspondence
\be
\sigma_i \to q^{-3/4}\big(q-(1+q) e_i\big),\quad i=1,2,\dots,d-1.
\ee
\end{thm}

Comparing with the third of eqn.\eqref{hacck} we see that the $\sigma_i$ are semi-simple and
\be\label{eig1}
\Big\{\text{eigenvalues of the monodromy generator $\sigma_i$}\Big\}\subset \Big\{q^{1/4},-q^{-3/4}\Big\}.
\ee
The transport of the $i$-th defect operator around the $(i+1)$ one corresponds to the square of $\sigma_i$ and has spectrum
\be\label{2pppiiitr}
\Big\{ q^{1/2}, q^{-3/2}\Big\}.
\ee
By arithmeticity of $tt^*$, in the present context requires these eigenvalues to be roots of unity, that is,
\be\label{eig2}
q\in\boldsymbol{\mu}_\infty.
\ee 
Hence the Hecke algebra representations which appear in $tt^*$ are the tricky ones at $q$ a root of 1.

\subsubsection{Examples}

\begin{exe}[Ising $n$-point functions]\label{ppoqwera} For all LG models whose $tt^*$ geometry  is given by the Ising $n$-point functions (\textbf{Examples 1}-\textbf{4}), the $R\to0$ limit of the $\cq$-theory is
the critical Ising theory i.e.\! the minimal Virasoro model 
$(p,q)=(4,3)$ with three primaries $1$, $\sigma$, and 
$\varepsilon$ (in the notation of \cite{BPZ}) of dimension
\be
\Delta_1=0,\qquad \Delta_\sigma=\frac{1}{16},\qquad\Delta_\varepsilon =\frac{1}{2}
\ee
and fusion rule
\be\label{fufffs}
\sigma\cdot\sigma= [1]+[\varepsilon].
\ee
The topological defect operator of the $w$-plane $\cq$-theory are identified with $\sigma$. Therefore, when we transport one of them around another one we get a matrix with eigenvalues
\be\label{iissseigen}
\Big\{e^{2\pi i(\Delta_\varepsilon-2\Delta_\sigma)}, e^{2\pi i(\Delta_1-2\Delta_\sigma)} \Big\}=
\Big\{ e^{2\pi i\, 3/8} , e^{-2\pi i/8}  \Big\}
\ee
which is of the form \eqref{eig2}\eqref{2pppiiitr} 
for $q=e^{2i\pi\, 3/4}$, i.e.\! for $\lambda=3/2$.
\end{exe}

We have recovered from $tt^*$ the known action of the braid group $\cb_d$ on the Ising $d$-point functions at criticality. In particular they are flat sections of the connection in eqn.\eqref{jasqw2} with $\lambda=3/2$.

\subsection{Chern-Simons, Jones polynomials, minimal models, {\rm\textit{etc.}}} 

The braid group actions which factor through representations of the Temperley-Lieb at $q$ a root of unity are ubiquitous in mathematical physics.
They describe the monodromy of the conformal blocks of two-dimensional $SU(2)$ current algebra at level $k$ \cite{KZ,kohnoB}. Due to the relation of 2d current algebra with 3d Chern-Simons \cite{wittenCS}, they also describe the braiding properties of Wilson loops in $SU(2)$ Chern-Simons theory, and hence are the representations relevant for the Jones polynomials \cite{jones} and the theory of the quantum groups \cite{QG}. 

In addition, they also describe the braiding properties of the $(p,q)$ Virasoro minimal models \cite{fro}.
In the
 Virasoro $(p,q)$ minimal models
the operator $\phi_{r,s}$ has dimension
\be
h_{r,s}=\frac{(pr-qs)^2-(p-q)^2}{4pq},\qquad 1\leq r\leq q-1,\quad 1\leq s\leq p-1.
\ee 
The braiding the operator $\phi_{1,2}$ (which for $(p,q)=(4,3)$ reduces to the spin field $\sigma$ of \textbf{Example \ref{ppoqwera}}) correspond to the Temperley-Lieb algebra with
\be\label{kawerhh}
q\equiv e^{\pi i\,\lambda} =e^{2\pi i\, q/p}.
\ee 
The fusion rule of $\phi_{1,2}$ is similar to eqn.\eqref{fufffs}
\be
\phi_{1,2}\;\cdot\phi_{1,2}=[1]+[\phi_{1,3}].
\ee
One has
\be
h_{1,1}-2\,h_{1,2}=1-3\,\frac{q}{2p},\qquad
h_{1,3}-2\,h_{1,2}= \frac{q}{2p},
\ee
so eqn.\eqref{iissseigen} generalizes to
\be\label{iissseigen2}
\Big\{e^{2\pi i(h_{1,3}-2 h_{1,2})},
e^{2\pi i(h_{1,1}-2h_{1,2})} \Big\}=
\Big\{e^{2\pi i q/2p}, e^{-6\pi i\, q/2p}\Big\}
\ee
corresponding to eqn.\eqref{2pppiiitr}
with 
$q =e^{2\pi i\, q/p}$.
\medskip

The Knizhnik-Zamolodchikov equations is also related to an integrable statistical model, the Gaudin model of type $\mathfrak{gl}(2)$ \cite{gaudin,opers}. This is discussed in the next subsection.

\subsection{Comparison with the asymmetric limit}

$tt^*$ geometry predicts that the 
monodromy representations defined by the asymmetric limit of the brane amplitude  and the UV Berry connection are ``essentially'' the same
(and isomorphic when the UV limit is regular).

For a very complete $tt^*$ geometry,
where $\pi_1(\mathscr{X})=\cb_d$,
the asymmetric limit monodromy yields a so-called \textit{homology representation} of the braid group, a.k.a.\! the Lawrence-Krammer-Bigelow (LKB) representation \cite{lawrence,big1,big2}, see also \cite{braid,kohnoN,kohnoN2,kohnoN3}. In facts, it is known that
the LKB representation is essentially equivalent to the monodromy of the Knizhnik-Zamolodchikov connection.
In this section we limit ourselves to sketch the relation between the two points of view on the monodromy.
\medskip

There exist explicit integral representations of the 
solutions to the Knizhnik-Zamolodchikov equations of the schematic form \cite{iperr,iperr2}
\be\label{kkasqwer}
\boldsymbol{\Psi}=\int_\Gamma e^{\lambda \Phi(z_i;\lambda)}\,A(z_i;\lambda)\,dz_1\wedge\cdots\wedge dz_N
\ee
where $\Phi$, known as the master function, has the same functional form as the superpotential $\cw$ but with ``renormalized'' couplings in general. $\Gamma$ is a basis of relative cycles (which may be chosen to be Lefschetz thimbles).
In the limit $\lambda\to\infty$ the integral may be evaluated by the saddle point. 
The saddle point equations coincide with the algebraic Bethe ansatz equations of an integrable model \cite{quasicla}, the Gaudin model of type $\mathfrak{gl}(2)$. 
$\Phi(z;\infty)$ is the corresponding Yang-Yang functional. The vector $A$ evaluated at the saddle point is a common eigenstate of the Gaudin Hamiltonians \cite{gaudin,opers}
\be
H_a=\sum_{b\neq a} \frac{s^\ell_{(a)}s^\ell_{(b)}}{w_a-w_b}.
\ee
The integral expression \eqref{kkasqwer} may be identified with the asymmetric limit brane amplitude, taking into account the redefinitions of the couplings in relating the different limits.

\subsection{Puzzles and {\it caveats}}

Some aspects of the previous discussion look a bit puzzling because they do not fit in the intuition one gets from the study of ``typical'' 4-\textsc{susy} models, i.e.\! Landau-Ginzburg models whose superpotentials  are \emph{entire} functions on $\C^n$. Except for a couple of instances, the very complete $tt^*$ geometries do not belong to  the ``typical'' class of LG models.
The very special $tt^*$ geometries have no non-trivial wall-crossing, and this aspect implies quite peculiar properties when $d>2$. Luckily, we have a nice theoretical laboratory to study  these issues, namely the Ising $d$-point functions. We know that the Ising functions exist and define a totally regular $tt^*$ geometry, so naively unexpected facts which do occur for Ising functions should not be regarded as ``strange'' but rather as archetypical of very complete $tt^*$ geometries not arising from entire superpotentials.
\medskip

Let us discuss the puzzling aspect in the
context 
of the $d$-point Ising function.
The corresponding $tt^*$ brane amplitudes are the solution to the Riemann-Hilbert problem in twistor space with Stokes matrix $S=1+A$, where $A$ is the strictly upper triangular $d\times d$ matrix
\be\label{matA}
A_{ij}=\begin{cases}2(-1)^{j-i} &\text{for } j>i\\
0&\text{otherwise.}\end{cases}
\ee 
Note the identity $S^{-1}=US U^{-1}$
with $U=\diag\{(-1)^i\}$, which implies 
$|S_{ij}|=|S^{-1}_{ij}|$, a manifestation of the fact that these $TT^*$ geometries have no wall-crossing.
As in \textbf{Example \ref{kasqwmmmc}}, the symmetric matrix 
\be\label{llasqw}
C=S^t+S,
\ee
is the Cartan matrix of the EALA  
$A_1^{(1,\cdots,1)}$
of nullity $\kappa=d-1$ and type $A_1$.
The lattice of imaginary roots of the EALA,
$\Gamma_\text{im}$, is identified with the radical sub-lattice of the integral symmetric  form $C$, hence  $\mathrm{rank}\,\Gamma_\text{im}=d-1$.
The action of $\mathrm{Weyl}(A_1^{(1,\dots,1)})$ on the rank-1 quotient lattice
\be
\Gamma_\text{root}/\Gamma_\text{im}
\ee 
yields a surjection $\overline{\,\cdot\,}\colon\mathrm{Weyl}(A^{(1,\dots,1)}_1)\to\mathrm{Weyl}(A_1)\cong\Z/2\Z$. The Coxeter element $\mathsf{Cox}\in\mathrm{Weyl}(A^{(1,\dots,1)}_1)$ is represented by the $d\times d$ matrix
\be
\mathsf{Cox}= -(S^t)^{-1}S,
\ee
(i.e.\! \emph{minus} the 2d quantum monodromy \cite{CV92}).
From eqn.\eqref{llasqw} it is clear that
$\mathsf{Cox}$ acts as the identity on $\Gamma_\text{im}$, while it acts as
$\pm1$ on $\Gamma_\text{root}/\Gamma_\text{im}$. Since $\det \mathsf{Cox}=(-1)^d$, we conclude that $\overline{\mathsf{Cox}}=(-1)^d$.
Thus for $d$ odd $\mathsf{Cox}$ is semi-simple with minimal polynomial $z^2-1$, and the radical of the skew-symmetric form\footnote{\ Under the 4d/2d correspondence \cite{Cecotti:2010fi}\!\!\cite{Cecotti:2011rv,Alim:2011ae}, the form $S^t-S$ is the 4d Dirac electro-magnetic pairing, and its radical is the lattice of the 4d flavor charges,
so its rank coincides with the rank of the 4d  flavor group.} $S^t-S$ has rank $1$; for $d$ even all eigenvalues of the Coxeter elements are $+1$ and $\mathsf{Cox}$ has a single non-trivial Jordan block of size 2. From the arguments at the end of \S.\,4.3 in \cite{CV92}
means that for $d$ odd the model behaves as a UV superconformal theory, and for $d$ even as an asymptotically-free QFT. This matches the physics of the first few $d$'s in terms of the 4d $\cn=2$ QFT which corresponds to the Ising correlators in the sense of refs.\cite{Cecotti:2010fi}\!\!\cite{Cecotti:2011rv}: in 4d
$d=1$ leads to a free hyper, $d=2$ to pure $\cn=2$ SYM with $G=SU(2)$, and $d=3$ to $SU(2)$ $\cn=2^*$ SYM.

For $d>3$ things become less obvious and we get the aforementioned puzzles. The 2d quantum monodromy is $H\equiv-\mathsf{Cox}$; its eigenvalues are identified with $e^{2\pi i q_R}$ \cite{CV92}, where $q_R$ are the $U(1)_R$ charges of the 
Ramond vacua of the SCFT emerging at the UV fixed point.  For $d$ even
we get $q_R=\tfrac{1}{2}\bmod 1$ for all Ramond vacua, while for $d$ odd there is in addition a single Ramond vacuum with  
$q_R=0\bmod1$.
The Ramond $U(1)_R$ charges should be distribute symmetrically around zero by PCT, so the natural conclusion is that we have 
$[d/2]$ Ramond charges $q_R=-\tfrac{1}{2}$, $[d/2]$ Ramond charge $q_R=+\tfrac{1}{2}$ and for $d$ odd one $q_R=0$.
For $d\geq4$ this looks odd since one expects the largest $q_R$ to have multiplicity 1 because it should correspond to a spectral flow of the identity operator.
However this  argument can be circumvented in several ways: e.g.\! one may think that some couplings get weak in the UV and the fixed point consists of several decoupled sectors; in this case in the UV limit we may get a distinct spectral flow of the identity for each decoupled sector.

We adopt the following attitude: we know for certain that the Ising $d$-point functions exist and are pretty regular; this proves beyond all doubt that the $tt^*$ geometry with Stokes matrix \eqref{matA} does exist and is well behaved, even if does not fit in the intuition from experience with LG models whose  superpotential is an entire function in $\C^N$.
\medskip

There is another issue which may look tricky.
Let us consider the one-dimensional subspace 
of $\mathscr{U}\subset\mathscr{X}$ given by
$w_i=\lambda\, w_i^0$ for some fixed (generic) $w_i^0$. The pulled-back connection
\be
\frac{d\lambda}{\lambda}\sum_{i<j} B_{ij}
\ee 
should agree with the connection along the ``RG flow''. For the $d$-point Ising functions the matrix
$\sum_{i<j} B_{ij}$ does not coincide with the UV limit of the new index $Q$ as one would expect. However, when comparing two connections we should content ourselves to check that they are gauge-equivalent, not identical. For flat connections this mean they need to have the same monodromy up to conjugacy.  In addition, we need to remember that to get a nice UV limit  we twisted the vacuum bundle by factors of the form 
\be
h_i^{1/2}=\mathrm{const.}\,\prod_{j\neq i}(w_i-w_j)^{1/2}\propto\lambda^{(d-1)/2},
\ee 
so that as $\lambda\to e^{2\pi i}\lambda$ we pick up an extra $(-1)^{d-1}$. Then consistency merely requires
\be
\textsf{eingenvalues}\Big[e^{2\pi i Q}\Big]=  (-1)^{d-1}\, \textsf{eingenvalues}\!\left[\exp\!\left(2\pi i \sum_{i<j}B_{ij}\right)\right],
\ee
which holds identically.

\subsection{Relation with the Vafa proposal for FQHE}

Let us compare the above discussion of the effective $\cq$ theory in the $w$-plane with the Vafa proposal for FQHE. As stated in the introduction around eqn.\eqref{opppq},
 the microscopic dynamics of the $N$ electrons should produce an effective QFT for the quasi-hole ``fields''.
 
Now we have a natural candidate for the effective macroscopic description expected on physical grounds,
namely the $\cq$ theory. 
The idea is that \eqref{opppq} and \eqref{iasqw1} should be identified.
The identification works  
 \emph{provided} the $tt^*$ geometry is very complete,
 so that the spaces in which we insert the operators
 $h(w)$ and $\co(w)$ are identified.
Thus, to close the circle of ideas, it remains to show that the Vafa models have very complete  $tt^*$ geometries
(and work out their specific details).  
Before going to that, we need to develop some other tool in $tt^*$ geometry specific to the LG model of the Vafa class.

\section{Advanced $tt^*$ geometry II}
\label{advanced2}

\subsection{Symmetry and statistics}\label{secsymm}

For simplicity, in this subsection we assume the target space of our LG model to be $\C^N$.\medskip

Suppose the superpotential is invariant under permutations of the chiral fields
\be\label{llllls2}
\cw(z_1,\cdots, z_N)=\cw(z_{\sigma(1)},\cdots,
z_{\sigma(N)}),\qquad \sigma\in\mathfrak{S}_N.
\ee
$\mathfrak{S}_N$ is a symmetry of the SQM, so the vacuum space carries a unitary representation of the symmetric group.
Hence the vacuum bundle $\mathscr{V}$ and the chiral ring $\mathscr{R}\subset\mathrm{End}\,\mathscr{V}$ have
parallel orthogonal decompositions
\be
\mathscr{V}=\bigoplus_{\eta\in \mathsf{Irrep}(\mathfrak{S}_N)}\mathscr{V}_\eta.\quad\qquad \mathscr{R}=\bigoplus_{\eta\in\mathsf{Irrep}(\mathfrak{S}_N)}\mathscr{R}_\eta.
\ee
The component associated to the trivial representation, $\mathscr{R}_\mathsf{s}$,
is a ring, while for all $\eta$, $\mathscr{R}_\eta$ is a 
$\mathscr{R}_\mathsf{s}$-module.
The linear isomorphism $\mathscr{R}\xrightarrow{\!\sim\!} \mathscr{V}$ becomes
\be
\mathscr{R}_\eta\xrightarrow{\!\sim\!} \mathscr{V}_{\mathsf{a}\cdot \eta}\quad \text{as $\mathscr{R}_\mathsf{s}$-modules,} 
\ee
where $\mathsf{a}$ is the sign representation.
This follows from the explicit form of the isomorphimsm
\be
\phi\mapsto \phi\; dz_1\wedge dz_2\wedge\cdots\wedge dz_N+\overline{Q}(\text{something}),
\ee
and the fact that $dz_1\wedge\cdots\wedge dz_N$ belongs to the sign representation.

We define the \emph{Fermi} (resp.\! \emph{Bose}) model to be 
the one obtained by restricting $\mathscr{V}$ to its symmetric (resp.\! antisymmetric)
component $\mathscr{V}_\mathsf{s}$ (resp. $\mathscr{V}_\mathsf{a}$). To call
``fermionic'' the model having symmetric wave-functions may look odd. To justify our definition let us count the number of ground states in an important special case.

\subsubsection{Special case: $N$ non-interacting copies} Suppose we have a one-particle superpotential $W(z)$ with $d$ vacua and let $f_j(z)$ ($j=1,\dots, d$) be a set of holomorphic functions giving a basis for the one-particle chiral ring. We consider $N$ non-interacting copies of the model
\be\label{kkkas12j}
\cw(z_1,\cdots, z_N)=\sum_{i=1}^N W(z_i).
\ee
The chiral ring of the $N$-particle model
is $\mathscr{R}=\otimes^N \mathscr{R}_1$
with $\mathscr{R}_1$ the one-particle chiral ring. Then
\be
\mathscr{R}_\textsf{a}=\wedge^N\mathscr{R}_1,\qquad 
\mathscr{R}_\textsf{s}=\odot^N\mathscr{R}_1.
\ee
If $\dim\mathscr{R}_1=d$, the number of anti-symmetric resp.\! symmetric elements of $\mathscr{R}$ is
\be\label{jjazsqw78}
\dim\mathscr{R}_\textsf{a}={d\choose N}\qquad\dim\mathscr{R}_\textsf{s}= {N+d-1 \choose N} 
\ee
which correspond, respectively, to Fermi and Bose statistics. Using the basic spectral-flow isomorphism $\mathscr{R}\cong\mathscr{V}$,
we get the linear isomorphisms
\be\label{fferebis}
\mathscr{V}_\textsf{F}\cong\wedge^N\mathscr{V}_1,\qquad 
\mathscr{V}_\textsf{B}\cong\odot^N\mathscr{V}_1,
\ee 
and the $tt^*$ metric, connection and brane amplitudes are the ones induced  from the corresponding one-particle
quantities. The group $\mathfrak{S}_N$ acts on the sub-spaces of $ \otimes^N\mathscr{V}_1$ as
\be
v_1\otimes v_2\otimes\cdots\otimes v_N\to \mathsf{sign}(\sigma)\, v_{\sigma(1)}\otimes v_{\sigma(2)}\otimes\cdots\otimes v_{\sigma(N)},\quad \sigma\in\mathfrak{S}_N.
\ee

\begin{rem} Eqns.\eqref{jjazsqw78} remain true if we add to the superpotential \eqref{kkkas12j} arbitrary
supersymmetric interactions (which do not change the behaviour at infinity in field space) since the dimension of the chiral ring is the Witten index $d$. 
\end{rem}

\subsubsection{The Fermi model chiral ring}
We return to the general case, eqn.\eqref{llllls2}.

 The elements of the $\mathscr{R}_\mathsf{s}$-module
 $\mathscr{R}_\mathsf{a}$ have the form
\be
\phi\, \Delta(z_i)\in\mathscr{R},\qquad\text{where }\Delta(z_i)=\prod_{i<j}(z_i-z_j)\ \ \text{and } \phi\in\mathscr{R}_\mathsf{s}.
\ee
The chiral ring of the Fermi model is then
\be
\mathscr{R}_\textsf{F}= \mathscr{R}_\textsf{s}\big/\ci_\textsf{Van}
\ee
where $\ci_\textsf{Van}\subset\mathscr{R}_\textsf{s}$ is the annihilator ideal of the Vandermonde determinant $\Delta(z_i)$.
We have the linear isomorphism
\be\label{fring}
\mathscr{R}_\textsf{F}\cong \mathscr{V}_\textsf{s}\equiv \mathscr{V}_\textsf{F},\qquad \dim \mathscr{R}_\textsf{F}={d\choose N}.
\ee

\subsection{$tt^*$ functoriality and the  Fermi model}

$tt^*$ functoriality \cite{tt*,CV92} yields a more convenient interpretation of the Fermi model.

\subsubsection{Review of $tt^*$ functoriality}
\label{Rtt*fun}
Suppose the superpotential map $W_x\colon \ck\to \C$ factorizes through  a Stein space $\cs$
for all values of the couplings $x\in\mathscr{X}$
\be\label{ttfact}
\xymatrix{\ck\ar@/^2pc/[rrrr]^{W_x}\ar[rr]_f && \cs \ar[rr]_{V_x}&&\C}
\ee
where $f\colon \ck\to\cs$ is a (possibly branched) cover independent of $x$. We wish to compare the $tt^*$ geometry over $\mathscr{X}$ of the LG model $(\cs,V_x)$
with the original one $(\ck,W_x)$.
Let\footnote{\ The $s_i$'s are coordinates on $\cs$.}   
\be
\psi=\phi\,ds_1\wedge\cdots\wedge ds_N+(\overline{\partial}+dV_x\wedge)\eta\in \Lambda^1(\cs)
\ee
be a vacuum wave-function of $(\cs,V_x)$. The pull-back
$f^*\psi$ is $(\overline{\partial}+dW_x)$-closed in $\ck$ and not $(\overline{\partial}+dW_x)$-exact, hence cohomologous to a vacuum wave-form\footnote{\ In general, the actual wave functions are not given by the pull back since the $D$-terms do not agree. In the case of one-field the vacuum Schroedinger equation does not contain the K\"ahler metric, and the exact wave-function on $\ck$ is the pull back of the one on $\cs$ \cite{iosqm}.} of $(\ck,W_x)$. Comparing $\overline{Q}$-classes, we see that the pulled-back vacuum is the spectral flow of the chiral operator \cite{tt*}
\be
f^*(\phi)\,\det[\partial s_i/\partial z_j]\in\mathscr{R}_\ck.
\ee 
The linear map 
\be
f^\sharp\colon\mathscr{R}_\cs\to \mathscr{R}_\ck, \quad \phi\mapsto f^*(\phi)\,\det[\partial s/\partial z]
\ee
 is an isometry of 
 topological metrics
\be\label{topiso}
\big\langle f^\sharp(\phi_1)\ f^\sharp(\phi_2)\big\rangle_\ck^\text{top}=\big\langle \phi_1 \ \phi_2\big\rangle^\text{top}_\cs
\ee
 compatible with the $\mathscr{R}_\cs$-module structures
\be\label{comrong}
f^\sharp(\phi_1\cdot\phi_2)= f^*(\phi_1)\cdot f^\sharp(\phi_2)\in\mathscr{R}_\ck.
\ee
Write
\be
\mathscr{R}_\ck=f^\sharp(\mathscr{R}_\cs)\oplus f^\sharp(\mathscr{R}_\cs)^\perp.
\ee
where $(\cdot)^\perp$ stands for the orthogonal complement in the $tt^*$ metric. 
 $tt^*$ functoriality is the statement that $f^\sharp\colon \mathscr{R}_\cs\to
 f^\sharp(\mathscr{R}_\cs)$ is an isometry also for the $tt^*$ metric. To show this fact, one has to checks two elements: 1) the two $tt^*$ metrics solve the same $tt^*$ PDEs, and 2) they satisfy the same boundary conditions.  
Since the classes in $\mathscr{R}_\ck$ of the operators $\partial_{x_a}\cw$ belong to the sub-space $f^\sharp(\mathscr{R}_\cs)\subset\mathscr{R}_\ck$, the first assertion follows from eqns.\eqref{comrong} and \eqref{topiso}. 
The boundary conditions which select the correct solution to these PDEs are encoded in the 2d BPS soliton multiplicities \cite{CV92}. The BPS solitons are the connected preimages of straight lines in the $W$-plane ending at critical points \cite{CV92,iqbal}. Since the map $W_x$ factorizes through $V_x$ (cfr.\! eqn.\eqref{ttfact}) so do the counterimages of straight lines,
and the counting of solitons agrees in the two theories.

\begin{defn} A \emph{$tt^*$-duality} between
two 4-\textsc{susy} theories is a Frobenius algebra isomorphism
 between their chiral rings
 \be
 \mathscr{R}_1\xrightarrow{\!\sim\!} \mathscr{R}_2
 \ee
which is an isometry for the $tt^*$ metric (hence for the BPS brane amplitudes). 
\end{defn}

$tt^*$ functoriality produces several interesting $tt^*$-dual pairs. See \S.\,\ref{tt8dual} for examples. The standard lore is that $tt^*$-duality implies equivalence of the full quantum theories for an appropriate choice of the respective $D$-terms. Thus $tt^*$ functoriality is a powerful technique to produce new quantum dualities.

\subsubsection{Application to Fermi statistics} \label{fermmmmqa}

If the superpotential $\cw(z_1,\cdots, z_N)$ is invariant under $\mathfrak{S}_N$, it can be rewritten as a function of the elementary symmetric functions
\be
\cw(z_1,\cdots, z_N)
\leadsto\cw(e_1,e_2,\cdots, e_N)
 \ee
where
\be
e_k:=\sum_{1\leq i_1 <i_2<\cdots < i_k\leq N} z_{i_1}z_{i_2}\cdots z_{i_k},\qquad e_0:=1.
\ee
The superpotential $\cw\colon \C^N\to \C$
factorizes through
the branched cover map of degree $N!$ 
\be 
E\colon\C^N\to \C^N,\qquad E\colon(z_1,\cdots, z_N)\mapsto (e_1,\cdots,e_N).
\ee
One pulls back the \textsc{susy} vacua of the LG model with superpotential $\cw(e_i)$
\be
E^* \Big(h(e_k)\,de_1\wedge de_N+\overline{Q}(\cdots)\Big)=h(e_k(z_j))\,\Delta(z_j)\, dz_1\wedge\cdots\wedge dz_N+\overline{Q}(\cdots),
\ee
getting non-trivial $\overline{Q}$-cohomology classes, hence vacua of the $\cw(z_k)$ theory up to $\overline{Q}$-trivial terms.
Then the pull back yields a correspondence 
\be
E^\sharp\colon |h\rangle_e\mapsto |h \Delta\rangle_z,\qquad h\in\mathscr{R}_e\equiv \C[e_1,\dots, e_N]/(\partial_{e_1}\cw,\cdots,\partial_{e_N}\cw),
\ee 
which is an isometry for the underlying TFT metric
\be
\langle h_1\;h_2\rangle^\textsf{top}_e=
\langle h_1\,\Delta\;h_2\,\Delta\rangle^\textsf{top}_z
\ee 
and
\be
E^\sharp(\mathscr{R}_e)\cong \mathscr{R}_\textsf{s}\big/\ci_\textsf{Van}\equiv \mathscr{R}_\textsf{F},
\ee
(cfr.\! eqn.\eqref{fring}).

In other words, $E^\sharp$ sets an equivalence between the TFT of the $\cw(e_k)$ model and the TFT of the Fermi sector of the $\cw(z_j)$ model. By the arguments in \S.\,\ref{Rtt*fun}, 
$E^\sharp$ is also isometry of $tt^*$ metrics and hence of brane amplitudes.

\subsection{Examples of $tt^*$-dualities}\label{tt8dual}

$tt^*$ functoriality relates the fermionic version of a $N$-field LG model to some other 2d (2,2) supersymmetric system.
In this subsection we present several examples of such $tt^*$-dualities. Only the first one will be referred to in the rest of the paper; the following examples may be safely skipped.

\begin{exe}[The $\nu=1$ model]\label{errample}
Consider the Fermi version of the LG model
\be\label{asferm}
\cw(z_1,\cdots,z_N)=\sum_{i=1}^N\left(\mu \,z_i+\sum_{a=1}^N m_a\,\log(z_i+x_a)\right)
\ee
($x_a$ all distinct)
which has 
$\dim\mathscr{R}_\textsf{F}=1$. It is clear from the discussion in section 2 that 
the Fermi model \eqref{asferm} with $m_a=-1$
describes the $\nu=1$ phase of the quantum Hall effect.

We combine the chiral fields $e_k=e_k(z_j)$ in the chiral-superfield valued polynomial
\be
P(y;e_k)=\sum_{k=0}^N e_{N-k}\,y^k,
\ee 
$y$ being an indeterminate.
The superpotential of the Fermi model then  reads
\be
\cw(e_1,\cdots,e_N)=\mu\, e_1+\sum_{a=1}^Nm_a\,\log P(x_a;e_k).
\ee
The chiral superfields $u_a\equiv P(x_a;e_k)$ are $N$ linearly independent linear combinations of the $N$ chiral superfields $e_k$; by a linear field redefinition we can take the $u_a$ to be the fundamental fields.
$e_1$ is a certain linear combination of the 
$u_a$; by rescaling the $u_a$'s we may write 
$e_1=u_1+\cdots+u_N+\mathrm{const}$. Then\footnote{\ When writing superpotentials, we usually omit additive field-independent constants.}
\be\label{jz129b}
\cw(u_1,\cdots, u_N)=\sum_{a=1}^N\Big(\mu\,u_a+m_a\,\log u_a\Big)
\ee 
i.e.\! the $\nu=1$ Fermi models is equivalent to $N$ non-interacting copies of the one-field ``Penner'' model $W(u)=\mu\,u+m\,\log u$.  
 Next, $tt^*$ functoriality with respect to the plane-to-cylinder cover map
\be
x_a\mapsto e^{x_a}\equiv \mu\,u_a
\ee 
sets a $tt^*$-duality between the $\nu=1$ Fermi system \eqref{asferm} and $N$  \emph{free} twisted chiral super-multiplets with twisted masses equal to the residues $m_a$ of the one-field rational differential $dW=(\mu+\sum_am_a(z+x_a)^{-1})dz$. The final superpotential is
\be\cw(x_1,\cdots, x_N)=\sum_{a=1}^N\Big(e^{x_a}+m_a\, x_a\Big),
\ee
whose $tt^*$ equations were explicitly solved in \cite{twistor,tt*3,Cecotti:2010fi}.
We shall return to this basic example below.
\end{exe} 

\begin{exe}[Polynomial $\nu=1$ models]\label{exepollly} We may consider $N$ non-interacting copies of other LG systems with Witten index $N$ (so that $\nu=1$) getting similar conclusion. For instance, we may take the one-field differential $dW$ to have a single pole of order $N+2$ at infinity
\be\label{kasqwert}
\cw(z_1,\dots, z_N)=\sum_{i=1}^N P_{N+1}(z_i)
\ee
where 
\be
P_{N+1}(z)= z^{N+1}+\sum_{k=1}^{N+1}t_k\, z^{N+1-k}
\ee 
is an arbitrary monic polynomial of degree $N+1$
which we take to be odd for definiteness.
The Newton identities yield
\be
\cw(e_1,\dots, e_N)=\sum_{k=1}^{N/2} e_k\Big(- (N+1)\,e_{N+1-k}+A_k(e_1,\cdots, e_{N-k})\Big)
\ee
for some polynomial $A_k(e_1,\cdots, e_{N-k})$ which depends on the $t_j$'s.
The field redefinition 
\be
u_k=-(N+1)\,e_{N+1-k}+A_k(e_1,\cdots,e_{N-k}),
\ee 
which has  constant Jacobian, reduces the Fermi model to $N$ copies of the free Gaussian theory
\be\label{kkkasqwerj}
\cw(e_1,\cdots, e_{N/2},u_1,\cdots, u_{N/2})=\sum_{k=1}^{N/2} e_k\, u_k
\ee
which has a single vacuum. That the $tt^*$ metric of the original Fermi model coincides with the one for the Gaussian theory is easily checked: the $tt^*$ metric of the $\nu=1$ Fermi model is the determinant of the one-field $tt^*$ metric (cfr.\! \eqref{fferebis});
 from the reality constraint \cite{tt*} $\det g$ is just
 the absolute value of the determinant of the topological metric $|\!\det\eta|$.
 $\eta$ may be set to 1 by a change of holomorphic trivialization \cite{frob}. The covering $E$ automatically implements such a trivialization. The wave-function of the unique vacuum of the Gaussian model \eqref{kkkasqwerj},
when written in terms of the original chiral fields $z_i$, has the form
\be\label{pppqwm}
\prod_{i<j}(z_i-z_j)\,dz_1\wedge\cdots \wedge dz_N+\overline{Q}(\cdots).
\ee 
This wave-function is cohomologous to the one obtained solving the Schroedinger equation in the original Fermi model, but not equal since the covering map $E$ implicitly involves a deformation of the $D$-terms. The $tt^*$ metric, i.e.\! the Hermitian structure of the vacuum bundle is correctly reproduced since it is independent of the $D$-terms. 
 \end{exe}

We stress that the Vandermonde factor in the wave-function \eqref{pppqwm} is produced by the cover map $E$, not by an interaction in the superpotential. This is physically correct, since this is the wave-function at $\nu=1$ of a $N$ \textit{non-interacting} fermions. In particular, $tt^*$ functoriality automatically yields the correct $\nu=1$
Laughlin wave-function \cite{Laughlin}.
 
\begin{exe} More generally, we may take the rational differential $dW$ to have a pole of order $\ell+2$ at infinity and $N-\ell>0$ simple poles in $\C$ with residues $m_a$; the corresponding Fermi model is reduced by
$tt^*$ functoriality to a non-interacting system of $\ell$ ordinary free massive chiral multiplets and $N-\ell$ twisted ones
 \be
 \cw(x_i)=\frac{1}{2}\sum_{a=1}^\ell x_a^2+\sum_{a=\ell+1}^N\Big(e^{x_a}+m_a\,x_a\Big). 
 \ee 
We may also consider the special cases $\ell=-1,-2$ (such models may be plagued by run-away vacua at $\infty$). If $\ell=-1$ and $dW$ has only simple poles of residues $m_a$, going through the same steps as in \textbf{Example \ref{errample}}, we get
\be
\cw(u_1,\cdots, u_N)=\sum_{a=1}^N m_a \log u_a+m_0\log\!\left(\sum_{a=1}^N u_a+\lambda\right).
\ee
Comparing this expression with eqn.\eqref{jz129b}, we see that already for $\nu=1$ assuming $dW$ to have a double pole at $\infty$ leads to a nice simplification (besides of making the model better defined).
\end{exe}

 \begin{exe}[$tt^*$ particle-hole duality]\label{parholeduality}
 We consider the case we have $N$ (super-)particles $z_i$ and $d=N+1$ (vacuum) one-particles states, so the Fermi statistics
 \eqref{jjazsqw78} yields $N+1$ vacua, which may be seen as single-hole states.  
 For simplicity, we consider the model
 \be\label{kasqwert2}
\cw(z_1,\dots, z_N)=\sum_{i=1}^{N} P_{N+2}(z_i),
\ee
where $P_{N+2}(z)$ is any polynomial of degree $N+2$.
Going through the same steps as in \textbf{Example \ref{exepollly}}, we get
(say for $N$ even)
\be
\cw=P_{N+2}(e_1)+\sum_{k=2}^{N/2}e_k\,u_k,
\ee
so, integrating away the free massive d.o.f., we get back the original one-particle model. The $tt^*$ geometry of the model with one hole is the same one as for the model with one particle. More generally, the $tt^*$ geometry of the Fermi model of $N$ copies of a LG model with $d$ vacua, is invariant under
\be
N\leftrightarrow d-N.
\ee  
 \end{exe}
 
 \begin{exe}[Grassmanian $\sigma$-models] In ref.\!\cite{CV92} it was shown that the Fermi model of $N$ copies of the $\sigma$-model with target $\bP^{M-1}$
 is $tt^*$-dual to the $\sigma$-model
 with target space the Grassmanian
 \be
 \mathsf{Gr}(M,N)=SU(M)\Big/U(N)\times SU(M-N).
 \ee 
 \end{exe}
 
 
 \subsection{Fermi statistics vs.\! Hecke algebras representations}\label{staHecke}
 
 We consider the Fermi model of $N$ decoupled LG systems
 \be\label{kasqwxx665}
 \cw(e_k)=\sum_{i=1}^N W(z_i)
 \ee
 where the rational differential $dW$ has $d\geq N$ zeros. In addition we assume that the one-field theory $W(z)$ yields a very complete symmetric $tt^*$ geometry, so the results of \S.\ref{lllz10n}
 apply.  
 
 The vacuum bundle of the $N$-particle model is
 \be
 \mathscr{V}_N\cong\wedge^N\mathscr{V}_1\to\mathscr{X},
 \ee
 and its UV Berry connection is just the one induced in the $N$-index antisymmetric representation by the one-particle UV Berry connection.
 It is convenient to  
 introduce the ``Grand-canonical'' bundle
 \be\label{jjjjxz}
\mathscr{W}:=\bigoplus_{N=0}^d \mathscr{V}_N\to \mathscr{X},\qquad\mathrm{rank}\,\mathscr{W}=\sum_{N=0}^d{d\choose N}=2^d.
 \ee
 The total number of states is $2^d$ since each of the $d$ one-particle (vacuum) states may be either 
empty or occupied. 

\begin{rem} In \eqref{jjjjxz} we added a direct summand $\mathscr{V}_0$ which does not correspond to any LG model
(the number of chiral fields $N$ is zero). This can be done without harm since
by the particle-hole duality (\textbf{Example \ref{parholeduality}}) the extra summand is $\mathscr{V}_0\cong\mathscr{V}_N$, i.e.\! the trivial line bundle.
\end{rem}

In \S.\ref{lllz10n} we associated a spin degree of freedom $s^{(j)}_\ell$ ($\ell=1,2,3$) to the $j$-th one-particle vacuum: spin down $\downarrow$ (up $\uparrow$) meaning that the $j$-th state is empty (resp.\! occupied). Then
 \be\label{llllcxz8}
 \textsf{fiber}\big(\mathscr{W}\big)\cong \boldsymbol{V}^{(1)}\otimes\cdots\otimes \boldsymbol{V}^{(d)}\equiv\boldsymbol{V}^{\otimes d},
 \ee 
 where $\boldsymbol{V}^{(j)}\cong\C^2$
 is the space on which the $s^{(j)}_\ell$ act. A vacuum with occupied states
 $\{j_1,j_2,\cdots, j_N\}$
 \be
\downarrow_{1}\otimes\cdots\otimes \downarrow_{j_1-1}\otimes\uparrow_{j_1}\otimes \downarrow_{j_1+1}\otimes\cdots\otimes
\downarrow_{j_N-1}\otimes \uparrow_{j_N}\otimes\downarrow_{j_N+1}\otimes\cdots\otimes\downarrow_N\qquad N>0 
 \ee
 corresponds (linearly) to the element of the $N$-particle chiral ring 
 \be
 \sum_{\sigma\in\mathfrak{S}_N} \mathsf{sign}(\sigma)\, E_{j_1}(z_{\sigma(1)})\,E_{j_2}(z_{\sigma(2)})\,\cdots\, E_{j_N}(z_{\sigma(N)})\in (\mathscr{R}_N)_\mathsf{a},
 \ee
 where $\{E_j(z)\}_{j=1}^d$ is the canonical basis of the one-particle chiral ring $\mathscr{R}_1$. Note that the operators $s^{(j)}_\ell$ for $\ell\neq3$ are not defined at the level of the single LG model with a definite number of chiral fields $N$. 
\medskip

 Comparing \eqref{llllcxz8} with eqn.\eqref{kzbig}
 \be
 \mathscr{W}\cong \boldsymbol{\cv},
 \ee
 and the linear PDEs  satisfied by the ``grand-canonical'' brane amplitudes $\boldsymbol{\Psi}$ is just the $\mathfrak{sl}(2)$ Knizhnik-Zamolodchikov equation up to twist by ``normalization'' factors.
 In $\mathscr{W}$ 
 one defines the operator number of particles as
 \be\label{kkasqw12}
 \hat N:=L_3+\frac{d}{2},\qquad\text{where } L_\ell=\sum_{j=1}^d s_\ell^{(j)}\quad \text{(generators of $\mathfrak{sl}(2)_\text{diag}$)}.
 \ee
The underlying one-particle model, having a very complete  symmetric $tt^*$ geometry, defines a Kohno connection acting on $\boldsymbol{V}^{\otimes d}$ that we argued has
the $\mathfrak{sl}(2)$ KZ form up to an overall twist, i.e.
\be\label{ovvverall}
\cd=d+\lambda\sum_{i<j} s^{(i)}_\ell s^{(j)}_\ell\, \frac{d(w_i-w_j)}{w_i-w_j}+\xi\sum_{i<j} d\log(w_i-w_j)
\ee
for some constants $\lambda$ and $\xi$. 
We shall see momentarily that the constant $\xi$ is related to $\lambda$.

The eigen-subbundles $\mathscr{V}_N\equiv \ker(\hat N-N)\subset\mathscr{W}$
 are preserved by parallel transport with $\cd$, and hence define a monodromy representation $\pi_1(\mathscr{X})$ (also denoted\footnote{\ When no confusion is possible,  we identify the monodromy representation with its representation space.} by $\mathscr{V}_N$) which is the one associated to the $N$-field Fermi theory \eqref{kasqwxx665}. For most $N$'s this representation is highly reducible. Indeed, the eigen-bundles
of the operator $L^2\equiv L_\ell L_\ell$ are also preserved by parallel transport.
So one has the monodromy invariant  decomposition
\be
\mathscr{W}=\bigoplus_{l=0\atop l=d/2\bmod1}^{d/2}\bigoplus_{m=-l}^l \mathscr{V}_{l,m},\qquad \mathscr{V}_{l,m}:=\ker\!\big(L^2-l(l+1)\big)\bigcap\,\ker\!\big(L_3-m\big),
\ee
and
\be
\mathscr{V}_N= \bigoplus_{l=|N-d/2|}^{d/2}\mathscr{V}_{l,m\equiv N-d/2}.
\ee
Since $\mathfrak{sl}(2)_\text{diag}$ centralizes the monodromy representation
\be
\mathscr{V}_{l,m}\cong \mathscr{V}_{l,m^\prime}\quad\text{for }-l\leq m,m^\prime\leq l,
\ee
in particular the monodromy representations $\mathscr{V}_N$, $\mathscr{V}_{d-N}$ are isomorphic. This is a manifestation of the particle-hole duality in Fermi statistics explicitly realized in terms of the $tt^*$-dualities described in \S.\,\ref{tt8dual}.  The non-zero eigenbundles $\mathscr{V}_{l,m}$ have ranks given by the 
 Catalan triangle
\be
\mathrm{rank}\, \mathscr{V}_{l,m}=
{d \choose \frac{d}{2}-l}-
 {d \choose \frac{d}{2}-l-1}\qquad\text{for }\left|\begin{array}{ll}
 0\leq l\leq d/2, &l=d/2\bmod1,\\
 -l\leq m\leq l, &l=m\mod 1.\end{array}\right.
\ee

The eigen-bundle $\mathscr{V}_{d/2,N-d/2}$ has rank 1, i.e.\! it contains a unique monodromy invariant vacuum $|d/2,N-d/2\rangle$.
$|d/2,N-d/2\rangle$ is a ``preferred'' vacuum for the Fermi model with $N$-fields. It is tempting to identify it with the one discussed in
\S.\,\ref{univacc}.
The fact that it is invariant under the monodromy representation is already a strong suggestion that this is the case.
\medskip

For a fixed number of particles $N$ the determinant of the brane amplitudes, $\det \Psi$, is a constant section of the line-bundle
\be
\det\wedge^N\mathscr{V}_1\cong \mathscr{V}_{d/2,N-d/2}\quad \Big(\cong \mathscr{V}_{d/2,d/2}\Big)
\ee
corresponding to the preferred vacuum
for the $N$-particle Fermi model.
The overall twist $\xi$ in eqn.\eqref{ovvverall} is fixed by the requirement that the preferred vacuum has trivial monodromy. This fixes $\xi$ in terms of $\lambda$
\be
\xi=-\frac{\lambda}{4}.
\ee
In other words, the normalized amplitudes $\boldsymbol{\Psi}_\text{norm}$ are related to the KZ ones $\boldsymbol{\Psi}$ by the formula
\be
\boldsymbol{\Psi}_\text{norm}= \frac{\boldsymbol{\Psi}}{\Psi_\text{priv}}
\ee
with $\Psi_\text{priv}$ a parallel section of the line-bundle $\mathscr{V}_{d/2,d/2}$.
In particular the normalized monodromy is trivial for the $\nu=1$ case.

 \subsection{Relation with the Heine-Stieltjes theory}\label{HStheory}
 
 We consider a LG model with $N$ chiral fields with superpotential differential
 \be\label{rrrtyuk}
 d\cw=2\beta\!\!\!\sum_{1\leq i<j\leq N} \frac{d(z_i-z_j)}{z_i-z_j}+\sum_{i=1}^N dW(z_i),
 \ee
 where $dW(z_i)$ is a rational differential with $d$ zeros and a pole of order $1\leq \ell\leq d+2$ at $\infty$. 
Generically, $dW$ has $p\equiv d+2-\ell$ simple poles
at finite points $\{y_1,\cdots, y_p\}\subset\C$ (all distinct), i.e.\!
\be
dW(z)=\frac{B(z)}{A(z)}\,dz,\qquad
\text{where }A(z)\equiv \prod_{s=1}^p(z-y_s),
\ee
 for some degree $d$ polynomial $B(z)$ coprime with $A(z)$. The LG model proposed by Vafa to describe FQHE has the form \eqref{rrrtyuk} with the residues of $dW$ equal $\pm1$ and $2\beta=1/\nu$.
 \medskip

 We think of this model as defined on the quotient K\"ahler manifold
 \be
 \ck=\Big\{(z_1,\cdots,z_N)\in \big(\C\setminus\{y_1,\cdots,y_p\}\big)^N\;\Big|\; z_i\neq z_j\ \text{for }i\neq j\Big\}\Big/\mathfrak{S}_N. 
 \ee
$\ck$ is affine (hence Stein). Indeed, the basic chiral fields are the elementary symmetric functions, 
$e_k=e_k(z_j)$; we identify the field configuration $(e_1,\cdots, e_N)$ configuration with the degree $N$ monic polynomial 
\be\label{stielpol}
P(z)=P(z;e_k)\equiv\sum_{k=0}^N(-1)^k e_{N-k}\,z^k,\qquad e_0=1.
\ee
Then
\be\label{wwhatkk}
\ck\cong\C^N\setminus S
\ee
where $S$ is the hypersurface (divisor)
\be
S\equiv \Big\{\textsf{discr}(P)\,\textsf{Res}(A,P)=0\Big\}\subset\C^N,
\ee
where $\textsf{discr}\big(P)$ and $\textsf{Res}(A,P)$ are the discriminant and the resultant of the polynomials seen as functions of the coefficients $e_1,\cdots, e_N$ of $P(z)$ for fixed $A(z)$.
\medskip

A vacuum configuration of the model \eqref{rrrtyuk} defined on the quotient manifold $\ck$ (i.e.\! up to $\mathfrak{S}_N$ action) is described by a degree $N$ monic polynomial $P(z)\equiv P(z;e_k)$ as in eqn.\eqref{stielpol} which satisfies the Heine-Stieltjes differential  equation
\be\label{kkkzq1y}
2\beta\, A(z)\,P^{\prime\prime}(z)+B(z)\,P^\prime(z)=f(z)\,P(z)
\ee
where $f(z)$ is a polynomial of degree
$d-1$. The degree $N$  polynomials $P(z)$ which solve this equation for some $f(z)$ are called \emph{Stieltjes polynomials;} to each of them there corresponds a degree $d-1$ polynomial $f(z)$, called its associated \emph{van Vleck polynomial.} The Heine-Stieltjes theory is reviewed in the  context of $tt^*$ in ref.\!\cite{twistor}. We refer to the vast literature \cite{heine1,heine2,heine3,heine4,heine5,heine6,heine7,heine8,heine9,heine10,heine11,heine12,heine13,heine14,heine15,heine16,heine17,heine18,heine19,heine20,heine21,heine22,heine23,heine24,heine25} for further details.

If $dW$ is a generic rational differential, with just simple poles in $\bP^1$, eqn.\eqref{kkkzq1y} is a generalized $d$-Lam\'e equation. The $d$-Lam\'e equation
 \cite{heine19} is the special case $\beta=1$ and $dW=d\log Q(z)$, where $Q(z)$ is a polynomial (which we may choose square-free and monic with no loss) of degree $d$. 
 Taking the same differential $dW$, but choosing $\beta=-1$, the superpotential $\cw$ in eqn.\eqref{rrrtyuk} becomes the Yang-Yang \cite{gaiottoknot,YY} functional (and its exponential the master function \cite{var2,var3}) of the $\mathfrak{sl}(2)$ Gaudin integrable model on $\boldsymbol{V}^{\otimes d}$,
the  Heine-Stieltjes equation is equivalent to the corresponding algebraic Bethe ansatz equations, and the roots of $P(z)$ are the Bethe roots \cite{gaiottoknot,crilev}.

The case most relevant for us is when precisely one of the poles in $\bP^1$ is double: then the Heine-Stieltjes equation is a  \emph{confluent}
generalized $d$-Lam\'e equation.\footnote{\ For instance, for $d=1$ (resp.\! $d=2$) one gets the confluent hypergeometric equation (resp.\! confluent Heun) instead of the hypergeometric (Heun) ODE.} The ODE is equivalent to the Bethe ansatz equation for the Gaudin model with an \emph{irregular} singularity \cite{gir1,gir2,gir3}\!\!\cite{gaiottoknot}.

Going to the confluent limit is very convenient, as we have already observed. In the Gaiotto-Witten language \cite{gaiottoknot}, passing to the confluent limit corresponds to breaking the gauge symmetry by going in the Higgs branch of the 4d $\cn=4$ gauge theory (``complex'' symmetry breaking \cite{gaiottoknot}). 
\medskip

A basic result of Heine-Stieltjes theory 
states that the number of solutions $(P(z),f(z))$ of eqn.\eqref{kkkzq1y} is (at most, and generically)
\be\label{azswq}
d_{d,N}\equiv {N+d-1\choose N}.
\ee
By definition, this is also the Witten index of the model $(\ck,\cw)$ hence, by eqn.\eqref{lllasqwp}, the dimension of the appropriate relative homology group.
\medskip

By construction \cite{twistor}, solving the Heine-Stieltjes equation \eqref{kkkzq1y} is equivalent to solving the equation $d\cw$ and considering the solutions modulo $\mathfrak{S}_N$. Explicitly,
\be
\partial_{z_i}\cw= \sum_{j\neq i}\frac{2\beta}{z_i-z_j}+\sum_a \frac{1}{z_i-w_a}-\sum_\ell\frac{1}{z_i-\zeta_\ell}=0
\ee
which is a generalization of the Algebraic Bethe Anzatz equation for the Gaudin model (the $z_i$'s are analogue to the Bethe roots). The Gaudin model arises from the semi-classical limit of the solutions to the Knizhnik-Zamolodchikov, and it is natural to expect that the relation remains valid in the present slightly more general context.

\begin{rem} In the (related) context \cite{matrix1,matrix2,matrix3} of large-$N$ matrix models the Heine-Stieltjes equation \eqref{kkkzq1y} is called the Schroedinger equation. 
\end{rem}

\subsection{The fermionic truncation}
\label{fermtruc}

Following Vafa \cite{cumrun}, we wish to interprete the SQM model \eqref{rrrtyuk} (defined on the quotient manifold $\ck$) as modelling $N$ electrons coupled to $d$ units of magnetic flux (produced by the one-particle superpotential form $dW$ as discussed in \S.\ref{basisoxxx}), while the Vandermonde coupling
\be
\cw=2\beta\sum_{i<j} \log(z_i-z_j)+\cdots
\ee
models the topologically relevant part of the  electron-electron Coulomb  interactions.
From eqn.\eqref{azswq} we see that, even if the model is of ``Fermi type'' in the technical sense of \S.\,\ref{secsymm},
we get the ``wrong'' counting of states:
eqn.\eqref{azswq} is the multiplicity for Bose statistics not for Fermi one. What happens is clear: 
for small but non-zero $\beta$, in a classical vacuum for \eqref{rrrtyuk} the $z_i$'s are near a classical vacuum of the one-field model;
several $z_i$ may take distinct values in the vicinity of the same one-field vacuum.
Since their values differ by $O(\beta)$,
such a vacuum corresponds to a non-zero
element of $\mathscr{R}_\textsf{a}$.

The obvious guess is that -- in order to get the correct FQHE phenomenology -- one has to consider only the subspace
\be
\mathscr{V}_\textsf{Fer}\subset \mathscr{V},\qquad\dim\mathscr{V}_\textsf{Fer}={d\choose N}
\ee 
of vacua which survive in the limit $\beta\to0$. In this limit all other vacua
$|\omega\rangle\in \mathscr{V}_\textsf{Fer}^\perp$ escape to the infinite end of  $\ck$, that is, they fall into the excised divisor $S$, cfr.\! eqn.\eqref{wwhatkk}.
The fermionic truncation from $\mathscr{V}$ to $\mathscr{V}_\textsf{Fer}$ is geometrically consistent 
if and only if it is preserved under parallel transport by the $tt^*$ flat connection, that is, if 
$\mathscr{V}_\textsf{Fer}$ is a sub-representation\footnote{\ Since the integral monodromy representation is not reductive in general, $\mathscr{V}_\textsf{Fer}$ needs not to be a direct summand of the monodromy representation. However the contraction of the monodromy given by the UV Berry monodromy is reductive, so $\mathscr{V}_\textsf{Fer}$ must be a direct summand of the UV Berry monodromy.} of the monodromy representation. Since the
flat $tt^*$ connection is the Gauss-Manin connection of the local system on $\mathscr{X}$ provided by the BPS branes (for fixed $\zeta\in\bP^1$), this is equivalent to the condition that the model has ${d\choose N}$ preferred branes which remain regular as $\beta\to0$ and span the dual space to $\mathscr{V}_\textsf{Fer}$.\medskip

Luckily, the fermionic truncation has already been studied by Gaiotto and Witten in a strictly related context, see \S.\,6.5 of \cite{gaiottoknot}. They show that preferred branes with the required monodromy properties do exist. We review their argument in our notation.
We assume that the rational one-form $dW$ has a double pole at infinity of strength $\mu$ and $d$ simple poles in general positions. Then 
 \be
 B(z)=\mu\,A(z)+\text{lower degree,}
 \ee
and eqn.\eqref{kkkzq1y} becomes
\be
2\beta\,A(z)\,P^{\prime\prime}(z)+\Big(\mu\, A(z)+\cdots\Big)P^\prime(z)=\mu\,\tilde f(z)\,P(z).
\ee
The monodromy representation is independent of $\mu$ as long as it is non-zero. One takes $\mu$ finite but very large (the reasonable regime for FQHE).
Up to $O(1/\mu)$ corrections the zeros of $P(z)$ coincide with zeros of $A(z)$ and hence with the zeros of $B(z)$.
The fermionic truncation amounts to requiring that their multiplicities are at most one, i.e.\! that the polynomials $P(z)$ and $P^\prime(z)$ are coprime.
In this regime, the product of $N$ one-particle Lefshetz thimbles starting at \emph{distinct} zeros of
$B(z)$ is approximatively a brane for the full interacting model; while the actual brane differs from the product of one-particle ones by some $O(1/\mu)$ correction, they agree in homology and this is sufficient for monodromy considerations.   
\medskip

Essentially by construction, the Fermi truncation is equivalent for the purpose of $tt^*$ monodromy to deleting the Vandermonde interaction from the superpotential (i.e.\! to setting $\beta=0$)
while inserting a chiral operator of the form
$\Delta(z_i)^{2\tilde\beta}$ in the brane amplitudes. Here $\tilde\beta$ is a kind of ``renormalized'' version of $\beta$.
To understand this operation we have preliminary to dwell into some other aspects of $tt^*$ geometry which we discuss next.

\subsection{$tt^*$ geometry for non-univalued superpotentials}\label{nonunivaluedppp}

The basic version of $tt^*$ geometry works under the assumption that the superpotential $\cw$ is an univalued holomorphic function $\ck\to\C$.
Supersymmetry only requires the one-form $d\cw$ to be closed and holomorphic, but not necessarily exact. When the periods of the differential $d\cw$ do not vanish, the topological sector of the SQM  is non-standard.
This aspect is more transparent in the (equivalent) language of the 
 2d TFT obtained by twisting the 2d (2,2) QFT with the same K\"ahler target $\ck$ and (multivalued) superpotential $\cw$. The 2d TFT is well-defined also when $d\cw$ is not exact, but now it is not always true that an infinitesimal variation $\delta x$ of the parameters entering in $\cw$,
\be
\delta S_\text{TFT}= \delta x\int\!\! d^2\theta\; \partial_x \cw +\overline{Q}\text{-exact,}
\ee
is equivalent to the insertion in the topological correlators of a 2-form topological observable $\int \partial_x\cw^{(2)}$, since $\partial_x\cw$ may be multivalued, and hence not part of the TFT chiral ring $\mathscr{R}$. Thus, while
the TFT exists, it does not define a structure of Frobenius manifold on the essential coupling space $\mathscr{X}$, and the dependence of the topological correlations on the parameters $x\in\mathscr{X}$ is not controlled by the Frobenius algebra $\mathscr{R}$.
Since $tt^*$ geometry is obtained by fusing together  the topological and anti-topological sectors, this means that the PDE's which govern the dependence of the $tt^*$ amplitudes on $x$ cannot be written remaining inside $\mathscr{R}$: one needs to enlarge $\mathscr{R}$ to a bigger Frobenius algebra.

\subsubsection{Abelian covers} 
We review the procedure in detail since the Vafa model of FQHE involves all possible subtleties in this story.
In this section we work in full generality: $\ck$ is any Stein (hence complete K\"ahler) field space endowed with a family of holomorphic superpotential one-forms $d\cw_{x}$, parametrized by $x\in\mathscr{X}$, which are closed but not exact.
 
An obvious way to get univalued superpotentials $\cw_{x}$ and reduce ourselves to ordinary $tt^*$ geometry, is to enlarge the model by replacing the field space $\ck$ by its universal cover $\widetilde{\ck}\xrightarrow{\;u\;}\ck$ endowed with the pulled back superpotential one-form $u^*d\cw_{x}$ which is automatically exact on $\widetilde{\ck}$. However, typically, this universal extension of the theory  introduces insuperable and unnecessary  intricacies. A more economic fix is to replace $\ck$ by its universal (Galois)  \emph{Abelian} cover $\ca$,
i.e.\! the cover $\ca\to\ck$ with deck group
the Abelianization $\pi_1(\ck)^\text{Ab}$ of $\pi_1(\ck)$ and 
\be
\pi_1(\ca)=[\pi_1(\ck),\pi_1(\ck)].
\ee 
If $\pi_1(\ck)^\text{Ab}$ contains torsion, we may further reduce the cover to $\ca/(\pi_1(\ck)^\text{Ab})_\text{tor}$. To keep the formulae simple, we assume $\pi_1(\ck)^\text{Ab}$ to be torsion-free,
and concretely define $\ca$ as the quotient of the space
of paths 
starting at a base point $\ast\in\ck$ by a suitable equivalence relation
\begin{gather}\label{sim1}
\ca:=\Big\{f\colon[0,1]\to \ck\ \text{continuous},\ f(0)=\ast\Big\}\Big/\sim\\
f\sim g\quad\Leftrightarrow\quad f(1)=g(1)\ \text{and }0=[g^{-1}f]\in H_1(\ck,\Z),\label{sim2}
\end{gather}
endowed with the projection
\be
\varpi\colon\ca\to\ck,\qquad\varpi\colon f\mapsto f(1)\in\ck.\label{cover}
\ee
The superpotentials $\cw_{x}$ are well-defined on $\ca$. $\ca$ is the smallest cover such that \emph{all} superpotentials are defined.
By construction, the cover \eqref{cover} is Galois with
  Galois group the Abelianization of the fundamental group
  \be
  \mathsf{Gal}(\ca/\ck)\equiv \pi_1(\ck)^\text{Ab}\cong \Z^{b_1(\ck)},\qquad b_1(\ck)\equiv\text{$1^\text{st}$ Betti number of $\ck$}.
  \ee 
  The deck group $\pi_1(\ck)^\text{Ab}$ acts freely and transitively on the pre-images of any point, i.e.\! $\ca\to\ck$ is a principal $\pi_1(\ck)^\text{Ab}$-bundle. $\ca$ is automatically Stein \cite{stein1}.
Since the first Betti number $b_1(\ck)>0$, the cover 
$\ca\to\ck$
 has infinite degree, which means that each vacuum of the original SQM defined on $\ck$ has infinitely many pre-images in $\ca$ which are distinct vacua for the Abelian cover SQM, which then has Witten index $\infty\cdot d$. Luckily, this additional infinity  in the number of vacua causes not much additional trouble.  
$\pi_1(\ck)^\text{Ab}$ acts as a symmetry of the covering quantum system, and hence its vacuum space $\mathscr{V}_\ca$ decomposes in the orthogonal direct sum of unitary irreducible representations of $\pi_1(\ck)^\text{Ab}$. The group is Abelian, and all its irreducible representations are one-dimensional.
The fiber of \eqref{cover} carries the regular representation of $\pi_1(\ck)^\text{Ab}$, and
each irreducible representation appears with the same multiplicity $d$.
Then we have an orthogonal decomposition of the vacuum bundle $\mathscr{V}_{\ca}\to \mathscr{X}$
into $\pi_1(\ck)^\text{Ab}$ eigen-bundles associated to the irreducible (multiplicative) characters of 
$\pi_1(\ck)^\text{Ab,}$
\be\label{decch}
\mathscr{V}_\ca=\bigoplus_{\chi\in \mathrm{Hom}(\pi_1(\ck)^\text{Ab},\, U(1))} \mathscr{V}_\chi,\qquad \mathrm{rank}\,\mathscr{V}_\chi=d\quad \text{for all }\chi.
\ee 
This orthogonal decomposition is preserved by parallel transport with the Berry connection $D$ (since $D$ is metric), but not in general by the flat connection $\nabla^{(\zeta)}$. 
There are two ways to remedy this.
The first is to consider the UV Berry monodromy representation.
This contraction of the $tt^*$ monodromy representation is unitary and metric, hence preserves the orthogonal decomposition
\eqref{decch}. The second in discussed in \S.\,\ref{gencov}.

Identifying $\pi_1(\ca)^\text{Ab}$ (modulo torsion) with $\Z^b$, we write the characters as
\be
\chi_{\vec\theta}\colon \vec n\mapsto e^{i\vec n\cdot\vec\theta},\qquad \vec n\in\Z^b
\ee
and call the states in the eigen-bundle $\mathscr{V}_{\vec \theta}\equiv \mathscr{V}_{\chi_{\vec \theta}}$ the $\vec\theta$-vacua \cite{tt*,twistor,tt*3}.

 \begin{exe}\label{hhhhasq1870} Consider the SQM with $\ck\equiv\C^\times$ and
 \be W(Z)=Z-m\log Z
 \ee where  $m\in\mathscr{X}\equiv\C^\times$.
 This is the basic model entering in the description of the $\nu=1$ phase of FQHE (cfr.\! \textbf{Example \ref{errample}}).
 The Abelian cover of $\ck$ is $\C$, $\varpi\colon X\mapsto e^X\equiv Z$, whose Galois group $\Z$ acts as $k\colon X\mapsto X+2\pi i k$; the corresponding characters are
 $\theta\colon k\mapsto e^{i\theta k}$, $\theta\in[0,2\pi)$. Since the model is free, its Witten index is $1$, and the  representations $\varrho^\theta\colon \pi_1(\mathscr{X})\equiv \Z\to \C^\times$ are one-dimensional. One finds $\varrho^\theta(s)=e^{is(\theta-\pi)}$ ($s\in\Z$).
 The vacuum $|\theta\rangle\in \mathscr{V}_\theta$
  is (up to normalization) the one corresponding to the chiral operator  $z^{\theta/2\pi}\in\mathscr{R}_\ca$. In particular, the brane amplitudes
   in character $\theta$ (which, for the present model, were computed explicitly in \cite{twistor,tt*3}) contain the insertion of $z^{\theta/2\pi}$, so that the effective mass parameter entering in the asymmetric limit amplitudes is $m_\text{eff}=m-i\zeta\theta/2\pi R$
\cite{twistor}. The so-called $\theta$-limit consists in taking the coupling in the superpotential to zero, $m\to0$, while keeping $m_\text{eff}$ fixed. 
 \end{exe}
 
 \subsubsection{General Abelian covers}\label{gencov}

Let $H\subset \pi_1(\ck)^\text{Ab}$ a subgroup, and let $\ca_H=\ca/H$. We have an Abelian cover
\be
\ca_H\to \ck,\qquad \mathsf{Gal}(\ca_H/\ck)=\pi_1(\ck)^\text{Ab}/H,
\ee
and we may consider the 4-\textsc{susy} SQM
with target space $\ca_H$ which is well-defined. One has
\be
\pi_1(\ca_H)=\mathrm{ker}\,\beta
\ee
where $\beta$ is the surjective group homomorphism
\be
\beta\colon \pi_1(\ck)\equiv\mathsf{Gal}(\widetilde{\ck}/\ck) \to \mathsf{Gal}(\ca_H/\ck).
\ee
The
 \textsc{susy} vacua of the LG theory formulated on $\ca_H$
 may be identified with the
$H$-invariant vacua of the universal Abelian covering theory, that is, 
\be\label{qw1234}
\mathscr{V}_H=\bigoplus_{\chi\colon \chi|_H=\text{trivial}}\mathscr{V}_\chi.
\ee
The $\ca_H$ model has its own (generalized) BPS branes, 
which lift to branes of the cover $\ca$ theory, and its vacuum-to-brane amplitudes are preserved by parallel transport with respect to the $tt^*$ Lax connection. Thus, even if each $\mathscr{V}_\chi$ may not be preserved by the brane monodromy representation $\mathsf{Mon}$,
we have one monodromy sub-representation $\mathsf{Mon}_H\subset \mathsf{Mon}$ for each subgroup $H\subset \pi_1(\ck)^\text{Ab}$.
This is an important condition on the monodromy representation $\mathsf{Mon}$.

In particular, we may choose $H$ to be of finite index in $\pi_1(\ca)^\text{Ab}$, so that $\mathsf{Gal}(\ca_H/\ck)$ is a finite Abelian torsion group.
In this case the theory on $\ca_H$ has finite Witten index
\be
d_H=[\pi_1(\ck)^\text{Ab}:H]\cdot d.
\ee
and we get a family of monodromy representations
\be
\varrho_H\colon \pi_1(\mathscr{X})\to GL(d_H,\Z).
\ee
To a sequence of finite-index subgroups
\be
\cdots \subset H_k \subset H_{k-1}\subset \cdots \subset H_1\subset H_0\equiv \pi_1(\ca)^\text{Ab}
\ee
there corresponds an inverse sequence of 
$tt^*$ monodromy sub-representations
\be
\mathsf{Mon}_{H_0}\subset \mathsf{Mon}_{H_1}\subset\cdots\subset \mathsf{Mon}_{H_{k-1}}\subset \mathsf{Mon}_{H_k}\subset\cdots
\ee
where $\mathsf{Mon}_{H_0}$ is the monodromy representation for the original model defined on $\ck$.

\subsubsection{Finite covers vs.\!\! normalizable vacua}\label{covcovvf}
It follows from the above that not all characters $\chi\in (\pi_1(\ck)^\text{Ab})^\vee$ are created equal.
Suppose $\chi$ is torsion, that is,
\be
\vec \theta\in (2\pi\,\bQ)^b,
\ee 
and let  $J_\chi\subset (\pi_1(\ck)^\text{Ab})^\vee$ be the finite cyclic group generated by $\chi$,  and $N_\chi=\mathrm{ker}\,\chi\subset \pi_1(\ck)^\text{Ab}$ the corresponding finite-index normal subgroup
\be
\pi_1(\ck)^\text{Ab}/N_\chi\cong J_\chi.
\ee
 In this case we may reduce from an infinite to a finite cover
\be\label{finitecov}
\varpi_\chi\colon\ca_\chi\equiv \ca/N_\chi\to \ck,\qquad \mathsf{Gal}(\ca_\chi/\ck)=J_\chi,\qquad
\deg \varpi_\chi=|J_\chi|.
\ee
Such a finite cover \eqref{finitecov} is much  better behaved that $\varpi$, e.g.\! if $\ck$ is affine $\varpi_\chi$ is a regular morphism of affine varieties.\footnote{\ See e.g.\! \cite{harris} page 124.}

From the physical viewpoint, torsion characters $\chi\in(\pi_1(\ck)^\text{Ab})^\vee$ have the special property that they allow a consistent truncation of the chiral ring $\mathscr{R}_\ca$ to a finite-dimensional ring $\mathscr{R}_\chi$ so that the $\vec\theta$-vacua $|\vec\theta, a\rangle$, $\vec\theta\in J_\chi$ become normalizable, while they are never normalizable for 
$\chi$ non-torsion. Normalizability of the ground state(s) is a basic principle in quantum mechanics.

 \subsubsection{$tt^*$ equations in $(\C\times S^1)^b$}
 The periods of $d\cw_x$ define an additive character of $\pi_1(\ck)^\text{Ab}\cong \Z^b$
 \be
 \cw_{x}(T_{\vec n}\,z)=\cw_{x}(z)+\vec n\cdot \vec\omega(x),
 \ee
 where $T_{\vec n}$ is the element of the deck group corresponding to $\vec n\in \Z^b$. We assume all components of $\omega_i\equiv \vec \omega(x)_i$ to be non-zero
 and $\bQ$-linearly independent
 (otherwise we consider a smaller Abelian cover and reduce to this case). We choose the local coordinates in $\mathscr{X}$ so that the first $b$ coordinates are the $\omega_i$'s, writing $t_a$ for the remaining ones such that $\partial_{t_a}\cw$ are well-defined holomorphic functions on $\ck$, representing elements of $\mathscr{R}_\ck$.
We write the character $\chi$ in the form
$\vec n\mapsto e^{i\vec n\cdot\vec \theta}$. 
 
We consider the rank-$d$ vector bundle $\mathscr{V}_{\vec \theta}\to\mathscr{X}$
for a fixed character $\vec \theta$, endowed with the $tt^*$ Hermitian metric $G(\vec\theta\,)$. In the canonical trivialization $G(\vec\theta\,)$ satisfies the reality condition
\be
G(-\vec\theta\,)^t=G(\vec\theta\,)^{-1}.
\ee 
As shown in \cite{tt*3},  the metric
$G(\vec \theta\,)$, seen as a function of the
variables 
\be
(\omega_i,\theta_i)\in \big(\C\times S^1\big)^b 
\ee 
for fixed $t_a$, satisfies the $3b$-dimensional analogue of the 3d non-Abelian Bogomolnji monopole equations with gauge group $U(d)$. Indeed the ``Higgs field'' in the $\omega_i$ direction, $C_{\omega_i}$, becomes an $U(w_k)$ covariant derivative in the $\theta_i$ direction
\be\label{corhyyyep}
C_{\omega_i}\leadsto \overline{D}_{\bar\vartheta_i}= \frac{\partial}{\partial \theta_i}+M_{\omega_i},\qquad
\overline{C}_{\bar\omega_i}\leadsto -{D}_{\vartheta_i}= \frac{\partial}{\partial \theta_i}-G\partial_{\theta_i}G^{-1}-GM_{\omega_i}^\dagger G^{-1}. 
\ee
At fixed $t_a$, the components of the $tt^*$ 
flat connection take the form
\be
D_{\omega_i}+\frac{1}{\zeta}\overline{D}_{\bar\vartheta_i}\equiv \boldsymbol{D}_{1,i}^{(\zeta)},\qquad \overline{D}_{\bar\omega_i}-\zeta\,D_{\vartheta_i}\equiv \boldsymbol{D}_{2,i}^{(\zeta)}.
\ee
Seeing the $\vartheta_i$'s as complex coordinates with real part $\theta_i$,
and introducing the new complex coordinates
$(\eta_i, \xi_i)$ ($i=1,\dots,b$)
\be
\eta_i= \omega_i-\zeta \bar \vartheta_i, \qquad \xi_i= \bar \omega_i+\frac{1}{\zeta}\,\vartheta_i,
\ee
which defines a $\bP^1$ family of complex structures parametrized by the twistor variable $\zeta$, and a flat hyperK\"ahler geometry with holomorphic symplectic structures
\be
d\eta_i\wedge d\xi_i= \frac{1}{\zeta}\,d\omega_i\wedge d\vartheta_i+\big(d\omega_i\wedge d\bar\omega_i+d\vartheta_i\wedge d\bar\vartheta_i\big)+ \zeta\, d\bar\omega_i\wedge d\bar\vartheta_i.
\ee 
One has 
\be
\boldsymbol{D}_{\alpha,i}^{(\zeta)}\, \eta_j=
\boldsymbol{D}_{\alpha,i}^{(\zeta)}\,\xi_j=0,\quad\alpha=1,2,
\ee
i.e.\! the first-order differential operators $\boldsymbol{D}_{\alpha,i}^{(\zeta)}$ are
of pure type (0,1) in complex structure $\zeta$ and the $tt^*$ Lax equations
\be\label{holohyper}
\boldsymbol{D}_{\alpha,i}^{(\zeta)}\,\Psi(\zeta)=0,\quad \alpha=1,2,
\ee
just say that the brane amplitudes $\Psi(\zeta)$ are holomorphic in complex structure $\zeta$ and independent of 
$\mathrm{Im}\,\vartheta_i$ \cite{tt*3}. The $tt^*$ equations then say that the curvature of the connection $\boldsymbol{D}^{(\zeta)}$ on the flat hyperK\"ahler manifold is of type (1,1) in all complex structures, i.e.\!
$\Psi(\zeta)$ is a section of a hyperholomorphic vector bundle \cite{tt*3}.
The hyperholomorphic condition, supplemented by the condition on translation invariance in $\mathrm{Im}\,\vartheta_i$, is equivalent to the higher dimensional generalization of the Bogomolnji monopole equations on
$(\R^2\times S^1)^b$.  

The $tt^*$ geometry decomposes into an Abelian $U(1)$ monopole and a non-Abelian $SU(d)$ monopole. 
The monopoles are localized at loci in parameter space $\mathscr{X}$ where the mass gap
of the 2d LG model closes. Thus each such locus carries an Abelian and a non-Abelian magnetic charge.
Restricted to the Abelian part, the $tt^*$ equations become linear; writing
$L(\vec\theta)=-\log\det G(\vec\theta)$, they read
\begin{align}\label{kkkqawe}
&\left(\frac{\partial^2}{\partial \omega_i\partial \bar \omega_j}+\frac{\partial^2}{\partial \theta_i\partial \theta_j}\right)L(\vec\theta)=0\\
&\frac{\partial^2}{\partial t_a\partial \bar \omega_j}L(\vec\theta)=\frac{\partial^2}{\partial \bar t_a\partial \omega_j}L(\vec\theta)=\frac{\partial^2}{\partial t_a\partial \bar t_b}L(\vec\theta)=0
\end{align} 
These equations hold in regions in parameter space where the model has a mass-gap; on the massless locus there are sources
in the \textsc{rhs} localized at trivial characters, that is, they are the loci where a non-zero abelian magnetic charge is present. In the $i$-th factor 3-space
of coordinates $\omega_i,\theta_i$ (all other fixed) this is a real codimension 3 locus.

We note that the equations \eqref{kkkqawe} are identical to the HKLR equations \cite{hklr} describing a hyperK\"ahler metric $\ch_n$ of quaternionic dimension $n$ with $n$ commuting Killing vectors $K_a$ such that their $Sp(1)$ orbits span $T\ch_n$.
For instance, for the model in  \textbf{Example \ref{hhhhasq1870}}  the K\"ahler manifold $\ch_1$ is the Hoguri-Vafa space \cite{ogurivafa} (a.k.a.\! periodic Taub-NUT). This is the target space of the GMN 3d $\sigma$-model obtained compactifing 4d $\cn=2$ SQED \cite{GMN1}, and the brane amplitudes $\Psi(\zeta)$ -- which are locally holomorphic functions in complex structure $\zeta$ -- coincide with the GMN holomorphic Darboux coordinates \cite{twistor,Cecotti:2010fi}.   
\medskip

The Abelian part of the Berry connection is
\be
A=\partial L(\vec\theta)=\partial_{w_i} L(\vec\theta)\,dw_i+\partial_{t_a}L(\vec\theta)\, dt_a.
\ee
The $tt^*$ relation\footnote{\ Here $\mathsf{A}$ is the full Berry $U(d)$ connection.} $[\mathsf{A}_{w_i}, C_{t_a}]=[\mathsf{A}_{t_a},C_{w_i}]$, together with eqn.\eqref{corhyyyep}, implies
\be
\partial_{\theta_i}\mathsf{A}_{t_a}=[\mathsf{A}_{t_a}, M_{w_i}]-[\mathsf{A}_{w_i}, C_{t_a}].
\ee
Taking the trace gives $\partial_{\theta_i}A_{t_a}=0$; since $A_{t_a}$ is odd in $\vec\theta$, we conclude that the $t_a$-components of the $U(1)$ connection vanish.

\subsubsection{The covering chiral ring $\mathscr{R}_\ca$}\label{covR}

 The chiral ring $\mathscr{R}_{\ca}$ of the (torsion-free\footnote{\ By the \emph{torsion-free} Abelian cover we mean eqns.\eqref{sim1}\eqref{sim2} where we replace $H_1(\ck,\Z)$ with $H_1(\ck,\R)$ in the definition of the equivalence relation $\sim$. 
The LG models relevant for FQHE have free $H_1(\ck,\Z)$, so the distinction is immaterial.}) universal  Abelian cover SQM has a simple form\footnote{\ In the following argument the assumption that $\ck$ is Stein is crucial.}. For a LG model with target a Stein manifold $\ck$ and a superpotential differential $d\cw$ with finitely many simple zeros, the chiral ring $\mathscr{R}$ is identified with the space of functions on the critical set $\{d\cw=0\}$.
This remains true for the LG model uplifted to the torsion-free Abelian cover $\ca$ of $\ck$. Let us sketch the construction. Since $\ck$ is Stein\footnote{\  See pages 445, 449, or 451 of \cite{GH}, or \textsc{theorem G} on page 198 of \cite{g}.}
\be
H^*(\ck,\C)\cong H^*_\text{DR}(\ck)\cong H^*_\text{DR}(\ck,\text{hol}).
\ee
Then we may find holomorphic one-forms
 $\varrho_k\in\Omega^1(\ck)$ ($k=1,\cdots, b$) whose classes generate $H^1(\ck,\Z)/\mathrm{tor}$. The critical set of $\cw$ in $\ca$ ($\equiv$ classical vacua in the Abelian cover model) is
 \be
 \mathsf{cri}_\ca=\Big\{l\colon[0,1]\to\ck\;\Big|\; l(0)=\ast,\ l(1)\in \{d\cw=0\}
 \Big\}\Big/\sim\quad \subset\ca.
 \ee
Adding to $\varrho_k$
 an exact term we may assume with no loss
 \be
 \int_l \varrho_k\in\Z\quad\text{for all }l\in\mathsf{cri}_\ca.
 \ee
 On $\ca$ there exist global holomorphic functions $h_k$ such that $\varrho_a=dh_k$.

Now let $\{\phi_a\}\in \mathscr{R}_\ck$  be holomorphic functions on $\ck$ which form a basis of the chiral ring for the original model, with $\phi_0=1_\ck$ and product table $\phi_a\phi_b={C_{ab}}^c\phi_c$. Clearly
the holomorphic functions on $\ca$
\be\label{jjasqwe6}
\Phi_a(\vec\theta\,)\equiv \varpi^*\phi_a\cdot \exp\!\big[i\, \vec\theta\cdot \vec h\big],\qquad a=1,\dots,w_\ck, \ \ \vec\theta\in [0,2\pi)^b, 
\ee
yield a topological basis of $\mathscr{R}_\ca$ diagonal in the characters of $H_1(\ck,\Z)/\mathrm{tor}$. The product table of $\mathscr{R}_\ca$ is then
\be
\Phi_a(\vec\theta\,)\cdot\Phi_b(\vec \varphi)= {C_{ab}}^c\,\Phi_c(\vec\theta+\vec\varphi).
\ee
From this it also follows that the UV Berry connection $A(\vec\theta)^\textsc{uv}$ is a piece-wise linear function of $\vec\theta$ \cite{twistor}. The discontinuous jumps of $A(\vec\theta)^\textsc{uv}$ correspond to gauge transformations, and the characters of the monodromy representation are continuous. For generic $\vec\theta$ the eigenvalues of the monodromy matrices are distinct, and hence no Jordan blocks are present; at characters where we have ``jumps'' typically a non-trivial Jordan blocks appear. 
\medskip

Let $H\subset \Z^b\cong \pi_1(\ck)^\text{Ab}$ be a subgroup and 
$J=\mathrm{ker}\,H$ the subgroup of characters which are trivial on $H$.
The chiral ring $\mathscr{R}_H$ of the model defined on the cover $\ca_H$ (cfr.\!
\S.\,\ref{gencov}) is spanned by the chiral  operators
\be
\Big\{\Phi_a(\vec\theta\,)\Big\}_{\vec\theta \in J}.
\ee

\begin{exe} For the model in \textbf{Example \ref{hhhhasq1870}},
one has $\phi_1=1$ and
$h=\log(z/m)$. Then the vacua has the form $m^{-\theta/2\pi}|z^{\theta/2\pi}\rangle$, and the brane amplitudes contain the insertion $(z/m)^{\theta/2\pi}$ as expected. 
\end{exe}

\subsubsection{A fancier language}\label{fancier}

For the sake of comparison with the literature on representation of braid groups and the Knizhnik-Zamolodchikov equation
\cite{braid,kohnoN} we state the above result in a different way. We write
$q_i=e^{i\theta_i}$ for $i=1,\dots, b$.
Clearly $\mathscr{R}_\ca$ is a module
over the ring $\C[\{q_i^{\pm1}\}]$ of Laurent polynomials in $q_1,\cdots,q_b$. The isomorphism\footnote{\ Recall that $\mathscr{R}_\ca^\vee\cong\mathscr{R}_\ca$ since $\mathscr{R}_\ca$ is a Frobenius algebra.} 
\be
\mathscr{R}_\ca\cong \mathscr{B}(\zeta)_\ca\equiv H_*\!(\ca,\ca_{x,\zeta},\Z)\otimes_\Z\C
\ee
 allows us to restrict the scalars to $\Z$. Thus

\begin{fact} $\mathscr{B}(\zeta)_\ca\cong\mathscr{R}_\ca\cong \mathscr{V}_\ca$ is a \emph{free}
$\Z[\{q_i^{\pm1}\}]$-module of rank $d$.
Then the $tt^*$ Lax connection defines a
group homomorphism 
\be
\varrho\colon \pi_1(\mathscr{X})\to \mathsf{Aut}_{\Z[\{q_i^{\pm1}\}]}(\mathscr{B}_\ca)
\ee
where 
\be
\mathsf{Aut}_{\Z[\{q_i^{\pm1}\}]}(\mathscr{B}_\ca)\subset GL\big(d,\Z[\{q_i^{\pm1}\}]\big)
\ee 
stands for the group of $\Z[\{q_i^{\pm1}\}]$-linear automorphisms which preserve
the bilinear intersection form between dual branes
\be
\mathscr{B}(\zeta)_\ca\otimes \mathscr{B}(-\zeta)_\ca\to\Z[\{q_i^{\pm1}\}]
\ee
(cfr.\! eqn.\eqref{bin1}).
\end{fact}

\section{$tt^*$ geometry of the Vafa {\rm 4-\textsc{susy}} SQM}\label{vvvmmm}

Now we have all the tools to analyze the
Vafa model of FQHE.

\subsection{Generalities}

For simplicity, we take the $N$ electrons to move in the plane $\C$ instead of the more rigorous treatment in which they move in a periodic box (i.e.\! a large 2-torus $E$). We write $z_i$, $x_a$ ($a=1,\dots, n$), and $y_\alpha$ ($\alpha=1,\dots,M$) for, respectively, the positions of the electrons, of the quasi-holes, and the support of the polar divisor $D_\infty\sim D$ which models the magnetic flux (cfr.\! \S.\ref{basisoxxx}). 
The points $d\equiv n+M$ points $\{x_a,y_\alpha\}\subset\C$
are all distinct. 

The Vafa model is the LG SQM with target space
\be
\ck_{d,N}=\Big\{(z_1,\cdots,z_N)\in \big(\C\setminus \{x_1,\cdots,x_n, y_1,\cdots, y_M\}\big)^N\;|\;z_i\neq z_j\ \text{for }i\neq j\Big\}\Big/\mathfrak{S}_N.
\ee
In the experimental set-up $N$ is very large, while $N/d=\nu$ and $n$ are fixed. Despite this, we shall keep $N$ arbitrary as our arguments apply both for $N$ small and large. 
\medskip

We have already noted in \S.\ref{HStheory} that $\ck_{d,N}$ is an affine variety
\be\label{whicchfkk}
\ck_{d,N}=\C^N\setminus S,\quad S=\big\{\mathsf{discr}(P)\,\mathsf{Res}(A,P)=0\big\},\quad A(z)=\prod_a(z-x_a)\prod_\alpha(z-\zeta_\alpha).
\ee
It is convenient to write $\ck_{d,N}=\bP^N\setminus T$, where $T$ is the obvious divisor. By Hironaka theorem we may blow-up the geometry so that
\be
\ck_{n,N}= \widehat{\bP^N}\setminus S_\textsc{snc},
\ee
with $S_\textsc{snc}$ a normal crossing divisor (see \cite{kohnoN} for details).  

The superpotential is
\be\label{vafamodellls}
\cw=\beta\sum_{i<j} \log(z_i-z_j)^2+\sum_i\!\left(\mu\, z_i+\sum_{a=1}^n\log(z_i-x_a)-\sum_{\alpha=1}^M \log(z_i-\zeta_\alpha)\right),
\ee
rewritten in terms of the elementary symmetric functions $e_k$
\be\label{vafamodellls2}
\cw=\beta \log\mathsf{discr}(P)+\mu\, e_1+\sum_{a=1}^n\log P(x_a)-\sum_{\alpha=1}^M \log P(\zeta_\alpha),
\ee
where
\be
P(z)=\sum_{k=0}^N (-1)^k\, e_{k}\,z^{N-k}.
\ee
We have introduced the coupling $\mu$ to make the problem better behaved. Note that, as long as $\mu$ is not zero, it can be set to 1 by a field redefinition. 
 
 The superpotential $\cw$ is not univalued in $\ck$. As discussed in \S.\ref{nonunivaluedppp}, we have two kinds of couplings: the $\omega$-type given by the residues of $d\cw$ at its poles, and the $t$-type given by the positions $x_a$, $\zeta_\alpha$.
The residues of the poles of $d\cw$ at $x_a$ and $\zeta_\alpha$ are frozen to the values $\pm 1$ by the argument in \S.\,\ref{herhermm}, and the corresponding couplings will play no role in the following discussion. The only relevant $\omega$-type coupling is $\beta$. 
Working in a periodic box, $\beta$ is frozen to the rational number $1/(2\nu)$ (cfr.\! \S.\,\ref{cobeta}); on the contrary, when the electrons move on $\C$, the SQM model makes  sense for an arbitrary complex $\beta$. The monodromy representation is independent of $\beta$, and we are free to deform it away from its physical value $1/(2\nu)$ to simplify the analysis. 

The non-frozen couplings are the $x_a$ and the $\zeta_\alpha$ which form a set of $d$ distinct points in $\C$ identified modulo permutation of equal ``charge'' ones. The manifold of essential couplings is then
\be\label{mmmbvxp}
\mathscr{X}= \cc_{n+M}\Big/\mathfrak{S}_n\times \mathfrak{S}_M\xrightarrow{\;\textsf{proj}\;}\cy_n,
\ee
where $\cy_n$ is the space defined in \eqref{qqquaspa}.
The
$\zeta_\alpha$'s are homogeneously distributed on $\C$, and their detailed distribution is not very important for our present purposes, so we mainly focus on the projection on $\cy_n$.

One has
\be
1\to\mathscr{P}_{n+M}\to \pi_1(\mathscr{X})\to \mathfrak{S}_n\times \mathfrak{S}_M\to 1.
\ee
$\pi_1(\mathscr{X})$ contains $\cb_n$ as a subgroup.
The UV Berry connection yields a family of  unitary arithmetic representations of 
$\pi_1(\mathscr{X})$; restricting to $\cb_n$ we get a monodromy representation 
\be
\varrho_\downarrow\!(\vec\theta\,)\colon \cb_n\to GL(\mathscr{V}_{\vec \theta}) 
\ee
parametrized by the characters $\vec \theta$ of $\mathsf{Gal}(\ca/\ck)$. 

Before the fermionic truncation the number of vacua with fixed $\vec \theta$ is
\be
d_{d,N}={N+d-1\choose N}
\ee 
which reduce to just ${d\choose N}$
after the truncation.

\subsection{Topology of the field space $\ck_{d,N}$}
One has
\be
\pi_1(\ck_{d,N})=\mathscr{B}(N, S_{0,d+1})
\ee
where $\mathscr{B}(n, S_{g,p})$ stands for the \textit{braid group in $n$ strings on the surface $S_{g,p}$} of genus $g$ with $p$ punctures\footnote{\ In this notation the standard (Artin) braid group is $\cb_n=\mathscr{B}(n,S_{0,1})$.}.
$\mathscr{B}(n,S_{0,p})$ has the following convenient presentation (\!\!\cite{presentationB} thm.\! 5.1):
\begin{align}
&\text{\bf generators:} &&\sigma_1,\sigma_2,\cdots,\sigma_{n-1}, z_1, z_2,\cdots, z_{p-1}\\
&\text{\bf relations:} &&\left\{\begin{aligned}
&\sigma_i\sigma_{i+1}\sigma_i=\sigma_{i+1}\sigma_i
\sigma_{i+1}, &\ \ & \sigma_i\sigma_j=\sigma_j\sigma_i\ \ \text{for }|i-j|\geq 2,\\
&z_j\sigma_i=\sigma_i z_j\ \ \text{for }i\neq1 &&
\sigma_1^{-1}z_j\sigma_1^{-1}z_j=z_j\sigma_1^{-1}z_j\sigma^{-1}_1\\
&\sigma_1^{-1}z_j\sigma_1z_l=z_l\sigma_1^{-1}z_j\sigma_1\ \ \text{for }j<l.
\end{aligned}\right.
\end{align}
The $\sigma_i$ generate a subgroup of $\mathscr{B}(n,S_{0,p})$ isomorphic to the Artin braid group $\cb_n$.
Then 
\be
\mathscr{B}(n,S_{0,p})^\text{Ab}=\text{the free Abelian group in the generators $\sigma, z_1,\cdots, z_{p-1}$}\cong \Z^p,
\ee
which corresponds to
$H^1(\ck,\Z)\cong\Z^p$ with generators (cfr.\! eqn.\eqref{whicchfkk})
\be
\frac{1}{2\pi i} d\log \mathsf{discr}(P),
\quad \frac{1}{2\pi i} d\log P(x_a),\quad  \frac{1}{2\pi i} d\log P(\zeta_\alpha).
\ee
\emph{A priori,} there is one angle
associated to each of these generators; let as call them
\be
\theta,\quad \phi_a,\quad \varphi_\alpha,
\ee
respectively.
If (as physically natural) we consider the quasi-holes and the magnetic-flux units to be indistinguishable we shall takes the corresponding angles to be all equal
$\phi_a\equiv\phi$ and $\varphi_\alpha\equiv\varphi$. In the formalism developed in \S.\,\ref{gencov}, this corresponds to taking the quotient group $\Z\oplus\Z\oplus\Z$ of $H_1(\ck_{d,N},\Z)$ dual to the subgroup of $H^1(\ck_{d,N},\Z)$ generated by the three differentials
\be
\frac{1}{2\pi i} d\log \mathsf{discr}(P),
\quad \frac{1}{2\pi i} d\log \prod_{a=1}^n P(x_a),\quad  \frac{1}{2\pi i} d\log\prod_{\alpha=1}^M P(\zeta_\alpha),
\ee
and considering the LG model on the Abelian cover $\ca_H$ where 
\be\label{kkkerna}
H\equiv \mathrm{ker}\,H_1(\ck_{d,N},\Z)\to \Z\oplus\Z\oplus\Z.
\ee 
In particular $\pi_1(\ca_H)\cong\Z^3$.
Therefore, \textit{a priori}, we have three angles $\theta$, $\phi$ and $\varphi$.
Setting $q=e^{i\theta}$, $t=e^{i\phi}$, and $y=e^{i\varphi}$ we conclude: 
\begin{fact} In the LG model with indistinguishable defects,
the BPS branes span a \emph{free} $\Z[q^{\pm1},t^{\pm1},y^{\pm1}]$-module of rank $d_{d,N}$. Normalizability of the ground states requires specialization to $q$, $t$ and $y$  roots of unit.
\end{fact}
 
However the physical FQHE is a much simpler quantum system, and further truncations are present. We shall dwell on this issue in \S.\!\ref{fqqsy}. Before going to that, we present a different application of the $tt^*$ geometry of a special case of the LG model \eqref{vafamodellls}.

\subsection{Homological braid representations as $tt^*$ geometries}

The theory of general homological braid representations \cite{lawrence,big1,big2,kohnoN}\!\!\cite{braid} is just a special topic in $tt^*$ geometry.
For the sake of comparison with the geometry of the Vafa model, we briefly review that story following \cite{big1,big2,kohnoN} but using $tt^*$ language. 
\medskip

There is a sequence of such monodromy representations $\mathsf{Mon}_N$ for the braid group $\cb_n$ labelled by an integer $N\in\bN$ \cite{lawrence}; 
for $N=1$ we get the Burau representation \cite{burau,kohno}, and for $N=2$ the Lawrence-Krammer-Bigelow one \cite{big1,big2}\!\!\cite{braid}.

$\mathsf{Mon}_N(\cb_n)$ is just the $tt^*$ (Lax) monodromy representation for the superpotential  \eqref{vafamodellls2} for $N$ electrons and $n$ quasi-holes with 
 $\mu=0$ (which makes things a lot less nice), $M=0$ and $\beta$ a real positive number, say 1. 
The quasi-hole are indistinguishable.
Since $M=0$, there are no angles $\varphi_\alpha$ and  eqn.\eqref{kkkerna} reduces to
\be
H\equiv \mathrm{ker}\,H_1(\ck_{n,N},\Z)\to \Z\oplus\Z.
\ee 
One defines the LG model on the Abelian cover $\ca_H$, so that the 
BPS-branes at given $\zeta$
\be
\mathscr{B}_\zeta\equiv H_\ast(\ca_H,\ca_{H,\zeta},\Z)\equiv H_N(\ca_H,\ca_{H,\zeta},\Z)
\ee
form
a $\Z[q^{\pm1},t^{\pm1}]$-module of rank $d_{n-1,N}$.\footnote{\ The shift $n\to n-1$ is due to $\mu=0$.}
The monodromy acting on the branes yields a braid group representation
\be\label{jjjasqw}
\cb_n\to GL\big(d_{n-1,N},\Z[q^{\pm1},t^{\pm1}]\big).
\ee 

Mathematicians focus on the two
dual modules of branes at 
$\zeta=+1$ and $\zeta=-1$.
The main character in the theory
is the non-degenerate pairing
\be
\langle\cdot,\cdot\rangle\colon \mathscr{B}_+\times\mathscr{B}_-\to\Z[q^{\pm1},t^{\pm1}]
\ee
corresponding to the TFT metric $\eta$
(cfr.\! eqn.\eqref{bin1}) given by the standard $tt^*$ formula already written in the original paper \cite{tt*}. In the present context, and for the special case $N=2$, it is called the Blanchfield pairing
\cite{blanch}\!\!\cite{big1,knotB} (for details, see e.g.\! \S.3.3.5 of the book \cite{braid}).

For \emph{generic} $q$ and $t$, the monodromy representation \eqref{jjjasqw} is
equivalent to a sub-representation of a $\mathfrak{sl}_2$ Kniznick-Zamolodchikov representation on the sub-bundle of the eigenbundle of the total angular momentum $L_3$ corresponding to the $N$ electron sector
(cfr.\! \S.\,\ref{staHecke}) of higher weight states, see \cite{kohnoN} for details.    

\subsection{The FQHE quantum system}
\label{fqqsy}

\subsubsection{Characters of $\mathsf{Gal}(\ca/\ck)$}

The FQHE quantum system is a particular version of the 4-\textsc{susy} LG model with superpotential $\cw$ in
\eqref{vafamodellls2}.
Quasi-holes and magnetic-flux units are indistinguishable, but we have still to fix the characters $\theta$, $\phi$, and $\varphi$.
\medskip

The sole purpose of the poles in $d\cw$ at the points $\zeta_\alpha$ is to mimic the external  magnetic field  $B$ \emph{via} the isomorphism in \S.\,\ref{basisoxxx}. The discussion in that section was done  entirely in $\ck$, without any mention of a non-trivial Abelian cover $\ca_H$, and hence referred to the trivial character $\varphi=0\bmod2\pi$. Therefore we set to zero the angles associated to the generators of $H^1(\ck,\Z)$ of the form $d\log P(\zeta_\alpha)/2\pi i$. 
This may look as a simplification, but it has a technical drawback. With this choice of character the \emph{genericity} condition in (say) ref.\!\cite{kohnoN} fails, and several standard results do not longer apply.

The quasi-holes may be though of as ``wrong-sign'' elementary magnetic fluxes, so it looks natural to expect that the characters associated to the generators $d\log P(x_a)/2\pi i$ should also be trivial,
$\phi=0\bmod2\pi$. Clearly, one may extend the analysis to $\phi\neq0$.
The previous \emph{caveat} apply to this character as well.

We remain with just one non-trivial angle $\theta$
associated to the Vandermonde coupling
$\beta$. From the considerations in \S.\ref{covcovvf} we expect $\theta$ to be  rational multiple of $2\pi$
\be\label{charcgar}
\theta=\pi\!\left(1+\frac{a}{b}\right), \quad a\in\Z, b\in \bN,\quad \gcd(a,b)=1,\quad
-b\leq a\leq b.
\ee
The $tt^*$ reality structure relates $-a$ to $a$, so if a pair $(a,b)$ corresponds to a quantum phase of FQHE it is natural to expect that a ``dual'' phase associated to $(-a,b)$ exists as well. We shall write $(\pm a,b)$ with $1\leq a\leq b$ to cover both phases at once. 
\medskip

The Abelian cover $\ca_{\theta}$ associated to the character \eqref{charcgar} is (cfr.\! eqn.\eqref{finitecov})
\be
\ca_{\theta}=\Big\{l\colon [0,1]\to \ck, \ l(0)=\ast\Big\}\Big/\sim_b,
\ee
where $\sim_b$ is the equivalence
\be
l\sim_b l^\prime\quad\Leftrightarrow\quad
l(1)=l^\prime(1)\ \text{and }\int_l  d\log \mathsf{discr}(P)- 
\int_{l^\prime}  d\log \mathsf{discr}(P)\in 4\pi i\,b \Z.
\ee

\subsubsection{Fermionic truncation vs.\!
$\theta$-limit} 

The model \eqref{vafamodellls2} is of the Heine-Stieltjes class. As discussed in \S.\ref{fermtruc},
to get the correct physical counting of states we need to consider its fermionic truncation, i.e.\! to keep only the states which do not escape to the excised divisor $S$ as $\beta\to0$.
Since the monodromy is independent of $\beta$ (as long as it is not zero), and its limit as $\beta\to0$ is smooth after the fermionic truncation, we may as well set $\beta=0$ while keeping track of the non-trivial topology of $\ck$ through its associated character $\theta$. In the language of
\cite{twistor} this is the ``$\theta$-limit''. Roughly speaking, in the $\theta$-limit the only effect of $\theta\neq0$ is to make a ``non-commutative deformation'' of the geometry with deformation parameter
\be
q\equiv e^{i\h}\leadsto e^{i\theta}.
\ee
The monodromy matrices then are valued in $\Z[q^{\pm1}]$, in agreement with \S.\,\ref{fancier} 
(see also the discussion in ref.\!\cite{gaiottoknot}).

\medskip

From eqn.\eqref{jjasqwe6} we see that
switching on a non-zero $\theta$
means
\be
|\phi_a\rangle \to |e^{i\theta\,h}\,\phi_a\rangle = e^{i\theta\,h}|\phi_a\rangle+\overline{Q}\text{(something)},
\ee
that is, it corresponds to inserting in the BPS brane amplitudes the chiral field
$e^{i \theta\, h}$.
For the model \eqref{vafamodellls2}, $e^{i\theta\, h}$ is proportional to
\be
\mathsf{discr}(P)^{\theta/2\pi}=\prod_{i<j}(z_i-z_j)^{\theta/\pi}.
\ee
Keeping into account the Jacobian of $\{z_i\}\mapsto\{e_k\}$, \S.\!\ref{fermmmmqa}, the vacuum wave-functions in terms of the $z_i$'s contain the factor
\be
\prod_{i<j}(z_i-z_j)^{1+\theta/\pi},\qquad 0\leq \theta \leq 2\pi.
\ee  
Comparing with the Laughlin wave-functions \cite{Laughlin} we are led to the identification
\be\label{nutheta}
\frac{1}{\nu}=1+\frac{\theta}{\pi}=2\pm \frac{a}{b}
\ee
which gives $1\leq 1/\nu\leq 3$. In particular, the minimal $b$-torsion character, $a=1$, yields the FQHE principal series \cite{cumrun}
\be\label{prinseries}
\nu= \frac{b}{2b\pm 1},\quad b\in\bN.
\ee
Although this series are the most natural LG quantum systems of the form \eqref{vafamodellls2}, it is by no means the only possibility in the present framework.

\subsection{The $tt^*$ geometry of the Vafa model is very complete}\label{kkkaqwu}

We are reduced to the fermionic truncation of the model \eqref{vafamodellls} which allows us to effectively put the coupling $\beta$ to zero.
Then, provided we may show that the $tt^*$ geometry of the one-field model
\be
W(z)=\mu\, z+\sum_{a=1}^n\log(z-x_a)-\sum_{\alpha=1}^M \log(z-\zeta_\alpha)
\ee
is very complete, we may apply the arguments of \S.\,\ref{staHecke} and conclude that the monodromy representation factors through a Hecke algebra
(in facts through the Temperley-Lieb algebra). 

Again we follow \cite{gaiottoknot}. The monodromy representation is independent of $\mu$,
and we choose it to be very large $\mu\gg 1$. The \textsc{susy} vacuum equations (the Bethe ansatz equations in the language of \cite{gaiottoknot})
\be
\mu= \sum_\alpha \frac{1}{z-\zeta_\alpha}
-\sum_a \frac{1}{z-x_a},
\ee
have $n+M$ solutions of the form
$z=x_a+O(1/\mu)$ or $z=\zeta_\alpha+O(1/\mu)$ and the critical values, rescaled by a factor $\mu^{-1}$, are
\be
\Big\{w_1,\cdots, w_{n+M}\Big\}=\Big\{x_1, x_2,\cdots, x_n,  \zeta_1,\zeta_2,\cdots, \zeta_M\Big\}+O\!\left(\frac{1}{\mu}\log \mu\right).
\ee
So that the cover $\cc_{n+M}$ of the coupling space 
$\mathscr{X}$ (cfr.\! eqn.\eqref{mmmbvxp}) is naturally identified 
with the cover $\cc_{n+M}$ of the critical value space. Since to get $\mathscr{X}$ we quotient out only the subgroup $\mathfrak{S}_n\times \mathfrak{S}_M\subset \mathfrak{S}_{n+M}$, we need to work on a cover of the actual critical value space $\cy_{n+M}\equiv\cc_{n+M}/\mathfrak{S}_{n+M}$, but this is immaterial for the monodromy representation of the subgroup $\cb_n$.

The commutative diagram \eqref{kkxz00irr} takes the form \be\label{kkxz00irr4}
\xymatrix{\cc_{n+M}\ar[rr]\ar@/^2.5pc/[rrrr]^p &&\mathring{\mathscr{X}}\simeq\cc_{n+M}/\big(\mathfrak{S}_n\times \mathfrak{S}_M\big)\ar[rr]_w&&\cy_{n+M}\equiv \cc_{n+N}/\mathfrak{S}_{n+M}}
\ee 
where all maps are canonical projections.
This shows that the one-field theory (hence the tensor product of $N$ decoupled copies of it) has a very complete $tt^*$ geometry.
\medskip

The ideas of section 4 lead to the conclusion that the UV Berry monodromy representation of $\pi_1(\cy_n)\equiv \cb_n$
 is given by the holonomy in $\cy_n$ of a flat $\mathfrak{sl}(2)$ Kohno connection
\be\label{conclusion}
\cd= d+\lambda(\theta)\,\sum_{i<j} s^{(i)}_\ell s^{(j)}_\ell\, \frac{d(w_i-w_j)}{w_i-w_j}
\ee
acting on the space $\boldsymbol{V}^{n+M}$, restricted to the subspace of total angular momentum 
\be
L_3=N-(n+M)/2.
\ee
In eqn.\eqref{conclusion},  $\lambda(\theta)$ is some piece-wise linear function of $\theta$.  In the context of actual FQHE the character $\theta$ is expected to be related to the filling fraction $\nu$ as in eqn.\eqref{nutheta}. 
In particular, the monodromy representation factors through a Temperley-Lieb algebra.

It remains to compute the function $\lambda(\theta)$.

\begin{rem} One expects a simple relation between the monodromy of the
Knizhnik-Zamolodchikov connection \eqref{conclusion} and the homological one associated to the asymmetric limit.
It is known that \emph{for generic angles} $\theta$, $\phi$ and $\varphi$ the monodromies associated to the $\mathfrak{sl}(2)$ Gaudin model with an irregular singularity at $\infty$ (i.e.\! with $\mu\neq0$) yield all monodromy representations of the $\mathfrak{sl}(2)$ eigenvectors \cite{gaiottoknot,gir3}. However here we have three major sources of difference with the situation studied in the math literature:
\begin{itemize}
\item[A.] the fermionic truncation: we consider a sub-module of $\mathscr{B}_\pm$ of ``small'' rank;
\item[B.] the angles are very non-generic. The math arguments do not apply;
\item[C.] the representation is twisted by the one-dimensional one given by the overall normalization factor $1/\tau(w_j)^N$ that has an important effect as \textbf{Example \ref{lllasq04}} shows. 
\end{itemize}    
\end{rem}

\subsection{Determing $\lambda(\theta)$}

Computing $\lambda(\theta)$ directly is hard and subtle.\footnote{\ See \textsc{appendix A} of \cite{twistor} for an example of how tricky the computation may be even in simple examples.} Therefore we shall take a different approach, namely try to fix it using the properties that it should have and consistency conditions. We fix the (discontinuous) function
$\lambda(\theta)$ mod 1. In order not to get confused by tricky issues of signs and bundle trivializations, we focus on the intrinsically defined quantity, $q(\theta)^2$, namely the ratio of the two distinct eigenvalues of
$\sigma_i^2\in\mathscr{P}_{n+M}$, i.e.\! of the operation of transporting one quasi-hole around another and getting back to the original position after a $2\pi$ rotation of their relative separation $w_i-w_j$.
For the connection \eqref{conclusion} one has
  \be
q(\theta)^2=\exp\!\big(2\pi i \,\lambda(\theta)\big).
\ee

Since the $tt^*$ geometry is very complete and symmetric between the quasi-holes,
we conclude that $\lambda(\theta)$ is a universal function which does not depend on $n$, $M$. Moreover we know that it must be piece-wise linear, i.e.\!
\be
\lambda(\theta)=C_1+ C_2\,\frac{\theta}{\pi}\bmod 1,
\ee
for some real constants $C_1$, $C_2$. We may assume $C_2>0$ by flipping the sign of $\theta$ if necessary.
Requiring $q(\theta)^2$ to satisfy the periodicity and ``reality'' conditions
\be
q(\theta+2\pi)^2=q(\theta)^2\qquad q(-\theta)^2=q(\theta)^{-2}, 
\ee
we get $2C_1=0\bmod 1$ and $2C_2=0\bmod1$. Imposing the same conditions on the ratio $q(\theta)$ of the eigenvalues of the braid generator $\sigma_i$ would give the stronger conditions $C_1=0\bmod 1$ and $C_2=0\bmod1$. The simplest solution to these conditions is
\be\label{kazq1t}
\lambda(\theta)=\frac{\theta}{\pi}\bmod1.
\ee
This identification is natural also from another point of view: we have that the monodromy is defined over $\Z[e^{\pm i\theta}]$, while from the KZ connection it is defined over $\Z[e^{\pm i\pi\lambda(\theta)}]$. 
Eqn.\eqref{kazq1t} just identifies these two Laurent polynomial rings.
\medskip

Comparing with eqn.\eqref{nutheta} we get
\be\label{rrqas}
q(\theta)^2=\exp\!\big(2\pi i \lambda(\theta)\big)=\exp(2\pi i/\nu).
\ee

\subsection{Comparison with Vafa's predictions}

\subsubsection{Allowed fractional filling levels $\nu$}\label{poq123}

First of all, let us see from the present viewpoint what singles out the principal series \eqref{prinseries} as ``preferred'' filling levels. Suppose the connection \eqref{conclusion} is an actually Knizhnik-Zamolodchikov connection for $SU(2)$ current algebra, with the level $\kappa$ properly quantized in integral units
\cite{KZ,wittenCS}. One has the identification\footnote{\ The minus sign in this formula arise from the minus sign in the \textsc{rhs} of eqn.\eqref{eig1}.}
\be\label{jjjawqg}
q(\theta)=-e^{\pm 2\pi i/(\kappa+2)},\qquad \kappa\in\Z.
\ee
Taking the square root of the two sides of
\eqref{rrqas} one has
\be
q(\theta)=e^{\pi i \lambda(\theta)}
\ee
and eqn.\eqref{jjjawqg} becomes
\be
\pm\frac{2}{\kappa+2}=1+\frac{\theta}{\pi}\bmod 2=\pm \frac{a}{b}\bmod2
\ee
which has solutions $a=1$ with $\kappa$ even and $a=2$ with $b$ and $\kappa$ odd.
The first case corresponds to the principal series with odd denominators
\be
\nu=\frac{b}{2b\pm1}\quad b\in\bN.
\ee
Since $\kappa$ is even, it is natural to think of the principal series as related to $SO(3)$ Chern-Simons rather than $SU(2)$ Chern-Simons. This is the more natural solution. But there are others.

The second case yields filling fractions with denominators divisible by $4$
\be
\nu=\frac{b}{2(b\pm1)},\quad b\ \text{odd}.
\ee  
On the other hand, we may consider the opposite (and less natural) solution to eqn.\eqref{rrqas}
\be
q(\theta)=-e^{\pi i \lambda(\theta)}
\ee
which implies
\be\label{kkkkasqwew}
1+\frac{2}{\kappa+2}=1+\frac{\theta}{\pi}\bmod 2=\frac{1}{\nu}\bmod2
\ee
that is,
\be
\nu=\frac{m}{m+2}\qquad m=\kappa+2\in\bN_{\geq 2}.
\ee
a series of filling fractions present in \cite{cumrun}
which contains the values of $\nu$ corresponding to the
Moore-Read
\cite{read1}
and the Read-Rezayi models \cite{read2}.
There is yet another possibility, namely we may replace  \eqref{kkkkasqwew} by
\be
3-\frac{2}{\kappa+2}=\frac{1}{\nu}\bmod2,
\ee 
which yields the solutions ($m=\kappa+2$)
\be
\nu=\frac{m}{3m-2},\qquad m\geq 2.
\ee

\begin{rem} Even if we do not know any compelling argument from the $tt^*$ side  to require $\kappa+2\in\Z$, this condition is certainly part of the definition of ``good'' Knizhnik-Zamolodchikov $\mathfrak{su}(2)$ connections, and we are pretty willing to believe that it is a necessary condition for consistency. Thus we conjucture that the above list of filling fractions is complete
as long as $\nu\leq 1$. 
\end{rem}

\subsubsection{Non-abelian statistics
(principal series)}

From the point of view of \S.\,2 of \cite{cumrun} 
the element  $\sigma_i^2$ of the pure braid group for the principal series has two distinct eigenvalues, in correspondence with the two different fusion channels of the $\phi_{1,2}$ operator in the minimal $(2n,2n\pm1)$ Virasoro model. The ratio of the two eigenvalues is
 \be
 q^2=\frac{\exp[2\pi i (h_{1,3}-2 h_{1,2}))]}{\exp[2\pi i (h_{1,1}-2 h_{1,2})]} =\exp(2\pi i\, h_{1,3})=\exp(2\pi i/\nu),  
 \ee
 which coincides with our equation \eqref{rrqas} deduced from the $tt^*$ geometry.
Thus we reproduce Vafa's result\footnote{\ Up to the interchange $\phi_{1,3}\leftrightarrow\phi_{3,1}$ in the last equation on page 6 of \cite{cumrun}.}.

\subsection{The emergent unique ground state} 

As discussed at the end of \S.\,\ref{staHecke}, we have a unique preferred  vacuum invariant under parallel transport by the UV Berry connection. We wish to identify it with the unique physical vacuum $|\textsf{vac}\rangle$ of the FQHE quantum system when all details of the Hamiltonian $H$ are taken into account, including the non-universal interaction $H_\text{int}$ (that is, the true vacuum is a topological trivial deformation of $|\Omega\rangle$). 

From the viewpoint of the spin-chain with state space the $2^{M^\prime}$-dimensional vector space $\boldsymbol{V}^{\otimes M^\prime}$ ($M^\prime\equiv M+n$), the preferred vacuum
is the state $|\Psi\rangle$ such that
\be\label{jjjjjx}
|\Psi\rangle\in \boldsymbol{V}^{\otimes M^\prime},\qquad L_3|\Psi\rangle=\tfrac{2N-M^\prime}{2}|\Psi\rangle
\ee
which maximizes
\be
\frac{\|L^2|\Psi\rangle\|}{\||\Psi\rangle\|}.
\ee
Between the states satisfying \eqref{jjjjjx},
$|\Omega\rangle$ is the most symmetric under permutations of the spin degrees of freedom. The most symmetric linear combinations of the idempotents $e_i$ is
their sum $1=e_1+\cdots +e_{w}$. However, we have twists by signs, so
we can conclude only that 
the preferred vacuum corresponds to an element $\rho$ of the chiral ring of the form
$\pm e_1\pm e_2\pm\cdots\pm e_w$ for some choice of sign. One has $\rho^2=1$, and if an usual LG model with intere superpotential this means $\rho=1$.
That this applies to the present case is less obvious.

Anyhow $|\Omega\rangle=|\rho\rangle$
is the most symmetric vacuum. As long as 
the interaction $H_\text{int}$ preserves the symmetry between the holes and the units of fluxes, it lifts the degeneracy keeping the most symmetric state as the ground state. So it is natural to think of $|\Omega\rangle$ as the true ground state of the FQHE system.

\section{Conclusions}

In this paper we studied the supersymmetric quantum many-body system proposed by Vafa as a microscopic description of the Fractional Quantum Hall Effect from the perspective of $tt^*$ geometry.

Albeit our arguments are not fully mathematically rigorous (and improvements are welcomed)
our ``exact''  methods lead to an elegant and coherent picture which agrees with
physical consideration from several alternative viewpoints. In particular they agree and strengthen the results of \cite{cumrun};
it also make stronger the case for the 4-supercharge Vafa Hamiltonian to represent the correct universality class of the fundamental many-electron theory.
Indeed we argued that any Hamiltonian describing the motion in a plane of many electrons coupled to a strong magnetic field are described (at the level of topological order) by Vafa's 4-\textsc{susy} \textit{independently of the details of the interactions between the electrons}. 
It is remarkable that one can show that the electron filling fractions $\nu$ of any such quantum system should be a rational number belonging to one of the series in \S.\,\ref{poq123}. Of course, this is a manifestation of the universal nature of the topological quantum phases.   
\medskip

It is well-known that 3d Chern-Simons is a good effective description of the FQHE. 
From our present perspective this is quite obvious: the geometric structures we found (Kohno connections, Hecke algebras and all that) are the essence of Chern-Simons theory. The nice aspect is that we started from the ``obviously correct'' quantum description of the FQHE systems in terms of the many-body Schroedinger equation describing $N$ electrons coupled to a strong magnetic field and interacting between them in some ``generic'' way, and ended up with the Chern-Simons-like structure as an ``exact'' IR description.  

\subsection*{Acknowledgments}

We thank Cumrun Vafa for many stimulating discussions about his model of FQHE.

\appendix

\end{document}